\documentclass[a4paper,fleqn,usenatbib,mn2e,article]{mnras}
\usepackage{newtxtext,newtxmath}
\usepackage[T1]{fontenc}
\usepackage{ae,aecompl}
\usepackage{amsfonts, natbib, txfonts,epsfig,rotating,txfonts,mathtools}

\usepackage{graphicx}	
\usepackage{amsmath}	
\usepackage{amssymb}	
\usepackage{pdflscape,tabularx}

\usepackage{footnote}

\title{The  structure of the \ion{Mg}{ii} broad line emitting region\\ in Type 1 AGNs}

\author[Popovi\'{c} et al.]{
Luka \v C. Popovi\'{c},$^{1,2}$\thanks{E-mail: lpopovic@aob.rs}
Jelena Kova\v{c}evi\'{c}-Doj\v{c}inovi\'{c},$^{1}$
\& Sladjana Mar\v ceta-Mandi\'c$^{1,2}$
\\
$^{1}$Astronomical  Observatory,  Volgina  7, 11060  Belgrade, Serbia\\
$^{2}$Department of Astronomy, Faculty of Mathematics, Univeristy of Belgrade, \\ Studentski Trg 16, 11000 Belgrade, Serbia\\
}

\pubyear{2017}

\begin{document}
\label{firstpage}
\pagerange{\pageref{firstpage}--\pageref{lastpage}}
\maketitle

\begin{abstract}
We investigate the structure of the \ion{Mg}{ii} broad line emission region in a  sample of 284 Type 1 active galactic nuclei (AGNs), through comparing the kinematical parameters of the broad \ion{Mg}{ii} and broad H$\beta$ lines. We found that the  
 \ion{Mg}{ii} emitting region has more complex kinematics than the H$\beta$ one. It seems that the \ion{Mg}{ii} broad line originates from  two subregions: one which contributes to the line core, which is probably  virialized, and the other, 
'fountain-like' emitting region, with outflows-inflows nearly orthogonal to the disc, which become suppressed with stronger gravitational influence. This subregion mostly contributes to the  emission of the \ion{Mg}{ii} broad line wings. The kinematics of the \ion{Mg}{ii} core emitting region is similar to that of  the H$\beta$ broad line region (seems to be virialized), and therefore  the Full Width at Half Maximum (FWHM) of \ion{Mg}{ii} still 
can be used for the black hole (BH) mass estimation in the case where the \ion{Mg}{ii} core component is dominant. However, one should be careful with using the \ion{Mg}{ii} broad line for the BH mass estimation in the case of very large widths (FWHM$>$ 6000 km s$^{-1}$) and/or in the case of strong blue asymmetry.

\end{abstract}


\section{Introduction} \label{sec:intro}

The broad emission lines (widths about several 1000 km s$^{-1}$) are one of the most important characteristics
of Type 1 
active galactic nuclei (AGN) spectra. They are originating in a broad line region (BLR) that is supposed 
to 
be close to the central black hole (BH) and consequently, one can assume that the 
BLR emission gas kinematics  is virialized, i.e. it is following the gravitationally driven rotation
\citep[see e.g.][etc.]{su00,ga09,ne15}. The gas motion in the BLR affects 
the line profile \citep{su00}, and, in the case of Keplerian-like motion, the 
broad line width 
(Full Width at Half Maximum -- FWHM) can be used for the
BH mass estimation \citep[for review see][]{pet14}.

There are several methods for the BH mass estimation \citep[direct and indirect, see e.g.][]{pet14}.
Among them, the reverberation method is  the most frequently used one.
Using reverberation, one can  determine  the BLR size, then assuming the virialization in the 
BLR and measuring the FWHM of a broad line, 
the BH mass can be obtained \citep[][]{pet93}.
Reverberation  was also used for establishing the relationships 
between the BLR size and the 
continuum luminosity \citep[see e.g.][]{vp06,be13}, which is widely used for the BH mass calculations from one epoch observations.

For this purpose, there are several 
relationships (in different wavelength bands)
between the black hole mass,
continuum luminosity and broad line widths, which are defined assuming the virialization in the  BLR 
\citep[see e.g.][etc.]{vp06,on08,vo09,wa09,tn12,ti13,me16,co17,me18}. 
Depending on the redshift of an AGN, there is a possibility to use different broad lines, and the most frequently used are H$\beta$,
\ion{Mg}{ii} and \ion{C}{iv} \citep[see][]{me16}.

The \ion{Mg}{ii} relationship for the BH mass estimation is derived from the H$\beta$ relationship 
\citep[see e.g.][]{on08,wa09,vo09,m2013,su14,me16} based on the reverberation measurements of the H$\beta$ BLR and 
the continuum luminosity at 5100 \AA. The \ion{C}{iv} line parameters, and consequently the estimated BH mass can be compared with
those obtained from the H$\beta$ and \ion{Mg}{ii} relationship \citep[see e.g.][]{me16,co17,me18}.
However, one of the problems of this method is the assumption that the BLR gas is virialized, 
particularly for the \ion{Mg}{ii} and \ion{C}{iv} line emitting regions \citep[see][]{ma12,le13}.
The broad \ion{C}{iv} and \ion{Mg}{ii} line profiles can be affected by some other effects, as 
e.g. outflows \citep[see][]{de12,le13}.
Therefore, there is a question of validity of using these lines and their robustness for the BH mass measurements.
 Since \ion{C}{iv} and \ion{Mg}{ii} lines are very important as BH mass estimators in high redshifted AGNs, it is of great importance to 
investigate the structure of their broad line emitting regions and to compare it with the H$\beta$ one for 
a number of AGNs. Here we will focus on the \ion{Mg}{ii} broad line emitting region, comparing 
the \ion{Mg}{ii} line shape  with the H$\beta$ line profile in order to explore similarities and differences between their emitting regions.

In principle, the \ion{Mg}{ii} 2800 \AA\ line can be directly compared with the 
H$\beta$ and the 
consistency of the obtained BH mass using these two lines can be checked 
\cite[see][]{wa09,m2013,m2013b,me16}. 
However, it seems that the \ion{Mg}{ii} emitting line region is more complex \citep[][]{kp15,jon16}, 
and in some cases it can be connected with outflows \citep[see, e.g.][]{le13}.
There is an indication that in the case of smaller widths of \ion{Mg}{ii} (FWHM$<$6000 km s$^{-1}$) \
there is a good correlation with the H$\beta$ width, and therefore the \ion{Mg}{ii} is equally good for 
BH mass estimation as H$\beta$ \citep{tn12}. 
 On the other hand, for FWHM \ion{Mg}{ii}$>$6000 km s$^{-1}$, the 
difference between the FWHM of these two lines seems to be  significant, and there 
is a question of validity of
using \ion{Mg}{ii} line for black hole mass measurements. 
\cite{s2000} identified two populations of AGNs according to different spectral properties: Population A (Pop. A), with FWHM of H$\beta$ $<$ 
 4000 km s$^{-1}$ and Population B (Pop. B), with FWHM H$\beta$ $>$ 4000 km s$^{-1}$. \cite{m2013} found that in Pop. B, the 
\ion{Mg}{ii} lines are about 20\% narrower than H$\beta$ and  they have more complex shape, in which one very broad, redshifted component is seen.

The aim of this work is to investigate the structure of the \ion{Mg}{ii} broad line 
emitting region and to compare its 
kinematics with the kinematics of the broad H$\beta$ emitting region, in order to check the validity of using the \ion{Mg}{ii} broad line as a BH mass estimator.

The paper is organized as following: In Sec. \ref{sec:sample} we describe the sample of AGNs and give the method of analysis.
In Sec.\ref{sec:results}  we explore the \ion{Mg}{ii} line profiles in the sample of AGNs and compare the line parameters with those obtained from H$\beta$. 
In Sec. \ref{sec:discussion} we discuss the obtained results and possible model for the \ion{Mg}{ii} BLR, and finally  in Sec. \ref{sec:conclusions} we outline our conclusions.

\section{The sample and method of analysis} \label{sec:sample}

\subsection{The sample}

To investigate the assumption of virialization in the \ion{Mg}{ii} and H$\beta$  broad line emitting 
regions, we
used already studied sample of
293 Type 1 AGN spectra from \citet{kp15}. 
This sample was selected from the Sloan Digital Sky Survey (SDSS), Data Release 7 and  {has}
S/N $>$ 25 (near H$\beta$ line), good pixel quality, redshift between 0.407 and 0.643 (in order to  {include} both Mg 
II and H$\beta$ lines), high redshift confidence, and  {no} absorption in the UV \ion{Fe}{ii} 
lines. 
 
Additionally, we excluded the objects with strong noise near the \ion{Mg}{ii} line, and with absorption lines which affect the \ion{Mg}{ii} profile. The final sample consists of 284 objects. 

\begin{figure}
\centering
\includegraphics[width=0.45\textwidth,angle=-90]{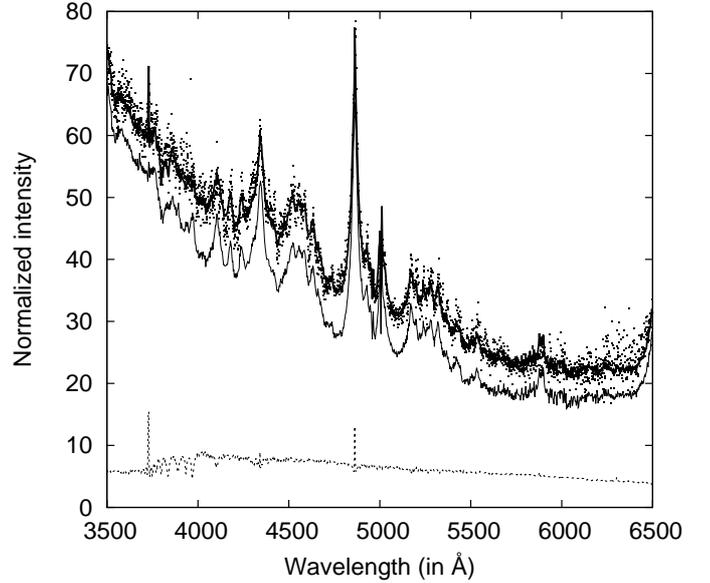}

\caption{ An example of the spectral decomposition to pure-host and pure-QSO contribution, using
Spectral Principal Component Analysis (object: SDSS J095912.93+445059.1). Dots -
observation, thin solid line - QSO, dotted line - host galaxy and thick solid line - model (QSO+host). }
\label{fig1}
\end{figure}

\begin{figure}
\includegraphics[width=0.45\textwidth]{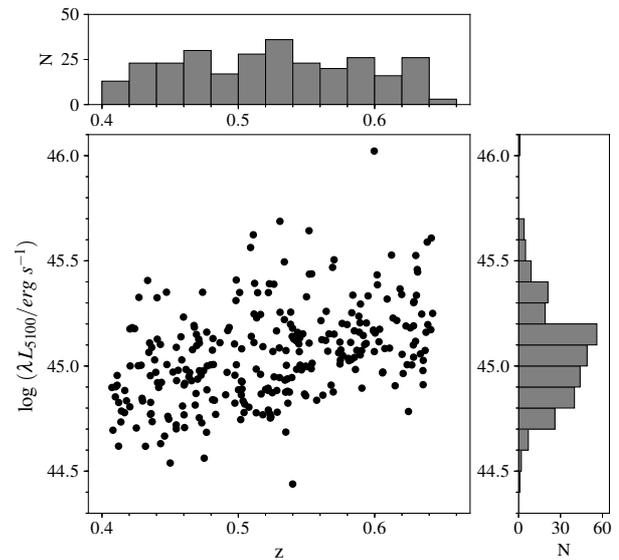}
\caption{ The AGN luminosity (after subtraction of the host galaxy contribution) vs. redshift of the sample and their 
distributions. 
 }
\label{fig2}
\end{figure}

\begin{figure}
\includegraphics[width=0.45\textwidth]{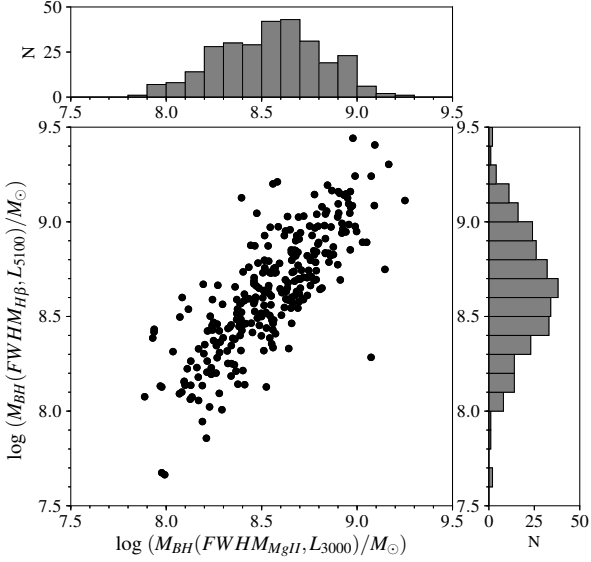}
\caption{ Estimated BH masses from the broad H$\beta$ line and continuum at 5100 \AA\ vs. those obtained from 
the \ion{Mg}{ii} broad line and continuum
at 3000 \AA\ and their distributions.
 }
\label{fig2a}
\end{figure}

\subsection{The analysis}

The spectra were corrected for Galactic extinction and cosmological redshift as described in \cite{kp15}. 
In \cite{kp15} the host galaxy contribution is was assumed to be negligible in this sample since majority of objects 
have high continuum luminosity (see their Fig 1). However, in order to estimate more precisely the continuum luminosity,
here we applied the Spectral Principal Component Analysis method \citep[see][]{co1995,yip04a,yip04b} to subtract the
host contribution \citep[see][]{vb2006} and obtain the pure AGN continuum. For the subtraction of the host-galaxy 
contribution, we followed the procedure described in more detail in \cite{l2017}.  
We found that in the majority of spectra the host-galaxy component is negligible \citep[as it was assumed in][]{kp15}. 
The significant host contribution was found only in 43 objects. For majority of these 43 objects, the host contribution
is smaller than 15\% at 5100 \AA. The example of spectral decomposition to pure-host and pure-AGN contribution is shown in Fig \ref{fig1}. 
After the host galaxy subtraction, the pure QSO continuum flux is measured at 5100 \AA, 
and continuum luminosity is  calculated using the formula given in \cite{Peebles93}, and
the cosmological parameters $\Omega_M$=0.3, $\Omega_\Lambda$=0.7 and $\Omega_k$=0, and Hubble constant 
$\rm H_{0}$=70 km s$^{-1}$ Mpc$^{-1}$. 
The  host-corrected continuum luminosity at 5100 \AA \ vs. redshift and their distributions are shown in Fig \ref{fig2}. 

The procedure for the fitting of the AGN spectra is given in \cite{kp15}.
After the continuum subtraction, we applied the multi-Gaussian emission 
line model, in which the complex line shapes are fitted with Gaussians of different widths, 
shifts and intensities, with the assumption that each Gaussian represents the emission
from one emission region \citep[see][]{pop04, kov10}.
 To reduce the number of free parameters in the fitting procedure, we assumed that the 
width and  shift of each 
Gaussian are connected with the kinematical properties of the emission region
where 
the corresponding
line component arises. As e.g. the narrow H$\beta$ and [\ion{O}{iii}] lines are assumed to be emitted from the same region, so the velocity dispersion of the Gaussian of the narrow H$\beta$ and [\ion{O}{iii}] are assumed to be same, as well their shifts. In this way, all lines or line components, which are supposed to originate from the same emission 
line region, are fixed to have 
the same widths and shifts \citep[see][]{pop04}.

For this research, it is necessary to extract the pure broad H$\beta$ and \ion{Mg}{ii} profiles from the complex AGN spectra, 
which is a very difficult task, with some level of uncertainties. The biggest problem is to remove the overlapping lines, 
as e.g. \ion{Fe}{ii} lines in the optical (around H$\beta$) and in the UV (around \ion{Mg}{ii}), and the narrow component of the H$\beta$ line.

The model of the line decomposition and the procedure of the spectral fitting in optical $\lambda\lambda$ 4000-5500 
\AA \ and UV $\lambda\lambda$ 2650-3050 \AA \ band are presented in detail in \cite{kov10}, and \cite{kp15}, and here we just 
recall the important steps for obtaining the broad  H$\beta$ and \ion{Mg}{ii} lines.

\begin{figure}
    \includegraphics[width=0.43\textwidth]{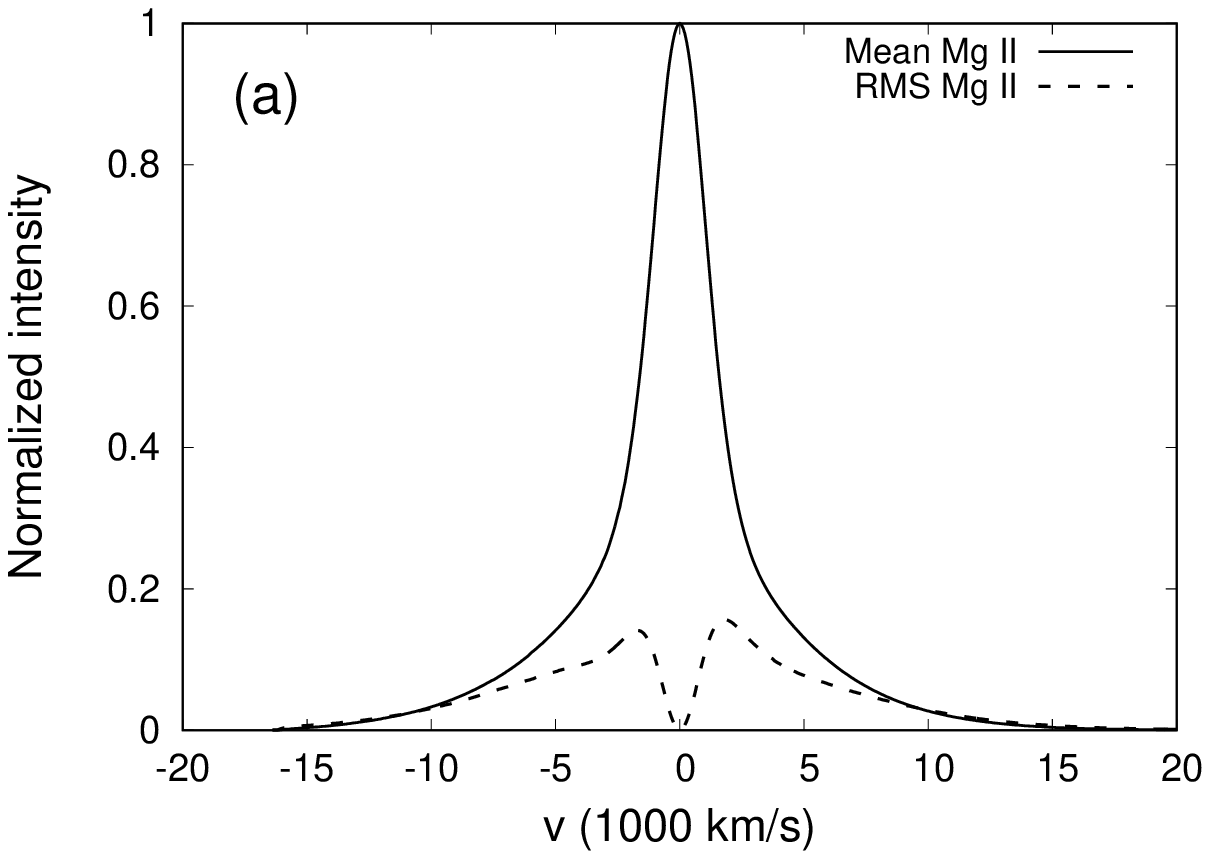}
    \includegraphics[width=0.43\textwidth]{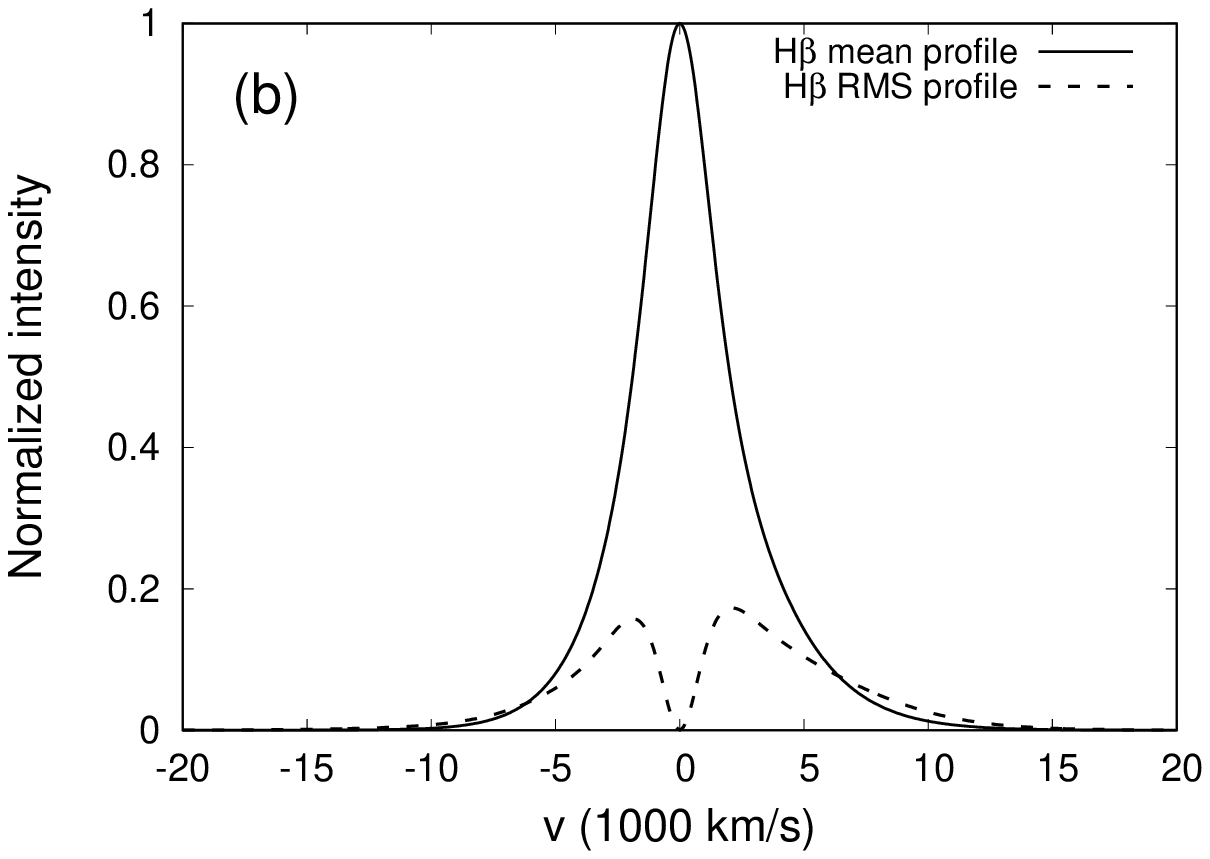}
\includegraphics[width=0.43\textwidth]{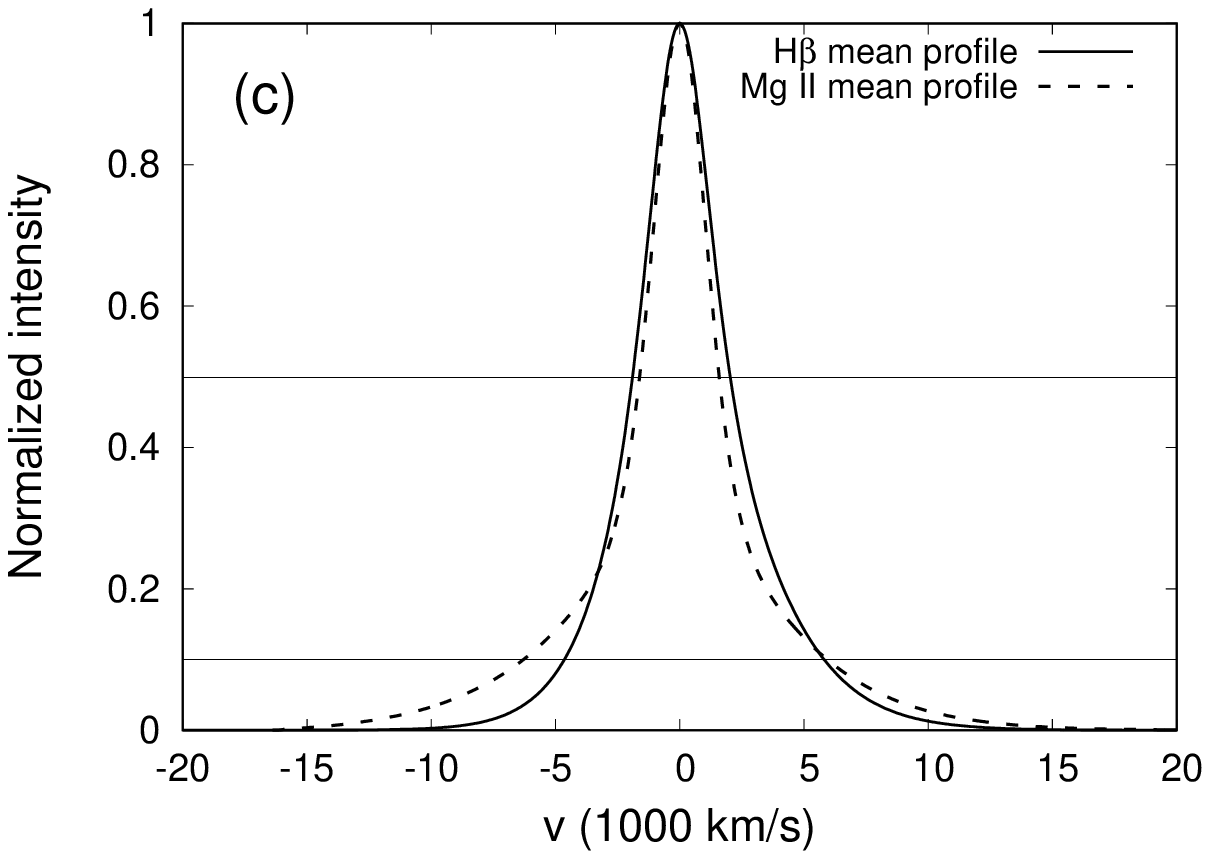}
\includegraphics[width=0.43\textwidth]{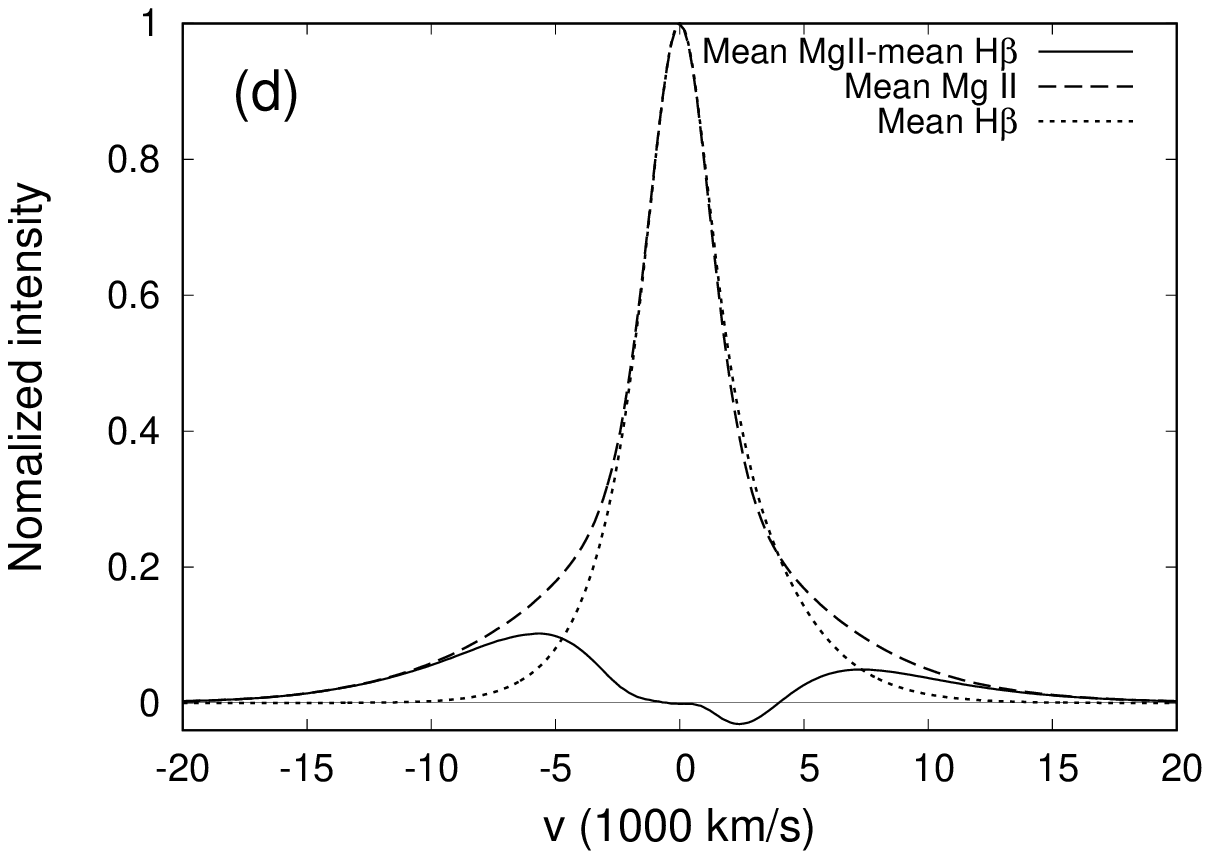}
\caption{ Panel (a): the mean (solid line) and RMS (dashed line) \ion{Mg}{ii} profiles of 284 AGNs. Panel (b): the same for H$\beta$. 
Panel (c): the comparison of the H$\beta$ (solid line) and \ion{Mg}{ii} (dashed line) mean profiles.
The horizontal solid lines indicate the widths at half and 10\% of the maximal intensity.
Panel (d): the \ion{Mg}{ii} broad line (dashed line) rescaled to have the same FWHM as the H$\beta$ (dotted line) and
their difference (solid line).}
\label{f01}
\end{figure}

\subsubsection{The pure broad H$\beta$ profile}

The broad H$\beta$ line overlaps with the numerous optical \ion{Fe}{ii} lines, [\ion{O}{iii}] 
$\lambda\lambda$ 4959, 5007 \AA \ doublet and narrow
H$\beta$ component.

In the line decomposition model, the complex optical \ion{Fe}{ii} bumps are fitted with the \ion{Fe}{ii} 
semi-empirical model\footnote{The 
template is available through Serbian Virtual Observatory (SerVo).
\url{http://servo.aob.rs/FeII\_AGN}.} presented in \cite{kov10} and \cite{sh2012}. 
This \ion{Fe}{ii} model enables very precise fitting of the \ion{Fe}{ii} bumps in the 
4000-5500 \AA \ spectral range \citep[see appendix in][]{kov10}. 
The Balmer lines  in the range $\lambda\lambda$ 4000-5500 \AA  \ (H$\delta$, H$\gamma$ and H$\beta$) are 
fitted with three components - 
one narrow and two broad components, 
one which comes from the Very Broad Line Region (VBLR) and fits the line wings, and the other
which comes from the Intermediate 
Line Region (ILR) and fits the core of the Balmer lines \citep[see][]{pop04, bo2006, hu2008}. 
The widths and shifts of each line component are thus the same for all considered Balmer lines,
which reduces the uncertainties in 
the H$\beta$ line decomposition. 

In some cases, the Balmer narrow lines are very difficult to resolve from the broad line component, especially in so-called 
Pop. A objects \citep[see][]{s2000}. On the other hand, the [\ion{O}{iii}] lines are well resolved in the majority of spectra. Therefore, 
the narrow Blamer lines are fixed to have the same width and shift as [\ion{O}{iii}] lines, since we are assuming that they arise in 
the same Narrow Line Region (NLR), and thus have the same kinematical properties. 
In this way, the number of free parameters in the fitting procedure
is reduced and more accurate fitting decomposition is achieved. 

Nevertheless, there is still some doubt about the non-uniqueness of the 
H$\beta$ line decomposition to the broad and narrow component.
\cite{Popovic11} tested the uniqueness of the H$\beta$ decomposition with this model, and discussed it in the Appendix A 
(in the same paper). As it can be seen from \cite{Popovic11} the method we applied can 
reasonable fit the narrow H$\beta$ component, and consequently the narrow H$\beta$ component can be 
subtracted in order to 
find the pure H$\beta$ broad component.

\begin{figure}
\centering
\includegraphics[width=0.47\textwidth]{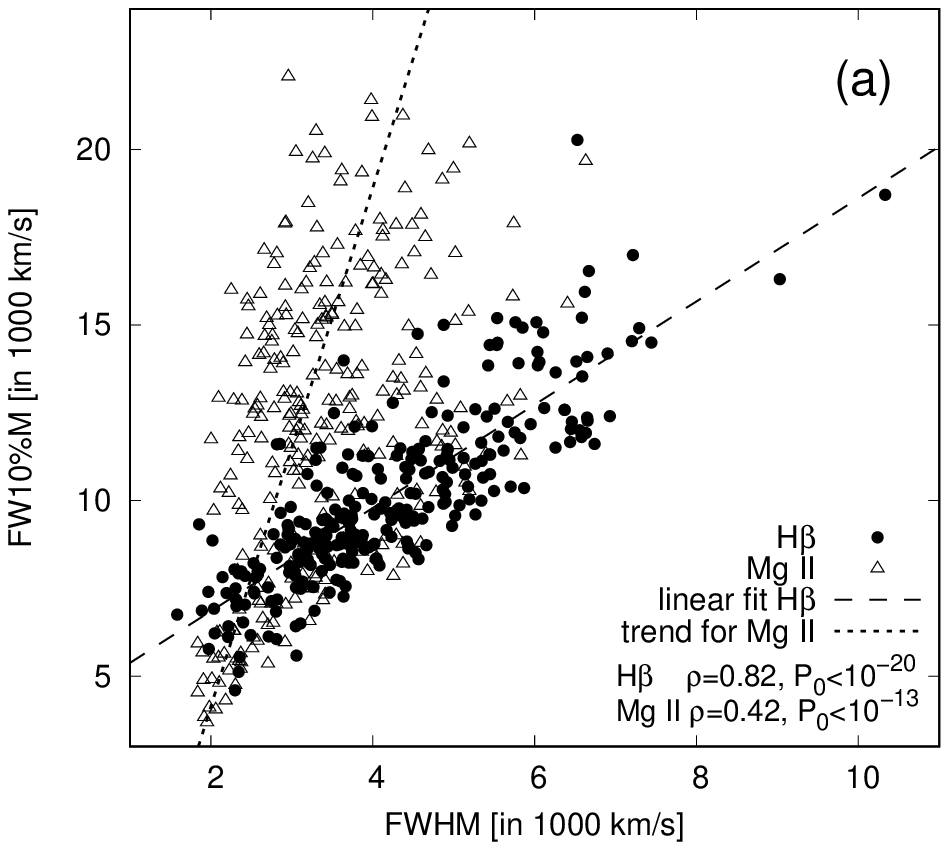}
\includegraphics[width=0.47\textwidth]{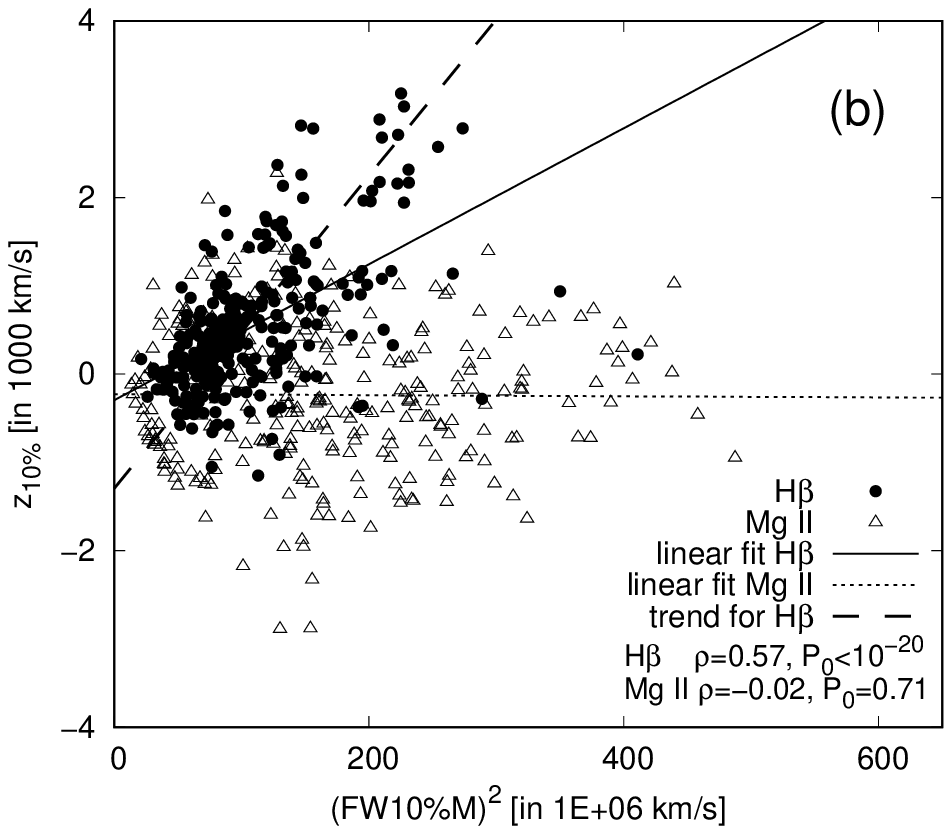}
\caption{Panel (a): the relationship between the widths of the line measured at 10\% and 50\% of the maximal line intensity (FW10\%M vs. FWHM). Panel (b): the relationship between the width and intrinsic shift measured at 10\% of the maximal line intensity (z$_{10\%}$ vs. (FW10\%M)$^{2}$). The triangles denote the \ion{Mg}{ii} and full circles H$\beta$ data. 
}
\label{f02}
\end{figure}

\subsubsection{The pure broad Mg II $\lambda$ 2800 \AA \ profile} \label{sec:2.2.2}

In the UV part of AGN spectrum, the \ion{Mg}{ii} line is contaminated with the Balmer continuum and numerous UV \ion{Fe}{ii} lines.
In addition, the \ion{Mg}{ii} line is a doublet with two components which overlap: 
$\lambda$ 2796 \AA \ and $\lambda$ 2803 \AA, that makes the extraction of the single broad profile even more complicated. The ratio of the component intensity is unknown, 
and it depends on the optical depth of the \ion{Mg}{ii} emitting region.
In the optically thin case, the ratio Int(\ion{Mg}{ii} $\lambda$ 2796 \AA)/Int(\ion{Mg}{ii} $\lambda$ 2803 \AA) is expected to be 2:1,
while in the case of fully thermalized lines it should be 1:1 \citep{la1997, m2013}.
For expected values of parameters in the AGN emission region (hydrogen density and ionization parameter) this ratio should be
close to 1:1 \citep{m2013}.
Taking into account that  the doublet transition wavelength separation is $\sim$ 8 \AA \ ($\sim$260 ${\rm km s^{-1}}$) and that the 
total line width is one order of magnitude larger, \ion{Mg}{ii} can be consi\-de\-red as a single line. Therefore, we assumed that  
\ion{Mg}{ii} profile and FWHM \ion{Mg}{ii} are not affected by doublet separation.

Note here that some authors apply the fitting procedure of the \ion{Mg}{ii} with broad and narrow line components 
\citep[see][]{md2004,wa09,sl2012}.  However, the \ion{Mg}{ii} abundance and the transition probability for the 
NLR conditions are too small, so the narrow \ion{Mg}{ii} component seems to be absent or very weak and can be neglected
\citep[][]{wa09,m2013, kp15}. Moreover, by visual inspection of the \ion{Mg}{ii} lines in our sample
we could not find any spectrum with the prominent narrow \ion{Mg}{ii} line. Therefore, we neglected the \ion{Mg}{ii} narrow line contribution 
to the broad \ion{Mg}{ii} component.

For the Balmer continuum, we used the model given in \cite{kov14}.

The UV \ion{Fe}{ii} lines around the \ion{Mg}{ii} are complex, and there are several theoretical and empirical
templates of the UV \ion{Fe}{ii} lines
\citep[see e.g. ][]{bk80,vw01,ts06,bv08,kp15}. We tested several templates and 
found that (see more 
details in Appendix A): (i)
the theoretical templates could not fit well the UV \ion{Fe}{ii} lines; 
(ii) the empirical templates are 
based on I Zw 1 \ion{Fe}{ii} emission, which is narrow-line Seyfert 1 galaxy, and one cannot expect that in the case of the
 broad line AGNs the 
ratio between the UV \ion{Fe}{ii} multiplets is the same as in the case of
 I Zw 1 \citep[as in the 
case of the optical \ion{Fe}{ii} lines, see][]{kov10,Popovic11}.

To solve this problem, we developed a semi-empirical model of the UV \ion{Fe}{ii} lines based on the 
model given in \cite{kp15}, where the \ion{Fe}{ii} multiplet intensities are 
free parameters, but the 
line ratios within one \ion{Fe}{ii} multiplet are fixed, taking that within a multiplet the intensity 
ratio of two lines is:

\begin{equation}
 {I_1\over I_2} $=$ {[{\lambda_2\over\lambda_1}]}^3 \cdot\frac{f_1}{f_2}\cdot
 {g_1\over g_2}\cdot e^{-[E1-E2]/kT}
 \label{eq:1}
\end{equation}
where $I_1$ and $I_2$ are intensities of the lines with the same 
lower level of transition, $\lambda_1$ and $\lambda_2$ are line wavelengths, 
$g_1$ and $g_2$ are corresponding statistical weights of the upper levels, 
and $f_1$ and $f_2$ are oscillator strengths, 
$E_1$ and $E_2$ are energies of the upper level of transitions,
$k$ is Boltzmann constant, and $T$ is the excitation temperature.

Taking that in our case the $\lambda_2\sim\lambda_1$, and $E_1\sim E_2$, an 
approximation can be used as: 

\begin{equation}
{I_1\over I_2}\sim {f_1\over f_2}\cdot{g_1\over g_2}
 \label{eq:2}
\end{equation}
as in the paper of \cite{kp15}. Here we use Eq. (\ref{eq:1}), to calculate the 
values of the line ratios within multiplets in Appendix A. In Appendix A we give
a detailed discussion about  
 the UV \ion{Fe}{ii} semi-empirical model as well as comparison of this model with
 theoretical and empirical templates. Additionally, using our model we calculated a grid of the UV \ion{Fe}{ii} templates in the  2650-3050 \AA \ range with different FWHM and shifts of the UV \ion{Fe}{ii} lines.\footnote{ The grid of the UV \ion{Fe}{ii} templates is available through SerVo: \url{http://servo.aob.rs/FeII\_AGN/link7.html}}.
 
 The \ion{Mg}{ii} line was fitted with two broad Gaussians: one which fits the core, 
 and one which fits the line wings. These two broad Gaussians reproduce well the \ion{Mg}{ii} 
 line profiles in our sample.

 \subsubsection{Measuring the broad line parameters and black hole masses}

 The \ion{Mg}{ii} and H$\beta$ broad line shapes have been reproduced from the best fit (from two Gaussians),
 then normalized to one, and 
 FWHM and Full Width at 10\% of the Maximal Intensity (FW10\%M) were measured.  The corresponding
intrinsic shifts (z$_{50\%}$ and z$_{10\%}$) are measured as well, as a centroid shift with respect to the broad 
line peak at
50\% and 10\% of Maximal Intensity, respectively \citep[see][]{jon16}.

The black holes masses (M$_{\rm BH}$) 
are estimated from relationships of \cite{vp06} which uses the 
H$\beta$ parameters, and \cite{vo09}
which uses the parameters of \ion{Mg}{ii} line. 
In Fig. \ref{fig2a} we show 
the  BH masses derived from H$\beta$ vs. those derived from \ion{Mg}{ii} line and their distributions. As it can be 
seen in Fig. \ref{fig2a}, there is strong correlation between the BH masses derived from these two lines, and their distributions
are very similar.

\subsubsection{The mean and RMS line profiles}

To explore the \ion{Mg}{ii} and H$\beta$ line profiles, we 
found the mean profiles of the normalized H$\beta$ and \ion{Mg}{ii} lines from the
sample and the root-mean-square -- RMS which indicates the difference
in the line profiles (see Fig. \ref{f01}ab). 
Comparing the mean \ion{Mg}{ii} and H$\beta$ line profiles
we found that the line cores are very similar, and that the difference is seen in the line wings (Fig. \ref{f01}c). 

The slight difference in the line cores is caused by a smaller FWHM \ion{Mg}{ii}. 
Therefore, to subtract two mean profiles, 
we first rescaled \ion{Mg}{ii} to have the same FWHM as H$\beta$ for each AGN (Fig. \ref{f01}d). As the result of subtraction, a double peaked feature appeared (see solid line in Fig. \ref{f01}d). It seems that there is 
an extra emission in the far (blue and red) \ion{Mg}{ii} wings compared to the H$\beta$ ones, with a bigger difference in the blue wings (see Fig. \ref{f01}d).

\subsection{Connections between the \ion{Mg}{ii} and H$\beta$ line widths/shifts -- virialization assumption}

 To test the virialization in the \ion{Mg}{ii} and H$\beta$ line emitting regions we used the
line parameters: intrinsic shifts and FWHMs of all AGNs in the sample. 
The FWHM can be related with  
the line dispersion sigma, which is the second moment or variance of the velocity distribution (as e.g. 
is the case for Gaussian-like profile). In particular
case where we assume a Keplerian-like motion, the FWHM is directly connected with the velocity and
with the emissivity weighted radius $R_{\rm FWHM}$, so the mass of the central black hole can be estimated from
\citep[see e.g.][]{pet14}:
\begin{equation}
M_{\rm BH}\sim {\rm FWHM}^2\cdot R_{\rm FWHM}
\label{eq:3}
\end{equation}
Therefore, in the virialization assumption,  the rotating velocity is strongly connected with the FWHM.

Assuming the Keplerian motion, one can write the following:

\begin{equation}
M_{\rm BH}$=$R_{\rm BLR} \cdot v^2/G
\label{eq:7}
\end{equation}
where $R_{\rm BLR}$ is the size of the rotating BLR and $v$ is the velocity of the rotation.

  Concerning the line intrinsic shift (asymmetry), it can be 
caused due to a number of effects, but in the case of full virialization, one can 
expect the connection between the intrinsic line shift and FWHM.

Namely, if we take that the BH mass can be measured using the gravitational shift ($z_g$) as \citep{liu17}: 
\begin{equation}
M_{\rm BH}\sim R_{\rm BLR}\cdot z_g\sim R_{\rm BLR} {\rm FWHM}^2,
\label{eq:8}
\end{equation}
this indicates that a good correlation should be present between $z_g$ and ${\rm FWHM}^2$. 
Since we measured the intrinsic shift at FWHM ($z_{50\%}$) and FW10\%M ($z_{10\%}$), in the case of virialization in the total line we expect:
\begin{equation}
z_{50\%}\sim {\rm FWHM}^2,\ \ \ \ \ \ \ \ \ z_{10\%} \sim {\rm FW10\%M}^2
\label{eq:9}
\end{equation}
where $z_{50\%}$ and $z_{10\%}$ are measured intrinsic shifts at FWHM and FW10\%M \citep[see][]{jon16}.

Using Eq. (\ref{eq:9}) we  explored the virialization in the \ion{Mg}{ii} and H$\beta$ emitting regions.

\begin{figure*}
\includegraphics[width=0.45\textwidth,angle=0]{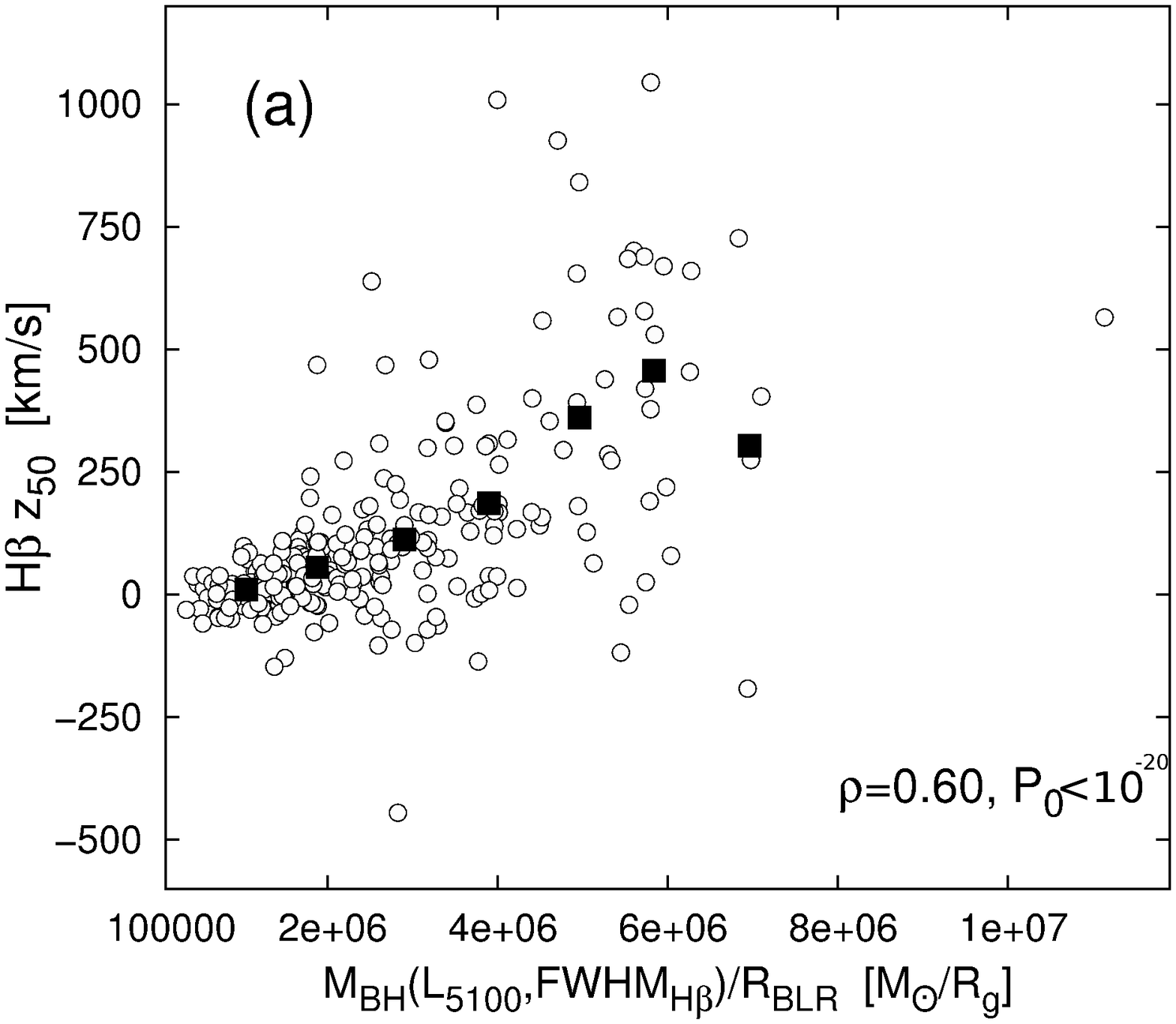}
\includegraphics[width=0.45\textwidth,angle=0]{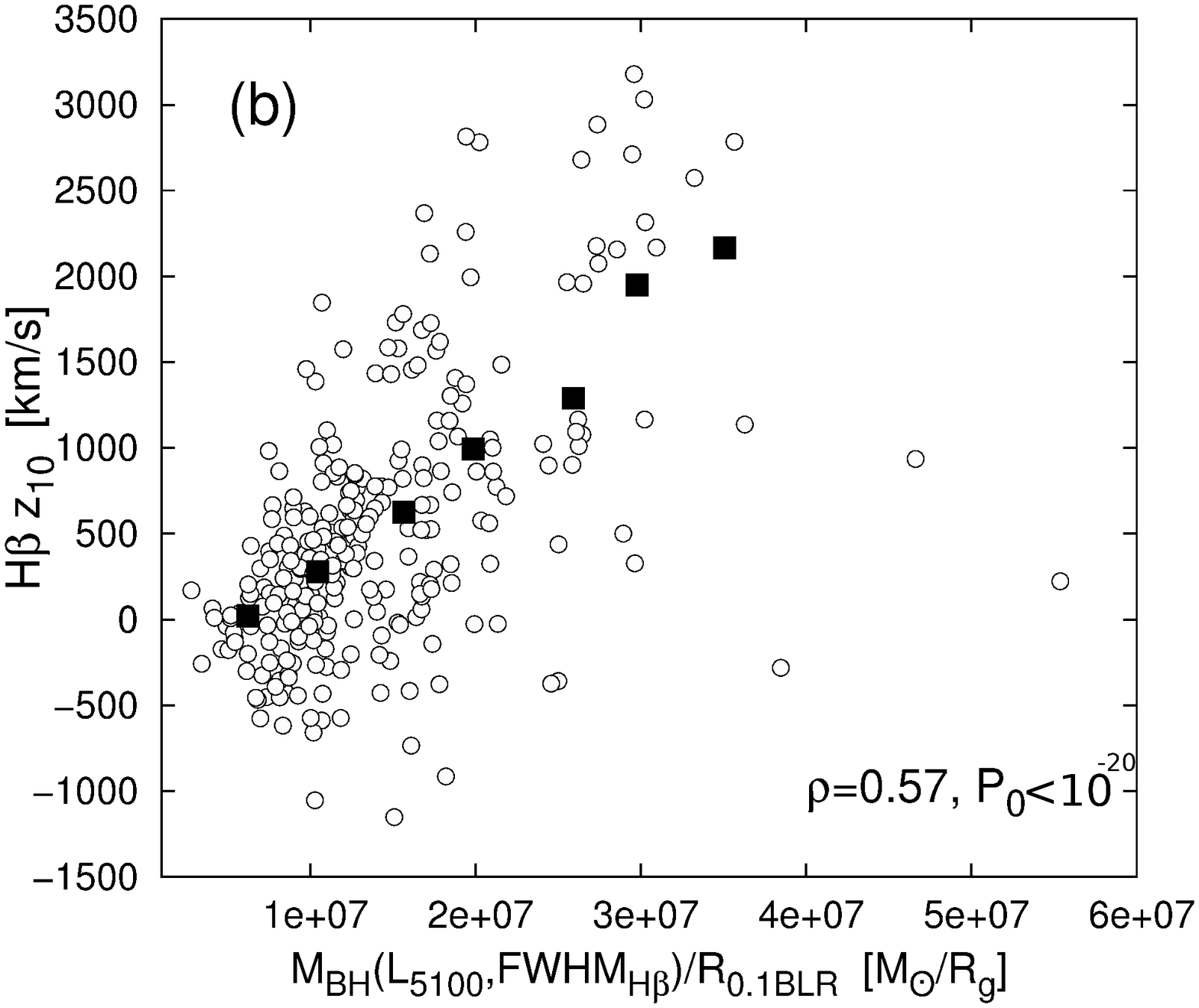}
\includegraphics[width=0.40\textwidth,angle=-90]{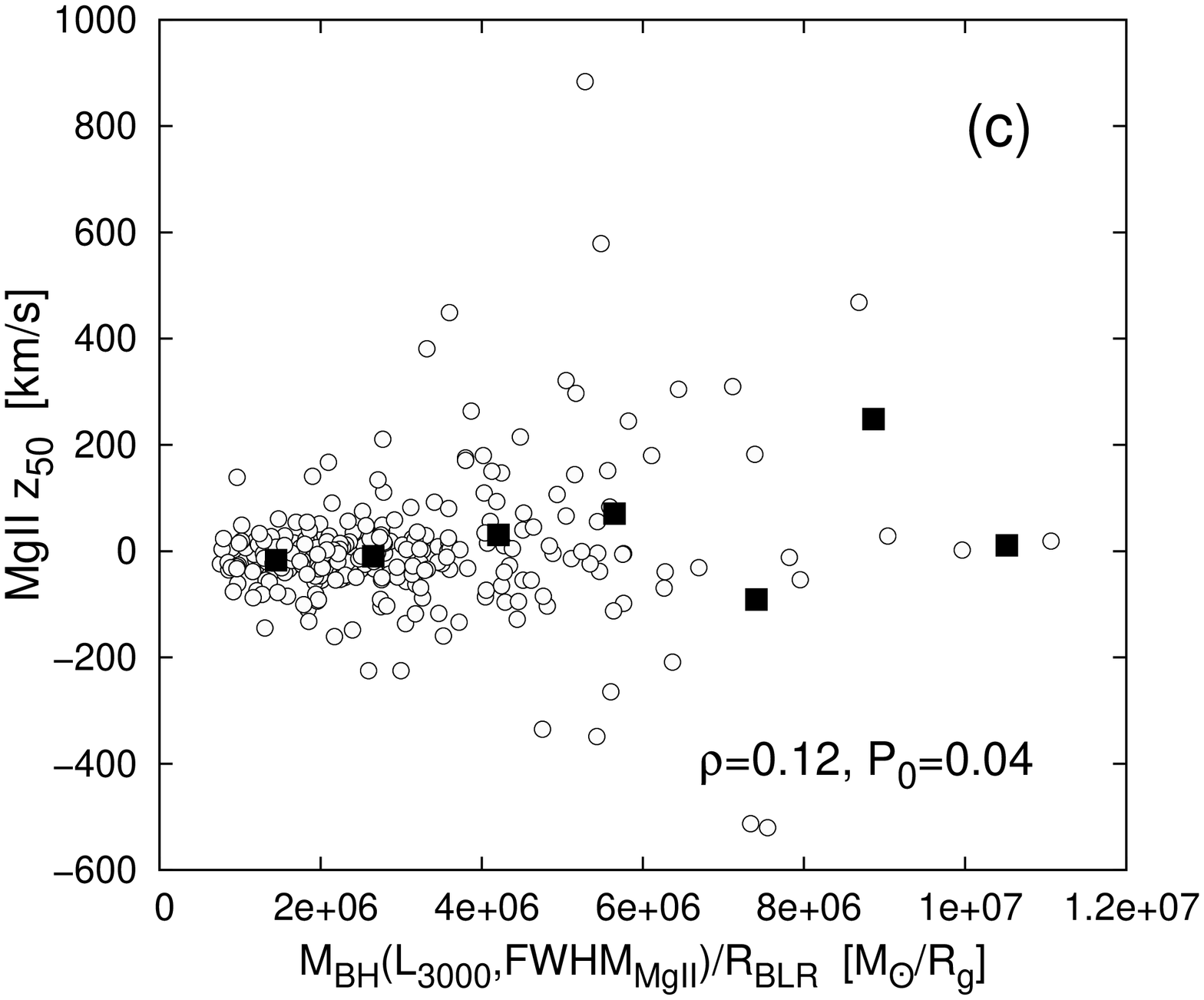}
\includegraphics[width=0.40\textwidth,angle=-90]{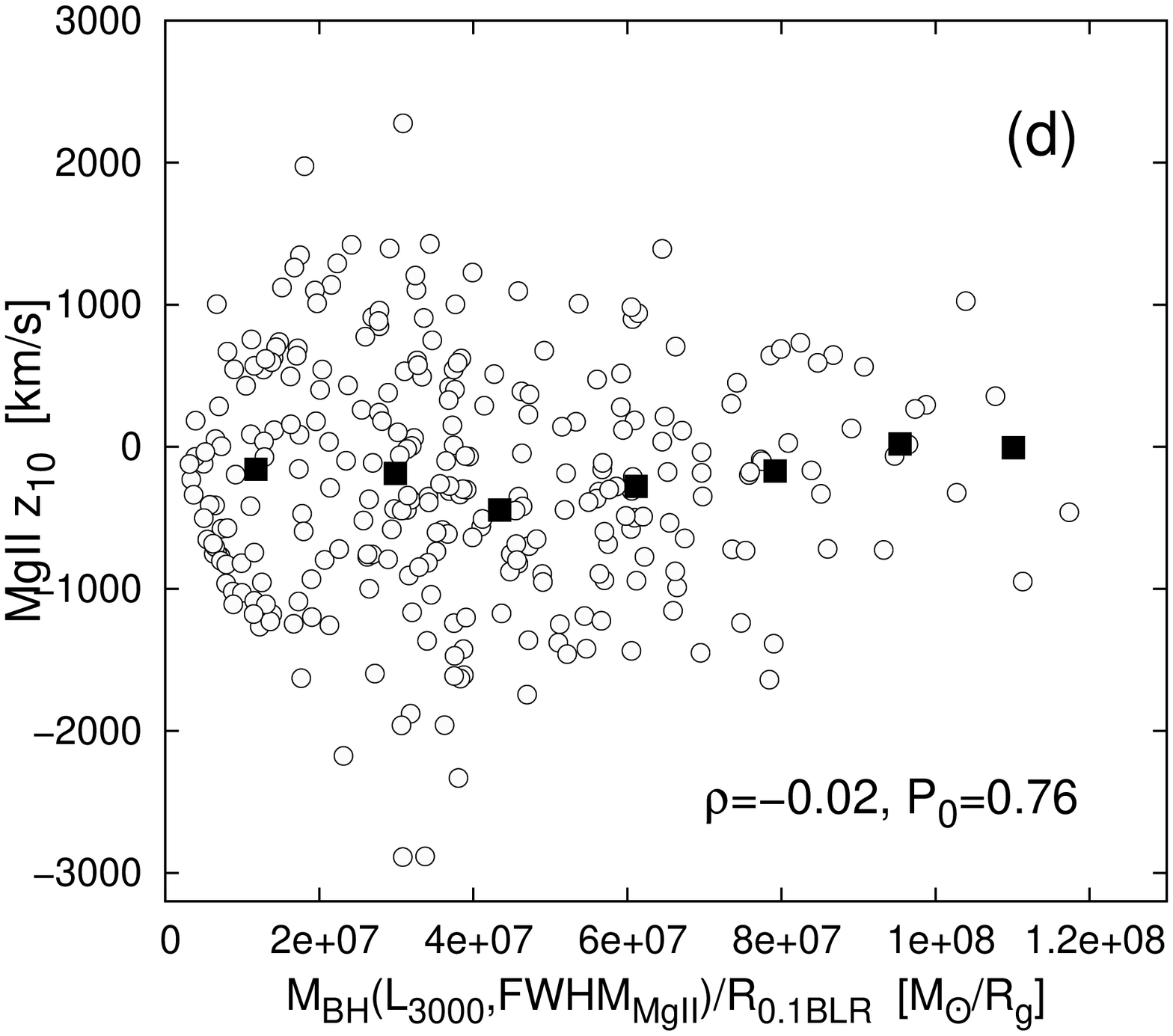}
\caption{ The intrinsic shifts at 50\% and 10\% of the maximal line intensity as 
the function of $M_{\rm BH}/R$ ratio for H$\beta$
(a, b) and \ion{Mg}{ii} (c, d). 
The binned values are denoted with black squares.}
\label{ff-z}
\end{figure*}

\section{Results} \label{sec:results}

 In order to investigate similarities and differences between the H$\beta$ and \ion{Mg}{ii} emitting regions,
we  compared the mean
and RMS line profiles of these broad lines (see \S 2.2.4). 

As it can be seen in Fig. \ref{f01}ab the  mean \ion{Mg}{ii} broad line has Lorentzian-like profile, 
while the mean broad H$\beta$ mostly has a Gaussian-like profile. The mean \ion{Mg}{ii} broad line
profile (normalized to one) is similar 
to H$\beta$ in the line core  (with slightly smaller FWHM), while in the line wings there is a large difference. The
\ion{Mg}{ii} has more extensive line wings (see Fig. \ref{f01}cd), and specially, the blue wing
 tends to be more intensive than the red in comparison with the H$\beta$ ones.

 The mean and RMS line profiles are taken from a sample that 
has the AGN luminosity spanning an order of magnitude from $\log(\lambda L_{5100})\sim$ 44.5 to  $\log(\lambda L_{5100})\sim$ 45.5 (see Fig. \ref{fig2}) and different black hole masses (see Fig. \ref{fig2a}). Therefore, some other effects may  
affect the broad emission line 
features such as a flux excess  in the red or/and blue side of line profile as e.g. photoionization effects in the BLR,
geometry and dynamics of the BLR, etc. However, in this paper we are {\it a priori} assuming the  virialization in the 
\ion{Mg}{ii} and H$\beta$
emitting regions. Therefore,
the FWHM should depend on the BH mass ($M_{\rm BH}$), and  BLR radius ($R_{\rm FWHM}$), i.e.
${\rm FWHM}\sim \sqrt{M_{\rm BH}/R_{\rm FWHM}}$. 

Taking a sample with different BH masses, one can 
expect that the RMS of normalized line profiles should have maxima around the velocities that 
correspond to the $\lambda_1$=$\lambda_0 - {\rm FWHM}/2$ and $\lambda_2$=$\lambda_0 + {\rm FWHM}/2$ 
(since the line maximum is 
normalized to one). Fig. \ref{f01}ab shows that the RMS has two maxima around  $\lambda_0 \pm {\rm FWHM}/2$ in
the case of both lines. 
Additionally, if  both broad lines originate from the region with similar kinematics, 
one can expect that the mean \ion{Mg}{ii} and H$\beta$  broad lines should have similar profiles,
as it is the case with their
line cores (see Fig. \ref{f01}cd).  It is clearly seen that there is a large difference in the 
line wings, where 
\ion{Mg}{ii} shows broader and more intensive line wings.

On the other hand, in the case that another geometry dominates the BLR, as e.g. inflow/outflow, one can expect that the RMS would have strong asymmetry (and peaks) in the far blue or red 
line wings. In other words, the mean and RMS profiles of the sample AGNs with different masses and BLR properties, can 
give some indications about the difference in the BLR geometry, especially if we compare mean profiles of two different
lines from the sample. Note here that in the case where the kinematics plays an important role in the line shapes, 
one can expect that in one spectrum the normalized \ion{Mg}{ii} and H$\beta$ broad line profiles should have similar (almost the same) shape.

 The \ion{Mg}{ii} RMS has two peaks (at $\sim$-1720 km s$^{-1}$ and $\sim$+1800 km s$^{-1}$) and is nearly symmetric,
while the H$\beta$ RMS has red asymmetry (two peaks at $\sim$-1860 km s$^{-1}$ and $\sim$+2080 km s$^{-1}$).
The distance between the RMS peaks in the both lines is very close to the FWHM of their mean profiles.

\subsection{The \ion{Mg}{ii} and H$\beta$ BLR kinematics} \label{sec:3.1}

  Since the FWHM of the broad H$\beta$ and \ion{Mg}{ii}
lines are in a good correlation \citep[see e.g.][]{kp15}, one can expect that their
 other parameters are also well correlated.
However, the correlation is absent in the case of FW10\%M between these two lines
\citep[see e.g.][]{jon16}. Here we explored correlation between kinematical parameters of Mg
II and H$\beta$ in more detail (see Fig. \ref{f02}).

We compared the FW10\%M vs. FWHM for both lines (Fig. \ref{f02}a). The H$\beta$ line 
parameters are in good correlation: $\rho$=0.82, $P_0<$10$^{-20}$,  where $\rho$ is Spearman correlation coefficient and $P_0$ is P-value of the null-hypothesis. In the case of the \ion{Mg}{ii} line the correlation between these parameters is weaker ($\rho$=0.42, $P_0$=10$^{-13}$).

As mentioned before, if  virialization is present in an emitting region, one can 
expect that the intrinsic shift will be in correlation with the corresponding width (see Eq. \ref{eq:9}).
In Fig. \ref{f02}b we showed the correlations between the intrinsic shifts and widths at 10\% of the line maximum. The expected 
correlation is present in the case of  H$\beta$ ($\rho$=0.57, $P_0<$10$^{-20}$), 
but in the case of \ion{Mg}{ii} there is no trend ($\rho$=-0.02, $P_0$=0.71). The big scattering of the points is present, showing randomly strong blue and red shifts in the wings.
This indicates that the \ion{Mg}{ii} wings probably originate in a region that is more connected with outflows/inflows than with the virial motion of the gas in an accretion disc.

 We should note that
 the relationship between the line profile width and centroid wavelength shift can be questioned, since the gravitational 
redshift is typically assumed to have a minimal effect on the line profile. To do an additional test of 
the connection between the intrinsic shift and the gravitational redshift, we explored the intrinsic shift as a function of the
$M_{\rm BH}/R_{\rm BLR}$ ratio (see Eq. \ref{eq:8}), using the 
BLR size from the $R-L$ relation \citep[given in][]{be13} and the 
single-epoch BH mass.

For the intrinsic shift at 10\% of the maximal intensity we rescaled the $R_{\rm BLR}$  as: 

\begin{equation}
R_{\rm 0.1BLR}$=$R_{\rm BLR}\cdot\Bigg(\frac{{\rm FWHM}}{{\rm FW10\%M}}\Bigg)^2.
\end{equation}

In Fig. \ref{ff-z} we plot  the intrinsic shift measured at FWHM ($z_{50\%}$) and at FW10\%M ($z_{10\%}$) as a function of the $M_{\rm BH}/R_{\rm BLR}$ for H$\beta$
(Figs. \ref{ff-z}ab) and \ion{Mg}{ii} (Figs. \ref{ff-z}cd). Also we show the binned values (black squares) in order 
to trace a trend between the intrinsic shift and the $M_{\rm BH}/R_{\rm BLR}$ ratio.  Figs. \ref{ff-z}ab show that 
the intrinsic shifts follow the expected correlation 
with the $M_{\rm BH}/R_{\rm BLR}$ for  H$\beta$. In the case of  the intrinsic shift at half of the maximal intensity for 
\ion{Mg}{ii}, there is some
indication of a weak trend   with the $M_{\rm BH}/R_{\rm BLR}$ only for $z_{50\%}>$0 (Fig. \ref{ff-z}c), but the intrinsic shift of \ion{Mg}{ii} wings shows no correlation  with the $M_{\rm BH}/R_{\rm BLR}$ (Fig. \ref{ff-z}d). This implies that there is no virialization in the \ion{Mg}{ii} wings. The 
values of the intrinsic shift in \ion{Mg}{ii} are in the range from -3000 km s$^{-1}$ to +2000
km s$^{-1}$ (see Fig. \ref{ff-z}d) which indicates high outflows/inflows, with similar probability. However, {\bf as $M_{\rm BH}/R_{\rm BLR}$ increases}, the range of intrinsic shift gets smaller (between $\pm$1000 km s$^{-1}$), i.e. this kind of motion is with smaller velocity. 
These results indicate some kind of 'fountain-like' motion that contributes to the extensive \ion{Mg}{ii} 
line wings, which becomes suppressed with stronger gravitational influence.

Additionally, we investigated the influence of the Eddington ratio to the gas velocities in the  'fountain-like' region. Therefore, we calculated  the $L_{5100}$/$M_{\rm BH}\big(H\beta\big)$ ratio (which is proportional to the Eddington ratio) for a subsample of 52 objects with the luminosities in the narrow range of 45.1$<\log(\lambda L_{5100})<$45.2, and we tested if there is a correlation with $z_{10\%}$, which is probably proportional to the 'fountain-like' region velocities. We found that the absolute value of $z_{10\%}$ (|$z_{10\%}$|) has weak correlation with $log \big(L_{5100}$/$M_{\rm BH}\big(H\beta\big)\big)$ ($\rho$=0.34, $P_0$=0.01), i.e the inflow/outflow velocities increase as Eddington ratio increases.

\begin{figure}
\includegraphics[width=0.45\textwidth]{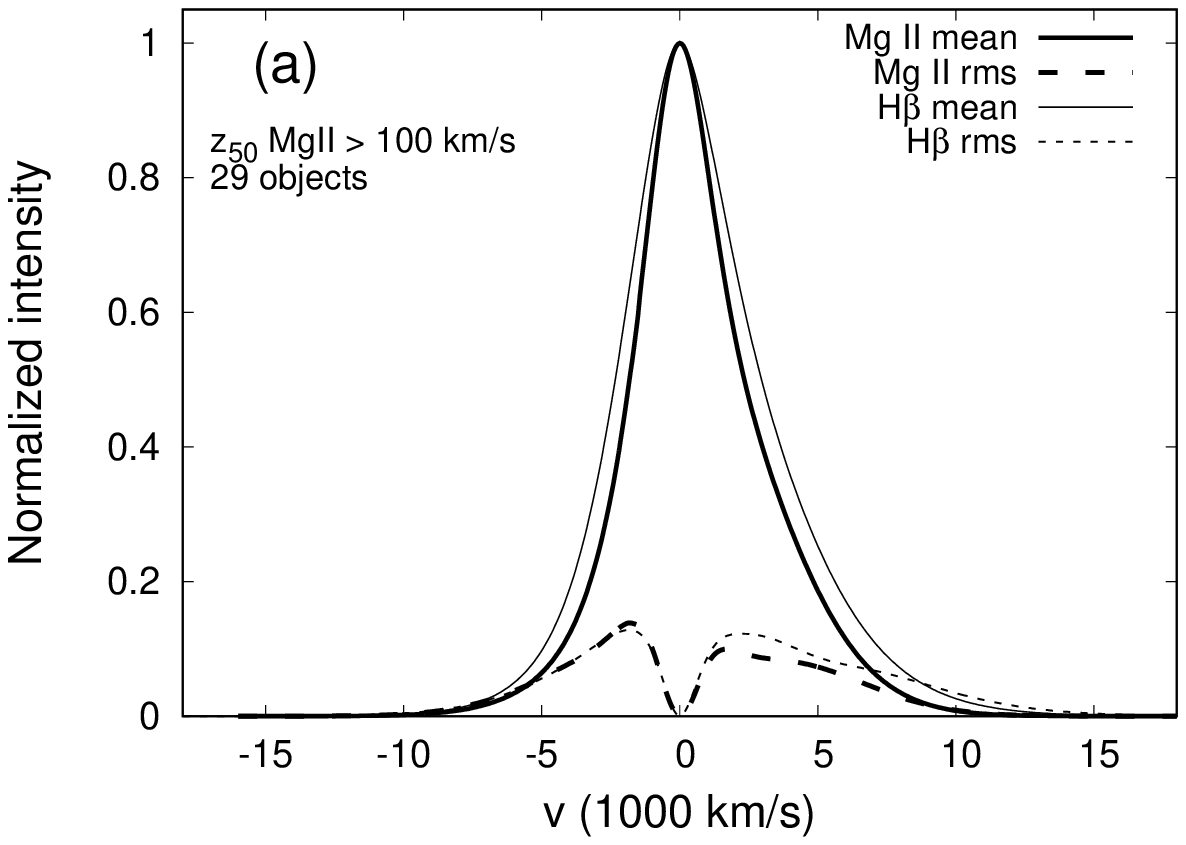}
\includegraphics[width=0.45\textwidth]{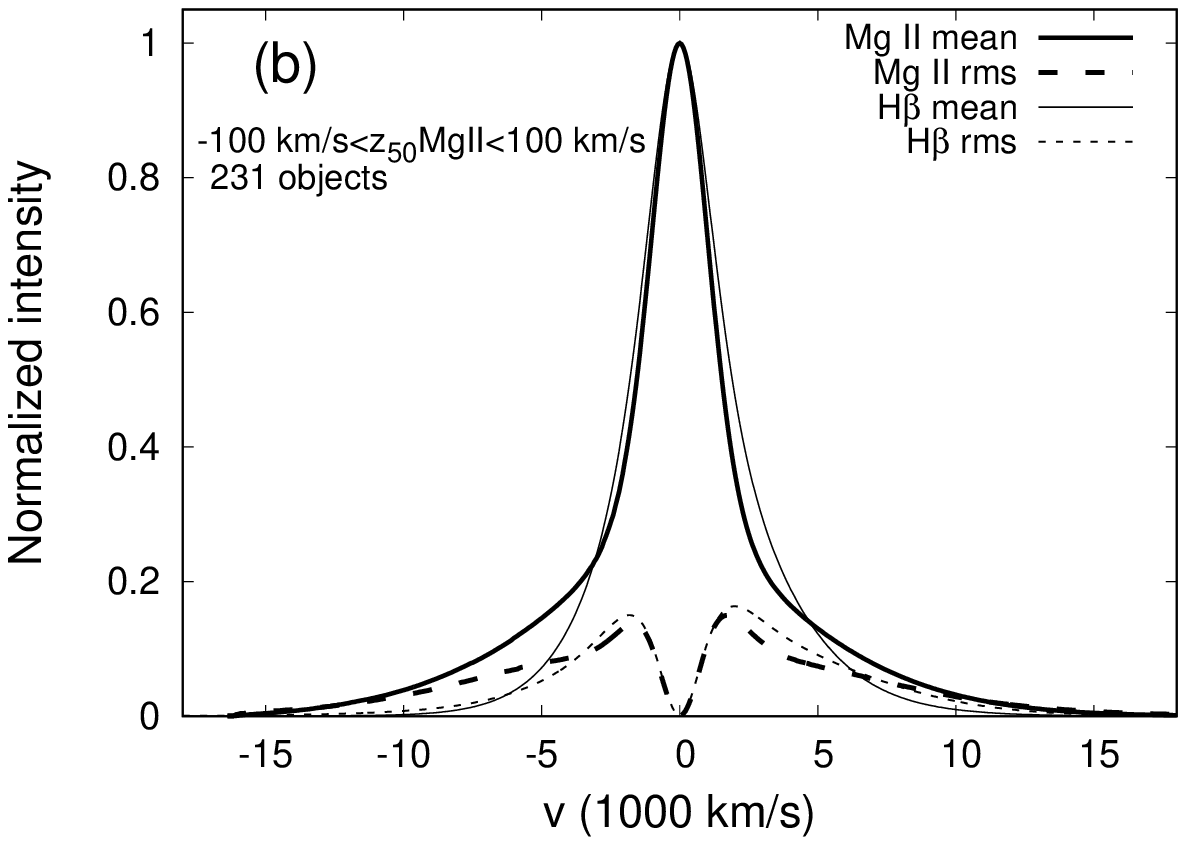}
\includegraphics[width=0.45\textwidth]{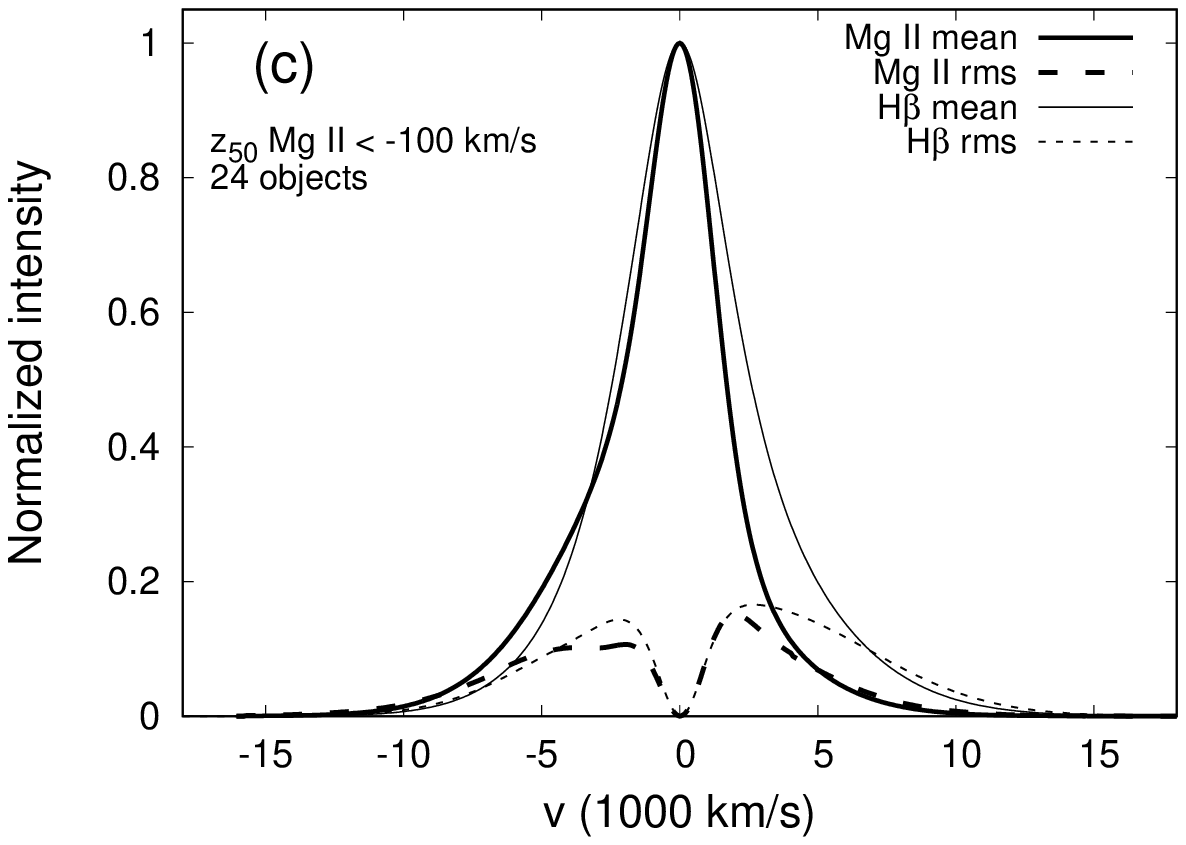}
\caption{ The mean (solid tick line) and the RMS (dashed tick line) \ion{Mg}{ii} broad line profiles for 
sub-samples of AGNs, for which \ion{Mg}{ii}: (a) has red asymmetry (29 AGNs with $z_{50\%}>100$km s$^{-1}$); 
(b) is almost symmetric (231 AGNs with $z_{50\%}$ between $\pm100$km s$^{-1}$); and 
(c) has blue asymmetry (24 AGN with $z_{50\%}<-100$km s$^{-1}$). The Mg 
II asymmetry is measured at FWHM. The corresponding mean H$\beta$ and RMS
profiles are shown with thin solid and dashed lines, respectively.
}
\label{f04}
\end{figure}

\begin{figure}
\includegraphics[width=0.45\textwidth]{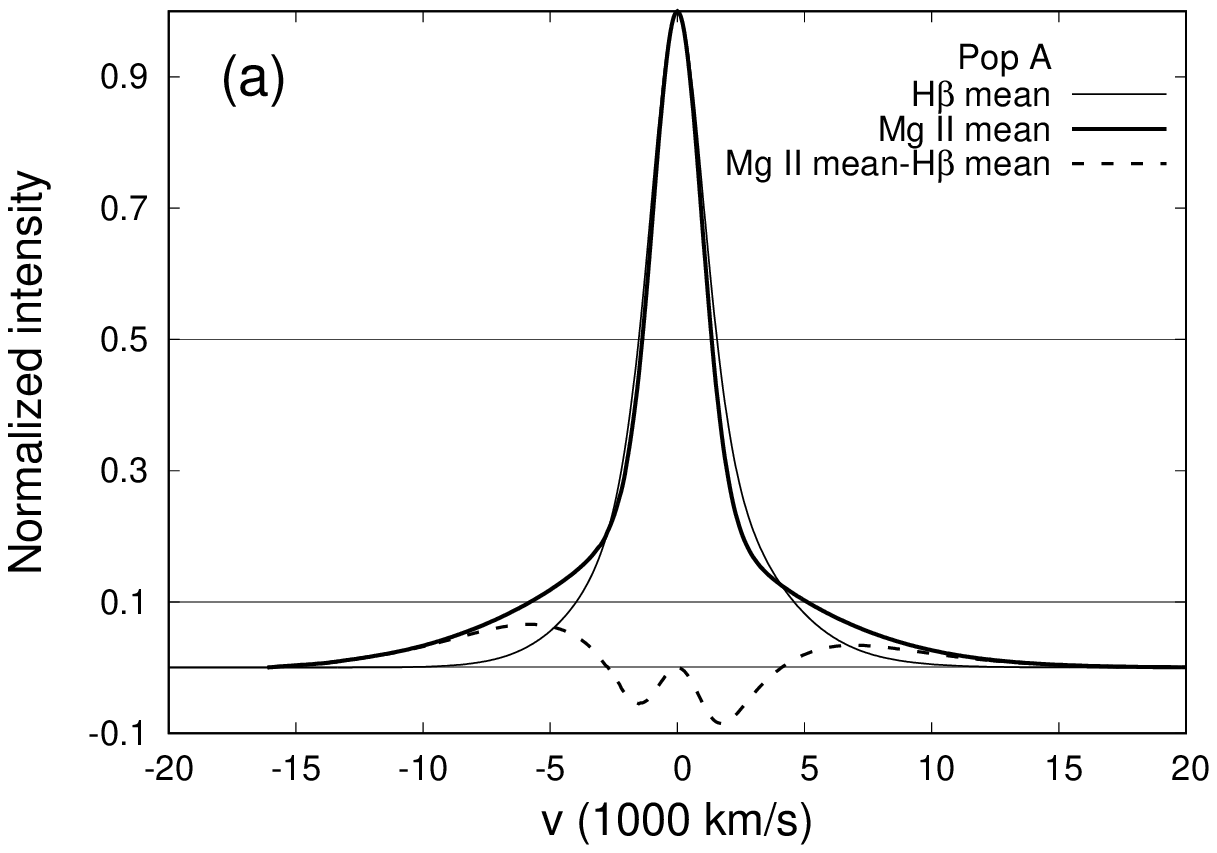}
\includegraphics[width=0.45\textwidth]{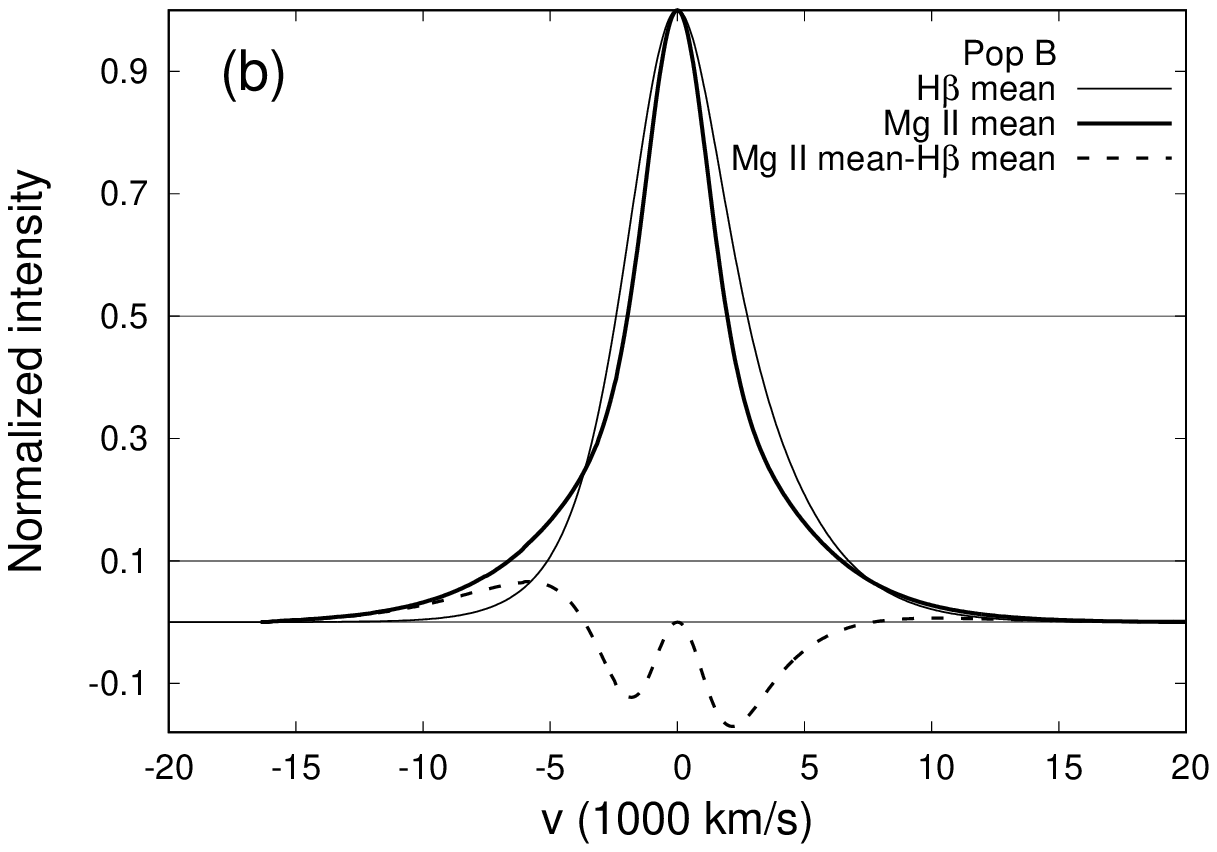}
\caption{ Comparison between the mean \ion{Mg}{ii} and H$\beta$  broad line profiles
and their difference (dashed line) for  the samples of (a) 149 Pop. A objects 
(FWHM H$\beta<$4000km s$^{-1}$) and (b) 135 Pop. B objects 
(FWHM H$\beta>$4000km s$^{-1}$). }
\label{f04a}
\end{figure}

\section{Discussion} \label{sec:discussion}

\subsection{Similarities and differences between \ion{Mg}{ii} and H$\beta$ emission line profiles}

First of all, let us give some general facts about the \ion{Mg}{ii} and H$\beta$ line profiles:

\begin{itemize}
  \item The core and FWHM of the mean profiles of \ion{Mg}{ii} and H$\beta$ seem to be very similar, while the difference is bigger 
  in the line wings. The \ion{Mg}{ii} line has more Lorentzian-like, and the H$\beta$ has Gaussian-like profile. 
 \item The H$\beta$ line shows virialization in the whole profile, i.e. the relationship between the kinematical parameters (widths and intrinsic shifts) obtained from the H$\beta$ BLR nearly follows those expected in case of the pure Keplerian motion. 
 On the other hand, the \ion{Mg}{ii} BLR does not follow this motion.
 \item It seems that \ion{Mg}{ii} shows virialization in the line core, while another effect 
 contributes to the line wings, making them very extensive.
\end{itemize}

There is a question how much an emitting region with  inflows and/or outflows contributes 
to the total \ion{Mg}{ii} emission.
At least in the wings, we can expect the contribution of the emitting region with 
some kind of vertical (to the disc plane) motion.

 Since the line profiles can indicate the inflow (red asymmetry) or outflow (blue asymmetry),
 we divided the sample into three
subsamples according to the \ion{Mg}{ii} asymmetry (i.e. z$_{50\%}$ measured at FWHM).
The first subsample contains AGNs with the \ion{Mg}{ii}
blue asymmetry ($z_{50\%}<-100$ km s$^{-1}$), the second with the red one 
($z_{50\%}>100$ km s$^{-1}$), and the third contains the AGNs with the symmetric \ion{Mg}{ii} profile (intrinsic shift, $z_{50\%}$, between these two values: $-100<z_{50\%}<+100$). We found that most of 
AGNs (231) have almost symmetric \ion{Mg}{ii} mean profile, showing extended wings in \ion{Mg}{ii} compared to H$\beta$,
and a big difference in the line wings of these two lines (see Fig. \ref{f04}b). The red asymmetry is present in 29 AGNs, and
as it can be seen in Fig. \ref{f04}a, the mean line profiles of \ion{Mg}{ii} and H$\beta$ are very 
similar (almost the same) for these objects. Finally, we found 24 AGNs with the significant blue asymmetry, and also with a big difference  between the mean \ion{Mg}{ii} and H$\beta$ line profiles (see Fig. \ref{f04}c).

Comparing the mean H$\beta$ and \ion{Mg}{ii} line profiles and their RMSs (Fig. \ref{f04}), one can conclude that in the case of the red asymmetry and symmetric \ion{Mg}{ii} profiles, the FWHM and RMSs are similar for both lines in the most of cases (see  Fig. \ref{f04}ab), and the \ion{Mg}{ii} line seems to be  a good estimator of the central BH mass. In the case of the strong blue \ion{Mg}{ii} asymmetry, the mean profiles of \ion{Mg}{ii} and H$\beta$ are quite different, and the \ion{Mg}{ii} line perhaps is not appropriate for BH mass estimation. However, we should note that we have statistically small samples of objects with strong blue and red asymmetry, so the result may be influenced by outliers.

 Let us here also recall the results obtained in \cite{m2013,m2013b}. They found a difference 
in the line
profiles of \ion{Mg}{ii} and H$\beta$ for Pop. A and Pop. B objects (see the definition of Pop. A/Pop. B in  Sec. \ref{sec:intro}). The red asymmetry was observed in the 
Pop. B objects in both broad lines (\ion{Mg}{ii} and H$\beta$) and they supposed that it may indicate an inflow in the BLR. Pop. A objects have more symmetric line profiles, but the 
blue-shift of \ion{Mg}{ii} relative to H$\beta$ indicates an outflow \citep[][]{m2013b}. Therefore, one can conclude that the \ion{Mg}{ii} line profiles in AGN show both outflows
(Pop. A objects) and inflows (Pop. B objects). 

We also compared the line profiles of Pop. A and Pop. B objects in our sample. We found that 
149 objects from the sample are Pop. A, with
the FWHM of H$\beta$ smaller than 4000 km s$^{-1}$, and 135 objects are Pop. B. 
The comparison of the \ion{Mg}{ii} and the H$\beta$ mean line profiles for both populations is presented  
in Fig. \ref{f04a}. It can be seen that the \ion{Mg}{ii} wings are more 
extensive than H$\beta$ wings, and that H$\beta$ is slightly broader than \ion{Mg}{ii} in both populations \citep[as it was noted in ][]{m2013,m2013b}. Also, the mean H$\beta$ and \ion{Mg}{ii} line profiles are different in these two populations. In Pop. A, the mean \ion{Mg}{ii} profile is Lorenzian-like, and it shows blue asymmetry in wings (velocities at 10\% of line intensity: -5780 km s$^{-1}$ to +5100 km s$^{-1}$, see Fig. \ref{f04a}a), while the mean H$\beta$ has red asymmetry (velocities at 10\% of line intensity: -3990 km s$^{-1}$ to +4590 km s$^{-1}$). However, if measured at 50\% of line intensity,  both lines are nearly symmetric (\ion{Mg}{ii}: -1380 km s$^{-1}$ to +1350 km s$^{-1}$, H$\beta$: -1520 km s$^{-1}$ to +1560 km s$^{-1}$). In Pop. B, the red asymmetry in the mean H$\beta$ line profile is much stronger compared to Pop. A (velocities at 10\% of line intensity: -5110 km s$^{-1}$ to +6740 km s$^{-1}$, and at 50\% of line intensity: -2410 km s$^{-1}$ to +2750 km s$^{-1}$), while the \ion{Mg}{ii} mean profile is almost symmetric with a slight asymmetry measured at 10\% of line intensity in the blue wing (-6620 km s$^{-1}$) compared to the red wing (+6470 km s$^{-1}$), and at FWHM the profile is symmetric (with FWHM/2 $\sim \pm$ 1960 km  s$^{-1}$).

 Similarly as \cite{m2013b}, we found that the mean H$\beta$ and \ion{Mg}{ii} profiles are nearly symmetric in Pop. A (measured at FWHM), and that the mean H$\beta$ profile has red asymmetry in Pop. B. However, in difference with \cite{m2013b}, we found  that emission in far wings of \ion{Mg}{ii} is present 
\citep[that is assumed as an iron contribution in][]{m2013b}, and we found no red asymmetry in Pop. B \ion{Mg}{ii} mean profile. Note here that in this sample, for the Pop. B objects, the FWHM \ion{Mg}{ii} is mostly in the range of 4000 km s$^{-1}<$ FWHM \ion{Mg}{ii} $<$ 6000 km s$^{-1}$. There are only three objects with FWHM \ion{Mg}{ii} $>$ 6000 km s$^{-1}$ (see Fig \ref{f02}a), while in the sample of \cite{m2013,m2013b}, there are dozens of objects like this \citep[see Fig 8 in][]{m2013}. For the mean \ion{Mg}{ii} profile of these three objects with FWHM \ion{Mg}{ii} $>$ 6000 km s$^{-1}$, we found the red asymmetry in the line wings (see Sec \ref{sec:4.2}), as \cite{m2013} found for Pop. B in their sample.

The possible \ion{Fe}{ii} emission contribution in the red and blue wings of the \ion{Mg}{ii} and H$\beta$ is discussed in Appendix A of this paper.

 \begin{figure}
\centering
\includegraphics[width=0.45\textwidth]{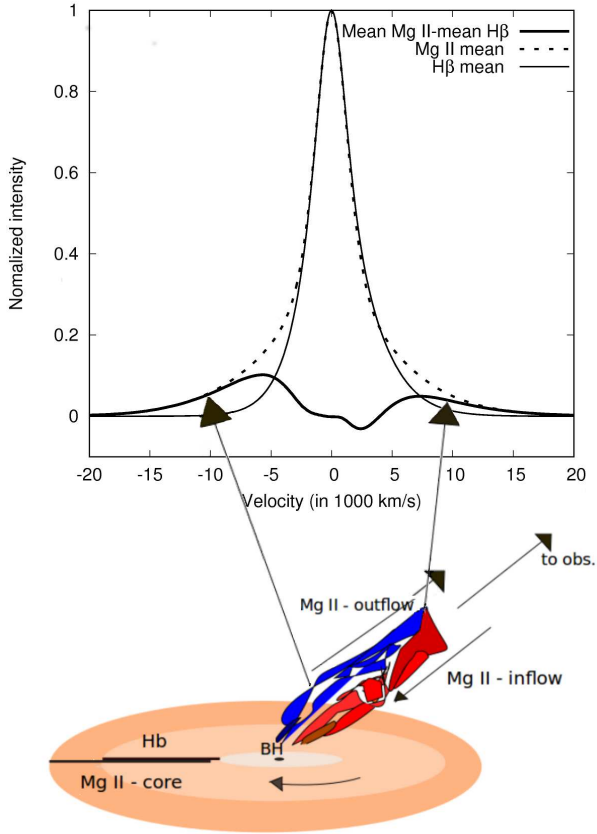}

\vspace{-10pt}
\caption{Upper panel shows the mean \ion{Mg}{ii} (dashed line) fitted with the H$\beta$
mean profile (solid thin line). Their difference is shown below as
double-peaked feature (solid tick line). Bottom panel presents a simple scheme
of \ion{Mg}{ii} and H$\beta$ emitting regions in AGNs, where the flow orthogonal to the
disc is caused by outflows and inflows. This motion contributes to the \ion{Mg}{ii}
line wings (see upper panel). The BLR with disc-like rotation contributes
to the \ion{Mg}{ii} line core and mostly to the total H$\beta$ broad line.}
\label{f-sch}
\end{figure}

\subsection{\ion{Mg}{ii} emitting region} \label{sec:4.2}

Comparing the mean \ion{Mg}{ii} and H$\beta$ line profiles in the total sample and  five sub-samples 
(see Fig. \ref{f01}cd
and Figs. \ref{f04} and \ref{f04a}), the large difference seems to be visible only in wings. The question is what can cause the difference only in the line wings? 
 We found that  the intrinsic shift in the \ion{Mg}{ii} wings can be 
 red or blue and that it is not in correlation with the $M_{\rm BH}/R_{\rm BLR}$ ratio 
 (see Fig. \ref{ff-z}d), which indicates that both, inflows and 
outflows could be present in the \ion{Mg}{ii} emitting region. To discuss a model for the \ion{Mg}{ii} emitting region, 
let us recall some results from the laboratory plasma investigations. As it was shown in \citet[][see Fig. 6 in their paper]{kur09}, the 
kinematics can affect the line profile in the case of 'fountain-like' motion with approaching and receding velocities, 
that in combination with the Doppler broadening can produce slightly asymmetric Lorentzian-like profiles.
Also, recently \cite{cz17} showed  that a flow with an orthogonal motion to the disc plane can produce Lorentzian-like profiles. The similar
results, with Lorentzian like profile was obtained by \cite{kz13}, considering some type of turbulent velocities in the model (vertical 
to the BLR rotation).

In Fig. \ref{f-sch} (upper panel) we fitted the  \ion{Mg}{ii} mean profile (dashed line) with 
the H$\beta$ one (solid line), 
and subtracted the best fit of H$\beta$ from \ion{Mg}{ii}
(thick double-peaked feature shown below).
As expected,  there is a double peaked structure emerging after the subtraction, 
showing the maxima at $\sim \pm 7000$
km s$^{-1}$. If there are some kind of outflows-inflows they should have high velocities, 
i.e. probably the outflows-inflows 
 originate very close to the central BH.
 One can also consider that this extra-emission may come from the 
fast motion of the material in the accretion disc, that is very close to the 
central BH. However, in that case we expect to have boosted blue wing and 
large red wing that is typical for the relativistic disc \citep[see e.g. the complex profile of NGC 3516, Figs. 7-10 in][]{po02}, which is not the 
case here (see Fig. \ref{f-sch}). Therefore, we  excluded this scenario. 
 
Taking into account the discussion above, we propose the following scenario for the
\ion{Mg}{ii} line origin (as it is shown in the simple 
scheme in Fig. \ref{f-sch}, bottom):

The line core is originating from the virialized disc-like BLR (represented as a disc in 
Fig. \ref{f-sch}-bottom panel), 
that may be at slightly larger radius than the H$\beta$ emission region (\ion{Mg}{ii} shows a slightly 
narrower FWHM than H$\beta$).
In the case where the emission from this region is dominant, the FWHM of \ion{Mg}{ii} 
represents the velocity of the emission 
gas rotation due to gravitational force, and the line can be used as very good BH mass 
estimator, as it is noted in \cite{tn12}. 
However, where the 'fountain-like' region is dominant (shown as outflows-inflows
in Fig. \ref{f-sch}-bottom), one can 
expect a broader \ion{Mg}{ii} line, and in that case, the virialization in the FWHM probably is not present, 
and therefore \ion{Mg}{ii} might not be suitable for the BH 
mass estimation \citep{le13}. This is in agreement with results obtained by \cite{tn12}, who
 found that the  \ion{Mg}{ii} can be used as a good BH mass estimator in the case of
FWHM$<$6000 km s$^{-1}$. 

\begin{figure}
\centering
\includegraphics[width=.45\textwidth]{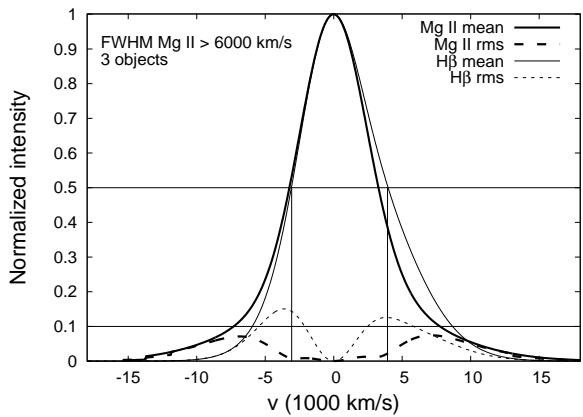}
\caption{ Mean line profiles of \ion{Mg}{ii} (tick line), and H$\beta$ (thin line) for 3 AGNs with FWHM \ion{Mg}{ii} $>$ 6000 km s$^{-1}$. Corresponding RMS profiles are denoted with tick dashed line (for \ion{Mg}{ii}) and thin dashed line (for H$\beta$). The vertical lines show positions of $\lambda_0\pm$FWHM/2, where it is expected to have RMS peaks due to the difference in BH masses.}
\label{f10}
\end{figure}

 As mentioned above, we found only 3 AGNs with \ion{Mg}{ii} FWHM larger than 6000 
km s$^{-1}$, for which we compared the mean H$\beta$ and \ion{Mg}{ii} 
line profiles (see Fig. \ref{f10}). 
As it can be seen, there is a difference between the
mean  H$\beta$ and \ion{Mg}{ii} lines and their RMS profiles. The  \ion{Mg}{ii} mean profile shows red asymmetry measured at 10\% of line intensity (-7300 km s$^{-1}$ to +7760 km s$^{-1}$) and it is nearly symmetric at FWHM (-3240 km s$^{-1}$ to +3270 km s$^{-1}$), while the H$\beta$ mean profile shows strong red asymmetry measured at both 10\% (-5940 km s$^{-1}$ to +8680 km s$^{-1}$) and 50\% of line intensity (-3090 km s$^{-1}$ to +3970 km s$^{-1}$).
Additionally, the differences (RMS) in far wings of the mean \ion{Mg}{ii} are without peaks which are usually seen near $\lambda_0\pm$FWHM/2 in the 
RMS of H$\beta$. 

Therefore, despite our limited number of objects with \ion{Mg}{ii} FWHM$>$6000, our analysis of these objects is in agreement with previous results. However, since our sample has no statistically significant number of AGNs with very broad \ion{Mg}{ii} lines, we could not find the strict constraint in the \ion{Mg}{ii} line parameters (width or asymmetry) which would indicate if \ion{Mg}{ii} is suitable (or not suitable) for BH mass estimation (see Appendix B).


\section{Conclusions} \label{sec:conclusions}

Here we explored the structure of the \ion{Mg}{ii}  line emitting region in a sample of 284 type 1 AGNs, by comparing the 
\ion{Mg}{ii} line parameters with the H$\beta$ ones.  We  investigated measured line properties 
and  tried to give the possible model of the  broad \ion{Mg}{ii} emitting region.
 
From our analysis we can conclude the following:

\begin{itemize}
\item   The mean \ion{Mg}{ii} line profile has a slightly asymmetric Lorentzian-like profile, and  very broad line wings.
The shape of the mean \ion{Mg}{ii} line profile and correlation between the line parameters indicate that there are two Mg 
II emitting regions: the one  similar to the H$\beta$ emitting region  
that is probably virialized (contributes to the line core), and the second emission region that seems to be 'fountain-like'.
 In order to explain the broad blue and red wings, the 'fountain-like' emission
 region should have both -- inflows and outflows,
with high velocity components orthogonal to the disc, which become suppressed with stronger gravitational influence.

\item If the virialized region  mostly contributes to the \ion{Mg}{ii} line flux, the  \ion{Mg}{ii} FWHM 
should be comparable with the H$\beta$ one,
and it can be used for BH mass estimation.
However, if the contribution of the 'fountain-like' region is more dominant, 
one can expect to have broader \ion{Mg}{ii} lines, with extensive wings.  This is especially seen in the case of very broad \ion{Mg}{ii} with FWHM $>$ 6000 km s$^{-1}$, or in the case of the \ion{Mg}{ii} with a strong blue asymmetry; in both cases we found that the FWHM probably does not represent the rotational motion of the emitting gas in the central BH gravitational field.
Therefore, one should be careful when using extremely broad or blue asymmetric \ion{Mg}{ii} line for BH mass estimation.

\end{itemize}

\section{Acknowledgments}

This work is a part of the project (176001) ''Astrophysical Spectroscopy of Extragalactic Objects''
supported by the Ministry of Education, Science and Technological Development of Serbia. 
We thank to the referee for very useful comments which significantly improved the paper.

\appendix
\section{The UV and optical F\lowercase{e} II line subtraction}

There are different UV and optical \ion{Fe}{ii} templates presented in the literature 
\citep[see e.g.][etc.]{bg92,vw01,ts06,bv08,kov10,kp15,me16}
which are used to account for the iron emission contribution to  the H$\beta$ and \ion{Mg}{ii} 
line profiles. For the subtraction of the optical \ion{Fe}{ii} lines around H$\beta$ we used the model given in \cite{kov10}. The differences between this \ion{Fe}{ii} model and models given in \cite{do08} and \cite{bv08} are explored and analysed in Appendix B of \cite{kov10}, where it is demonstrated that this template can well reproduce the \ion{Fe}{ii} optical emission around the H$\beta$ line. \cite{ba15} compared this template with the ones of \cite{bg92} and \cite{VC2004}, and they found that it returns the best $\chi^2$ values. Since the optical \ion{Fe}{ii} model given in \cite{kov10} is analysed in number of papers, we will not consider it here in more detail.

For subtraction of the UV \ion{Fe}{ii} lines we used the UV \ion{Fe}{ii} model given in \cite{kp15}, which has been improved for the purpose of this investigation. Therefore, here we give more details about improved version of that model.
 
\subsection{The improved UV \ion{Fe}{ii} model in the $\lambda\lambda=$2650-3050 \AA\ wavelength range}

In order to test the model of UV \ion{Fe}{ii} given in  \cite{kp15}, we fitted set of spectra with different  width of emission lines. We found that in the case of type 1 AGNs, the model fits very well the \ion{Fe}{ii} lines around \ion{Mg}{ii}. In order to test the UV \ion{Fe}{ii} model for Narrow Line Seyfert 1 (NLSy1) galaxies, we used the UV spectra of I Zw 1 and Mrk 493. These two NLSy1 galaxies have well resolved UV spectrum, observed with FOS spectrograph of Hubble Space Telescope\footnote{The spectra are downloaded from: \url{https://archive.stsci.edu/hst/search.php}}. We found that in the case of NLSy1 galaxies, where  \ion{Fe}{ii} lines are strong and narrow enough to be distinguished from \ion{Mg}{ii}, this UV \ion{Fe}{ii} model has a lack of lines at $\sim$ 2825--2860 \AA, $\sim$ 2650--2725 \AA \ and $\sim$ 3000 \AA.

We improved the model by adding the missing \ion{Fe}{ii} lines. We add the multiplet 78 \citep[see][]{bk80}, which covers the region at $\sim$ 3000 \AA. We could not identify with confidence the rest of emission at $\sim$ 2825--2860 \AA \ and 2690--2725 \AA, since there are numerous \ion{Fe}{ii} lines in that range (several lines at each \AA), and they all have high energy of excitation, which is not theoretically expected, since the observed multiplets 60-63 and 78 arise from relatively small energy excitation levels. Since we could not identify the lines, we {\it a priori} included them in our template as 'I Zw 1 lines', which probably originate from the high excitation levels.
We added two Gaussians at $\lambda$2720 \AA \ and $\lambda$2840 \AA, which represent emission of additional 'I Zw 1 lines', assuming that Gaussian intensity ratio is following the one obtained from the best fit of I Zw 1.

 The similar procedure was done in \cite{kov10} with the optical \ion{Fe}{ii} model, where the small amount of the \ion{Fe}{ii} emission, which could not be theoretically explained, was added empirically from I Zw 1 spectrum.

 We found that in both galaxies, I Zw 1 and Mrk 493, there is the line peak present at 2670 \AA, which is probably emission of the Al II 2670 \AA \ line, also noted by \cite{vw01} in their analysis of the I Zw 1 spectrum. Therefore, to fit the \ion{Mg}{ii}$+$\ion{Fe}{ii} spectral region, we additionally consider the Al II 2670 \AA \ emission line contribution.

\begin{figure}
\centering
\includegraphics[width=0.50\textwidth]{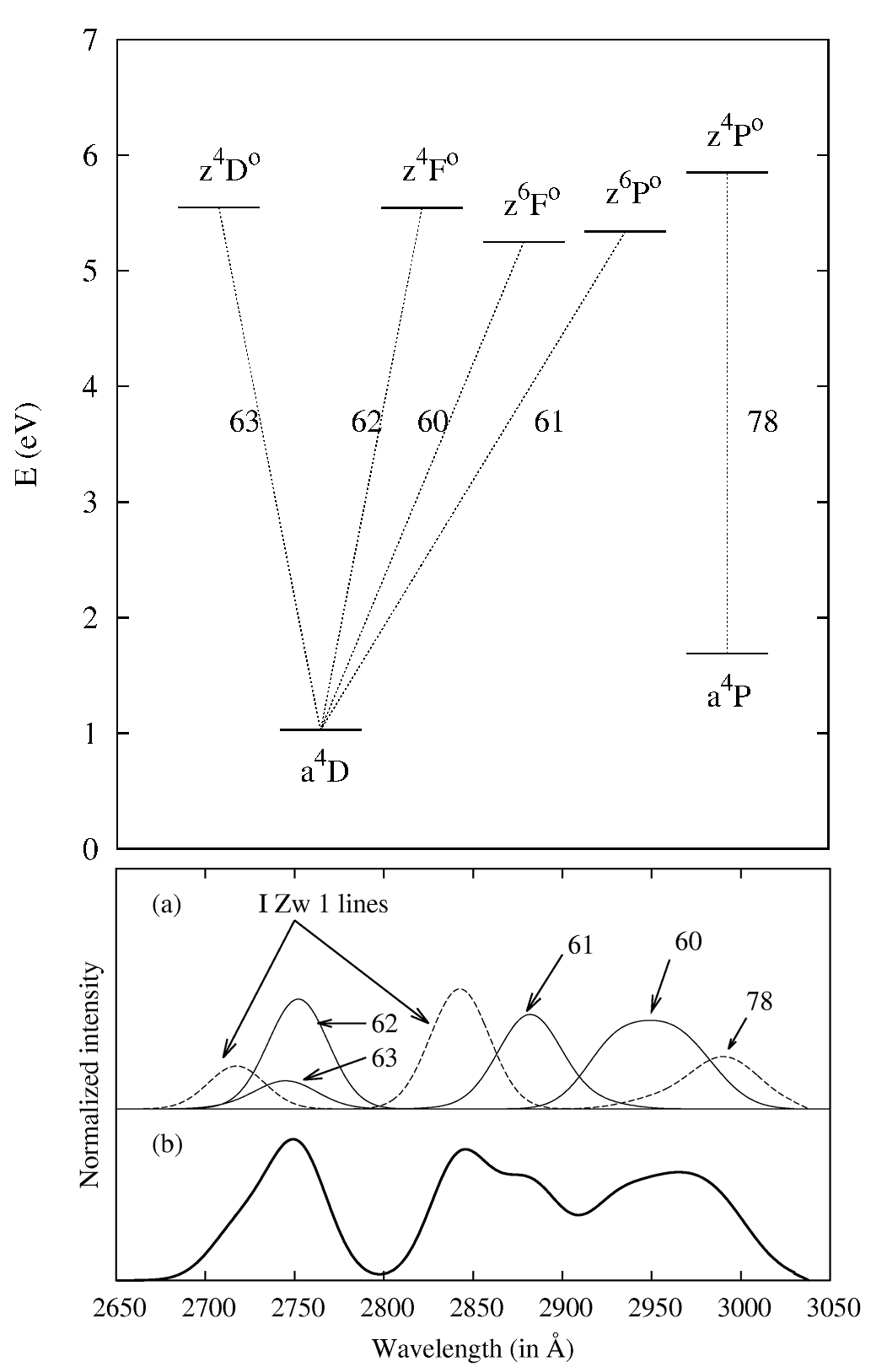}

\caption{The Grotrian diagram of multiplets included in the UV \ion{Fe}{ii} model \citep{kp15}, with additional multiplet 78 (top). The example of the multiplet emission calculated for T = 10000 K, Doppler width w = 1200 km s$^{-1}$ and arbitrary intensities of multiplets (a) and total improved UV \ion{Fe}{ii} semi-empirical model (b). The emission included in an improved version of model is denoted with dashed line (a). }
\label{GR}
\end{figure} 


Finally, the improved UV \ion{Fe}{ii} model consists of the 5 multiplets: 60 (a${\ }^4D$ - z${\ }^6F^o$), 
  61 (a${\ }^4D$ - z${\ }^6P^o$ ), 62 (a${\ }^4D$ - z${\ }^4F^o$), 63 (a${\ }^4D$ - z${\ }^4D^o$) and 78 (a${\ }^4P$ - z${\ }^4P^o$), and empirically added 'I Zw 1 lines' represented with Gaussians at $\lambda$2720 \AA \ and $\lambda$2840 \AA, which relative intensities are fixed to the value obtained from  the best fit of I Zw 1 \ion{Fe}{ii} lines in the UV (see Table \ref{tbl-1}, last row). The multiplets used for the UV \ion{Fe}{ii} model are shown in Grotrian diagram in Fig. \ref{GR}, with additional multiplet 78. The example of multiplets emission, with empirically added 'I Zw 1 lines' is shown in Fig. \ref{GR}a, and the shape of total UV \ion{Fe}{ii} model is shown in \ref{GR}b.
  
  We assume that the widths and shifts of all UV \ion{Fe}{ii} lines are the same, since they probably originate from the same emission region, so the improved UV \ion{Fe}{ii} model has 9 free parameters: width, shift, intensity of 'I Zw 1 lines', intensities for each of 5 multiplets, and the temperature, which is used to calculate the relative intensities of the lines within each multiplet (see Eq. (\ref{eq:1}), Sec \ref{sec:2.2.2}). The relative intensities of the lines within each multiplet were calculated for the T = 5000 K, 10000 K and 50000 K, normalized to the strongest line in the multiplet and shown in Table \ref{tbl-1}. For calculation of the relative intensities with Eq. (\ref{eq:1}), we used the updated atomic data from National Institute of Standards and Techology (NIST) Atomic Spectra Database\footnote{\url{https://physics.nist.gov/asd}} and Kurucz atomic database \citep{kur94}.
  
\begin{figure}
\centering
\includegraphics[width=0.40\textwidth]{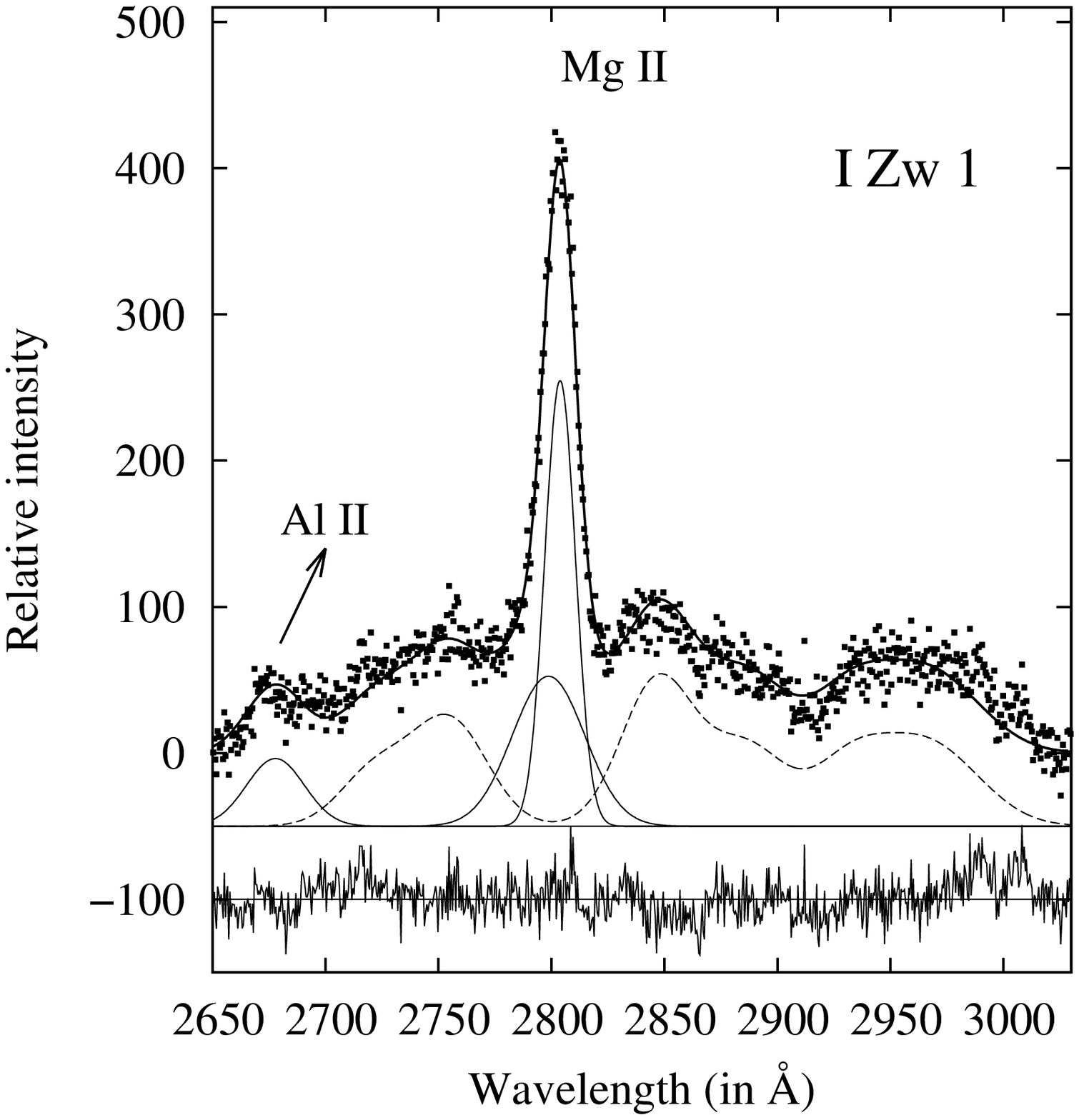}
\includegraphics[width=0.40\textwidth]{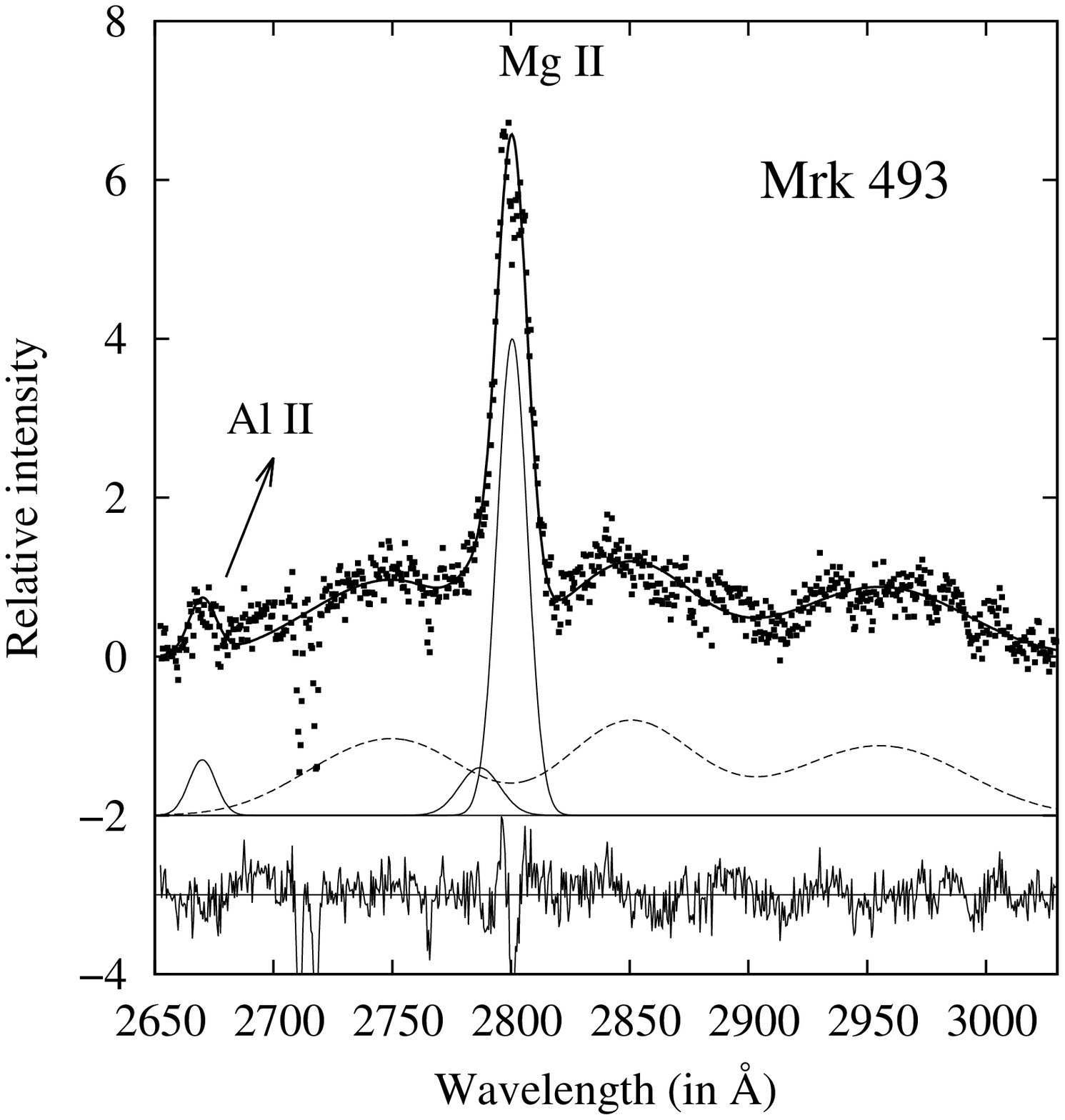}
\caption{ The fit of NLSy1 galaxies I Zw 1 (top) and Mrk 493 (bottom) with the improved UV \ion{Fe}{ii} model, denoted with dashed line.}
\label{IZw1}
\end{figure}

The fits of I Zw 1 and Mrk 493 with improved UV \ion{Fe}{ii} semi-empirical model are shown in Fig \ref{IZw1}. As it can be seen, the improved  UV \ion{Fe}{ii} model can fit very well the UV \ion{Fe}{ii} lines in NLSy1 galaxies.

 \citet{kov10} have found that the flux ratio of the \ion{Fe}{ii} multiplets is not constant, i.e. it can vary from object to object. In this work, we found that the ratio of the UV \ion{Fe}{ii} multiplets slightly differs in two observed NLSy1 galaxies: I Zw 1 and Mrk 493. Therefore, the UV \ion{Fe}{ii} model with free parameters for multiplet intensity enables fitting the objects with different strength of \ion{Fe}{ii} multiplets, which is not possible with other UV \ion{Fe}{ii} templates based on I Zw 1 \ion{Fe}{ii} emission lines, or theoretically calculated, with fixed relative intensities of UV \ion{Fe}{ii} lines. This is an advantage of the UV \ion{Fe}{ii} semi-empirical model that we have developed.

We compared this improved UV \ion{Fe}{ii} model with theoretical models given in \cite{SP03} and \cite{bv08} and empirical UV \ion{Fe}{ii} templates based on the spectrum of I Zw 1, and presented in \cite{ts06} and \cite{vw01}. The applied version of theoretical model from \cite{SP03} was calculated for photoionization model A of the BLR, with included Ly$\alpha$ and Ly$\beta$ pumping, and the applied version of theoretical model from \cite{bv08} was calculated for log[$\mathrm{n_H}$/(cm$^{-3}$)]=11, [$\xi$]/(1 km s$^{-1}$)=20 and log[$\mathrm{\Phi_H}$/(cm$^{-2}$ s$^{-1}$)]=20.5.

\begin{figure*}
\centering
\includegraphics[width=0.42\textwidth]{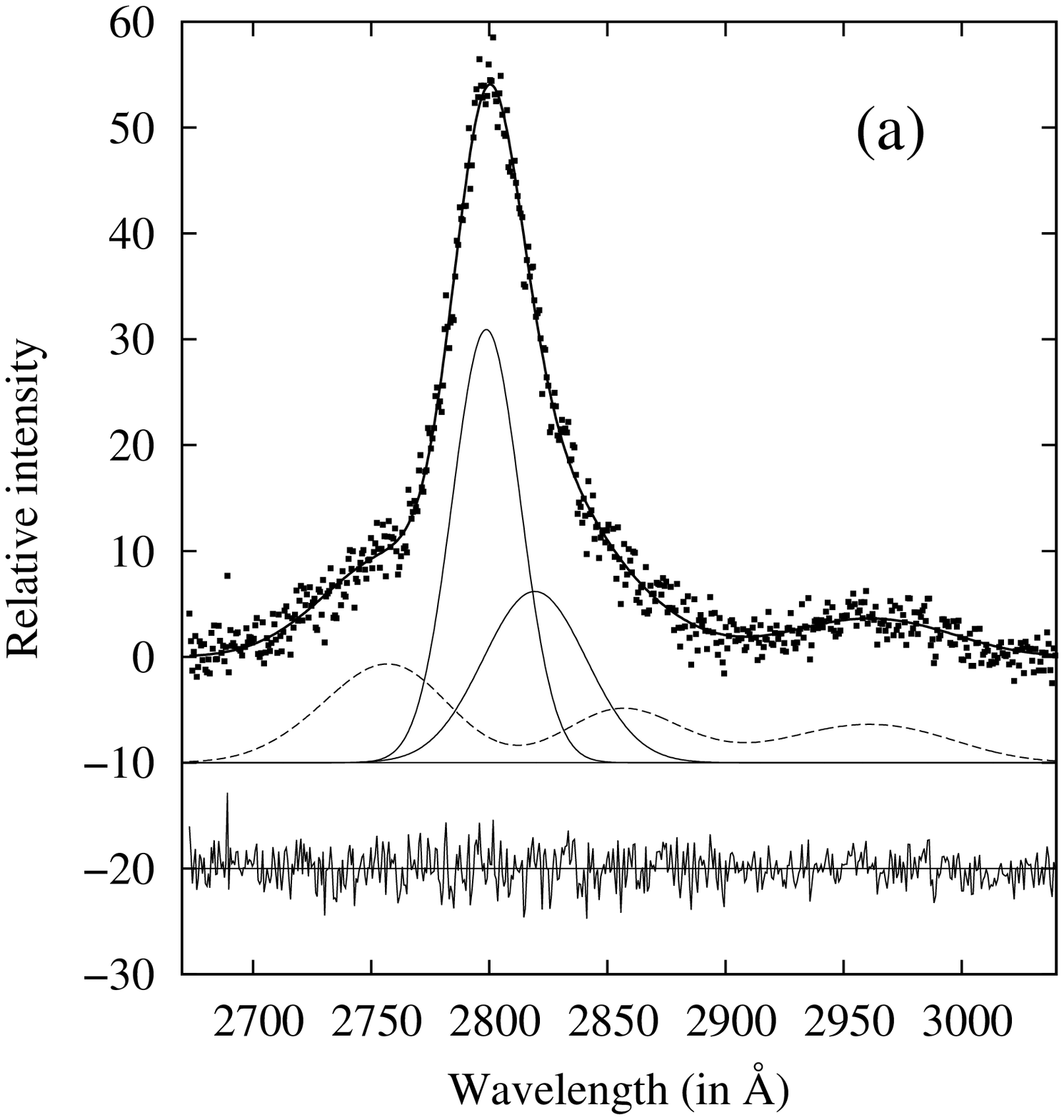}
\includegraphics[width=0.42\textwidth]{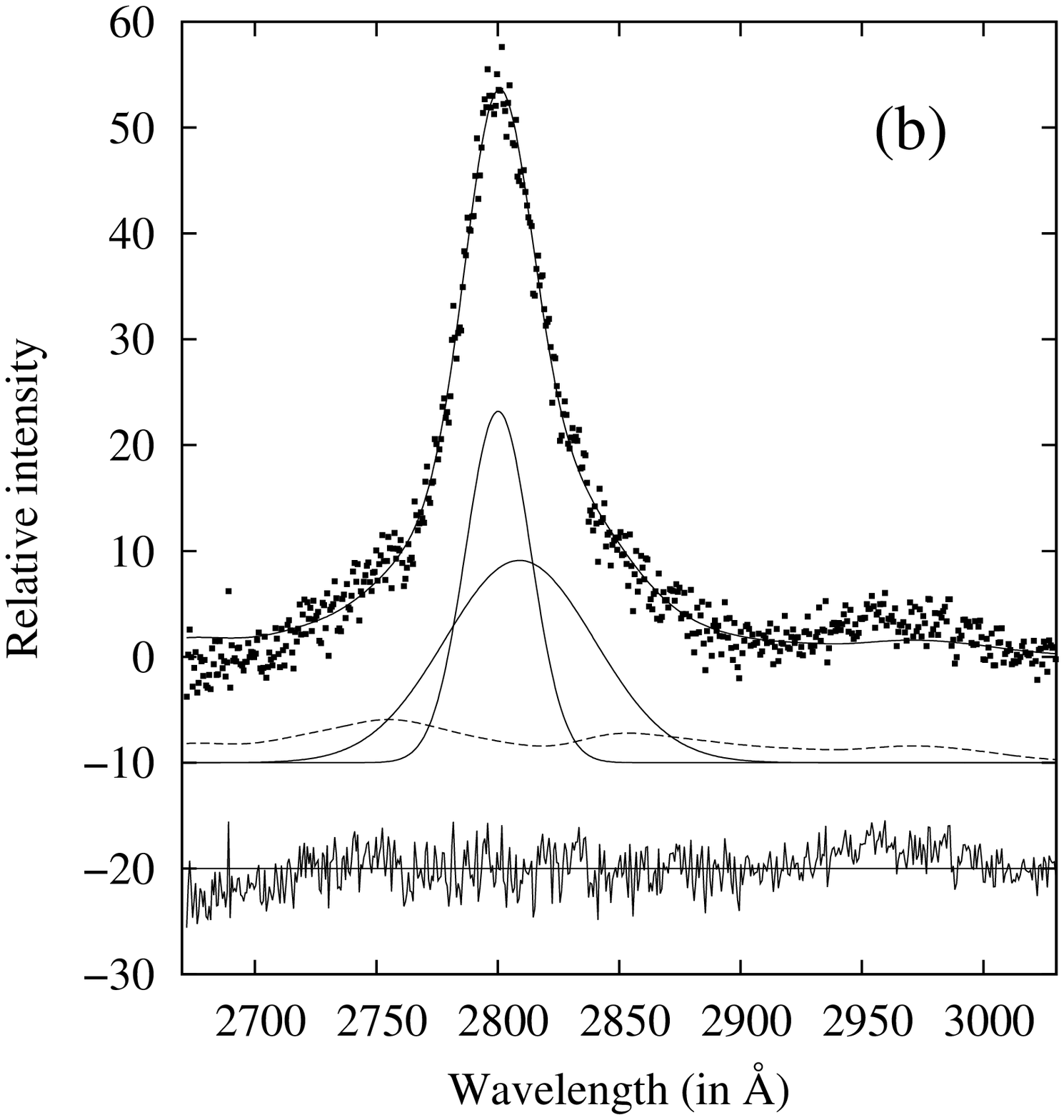}
\includegraphics[width=0.42\textwidth]{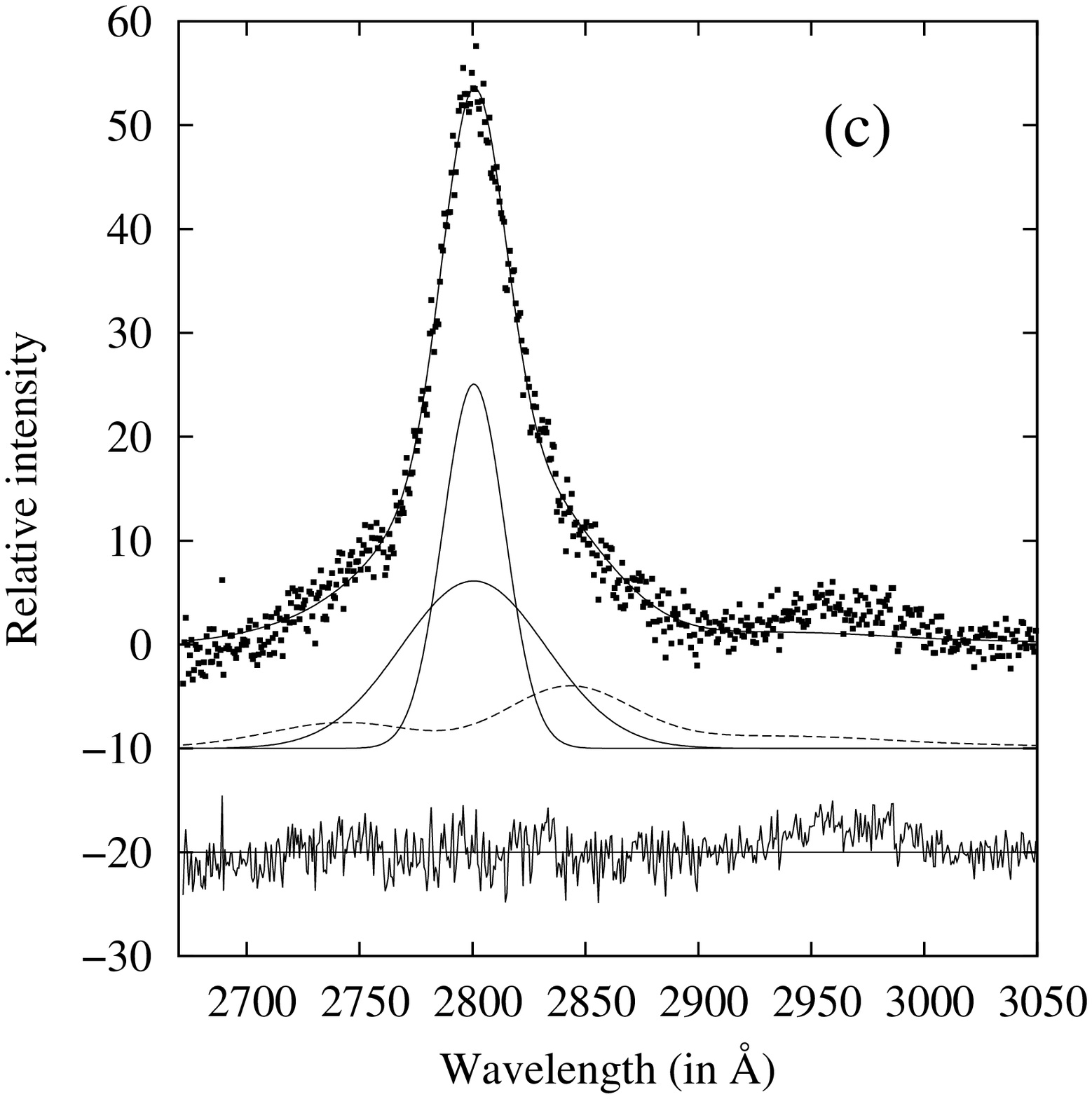}
\includegraphics[width=0.42\textwidth]{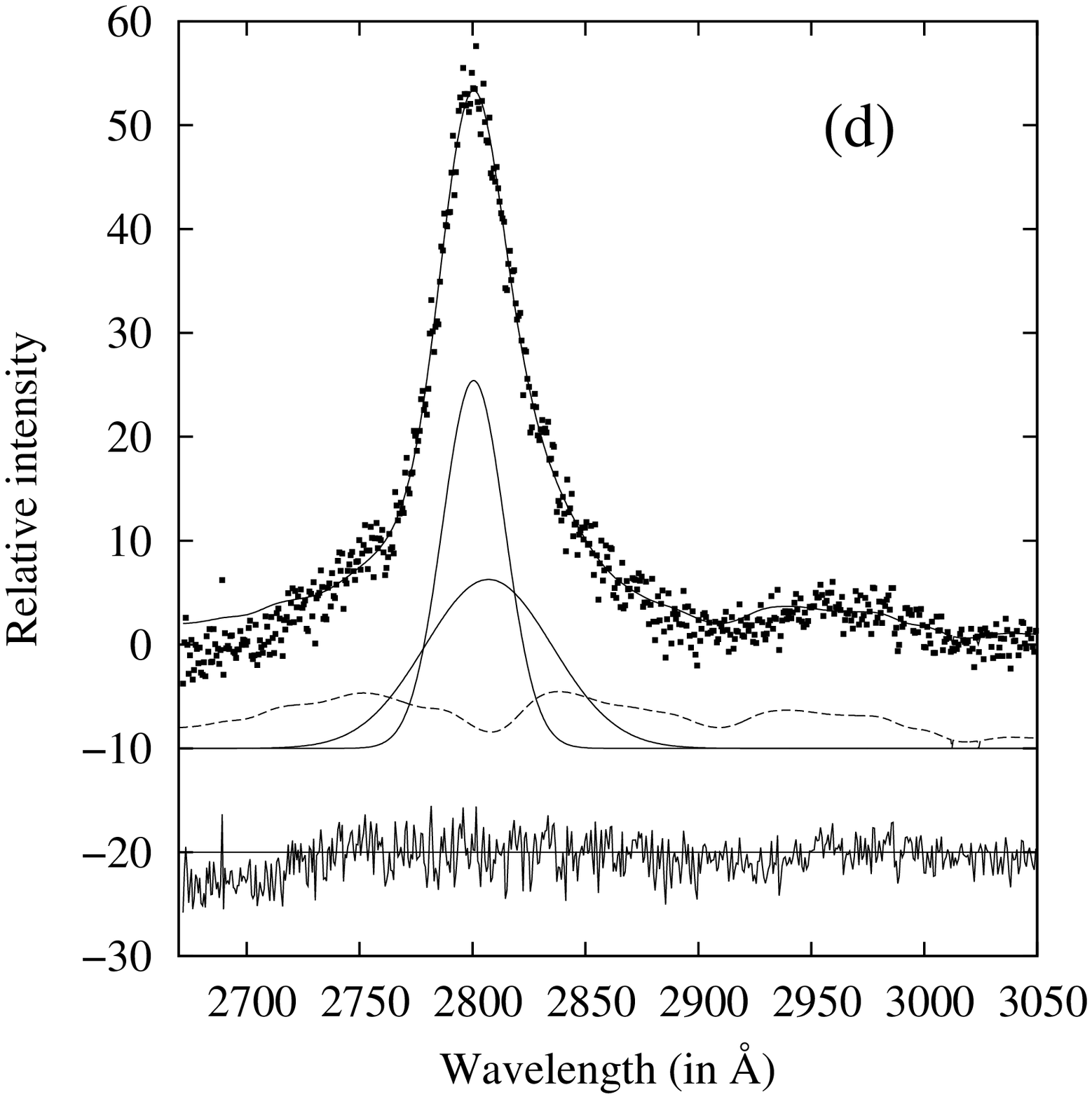}
\includegraphics[width=0.42\textwidth]{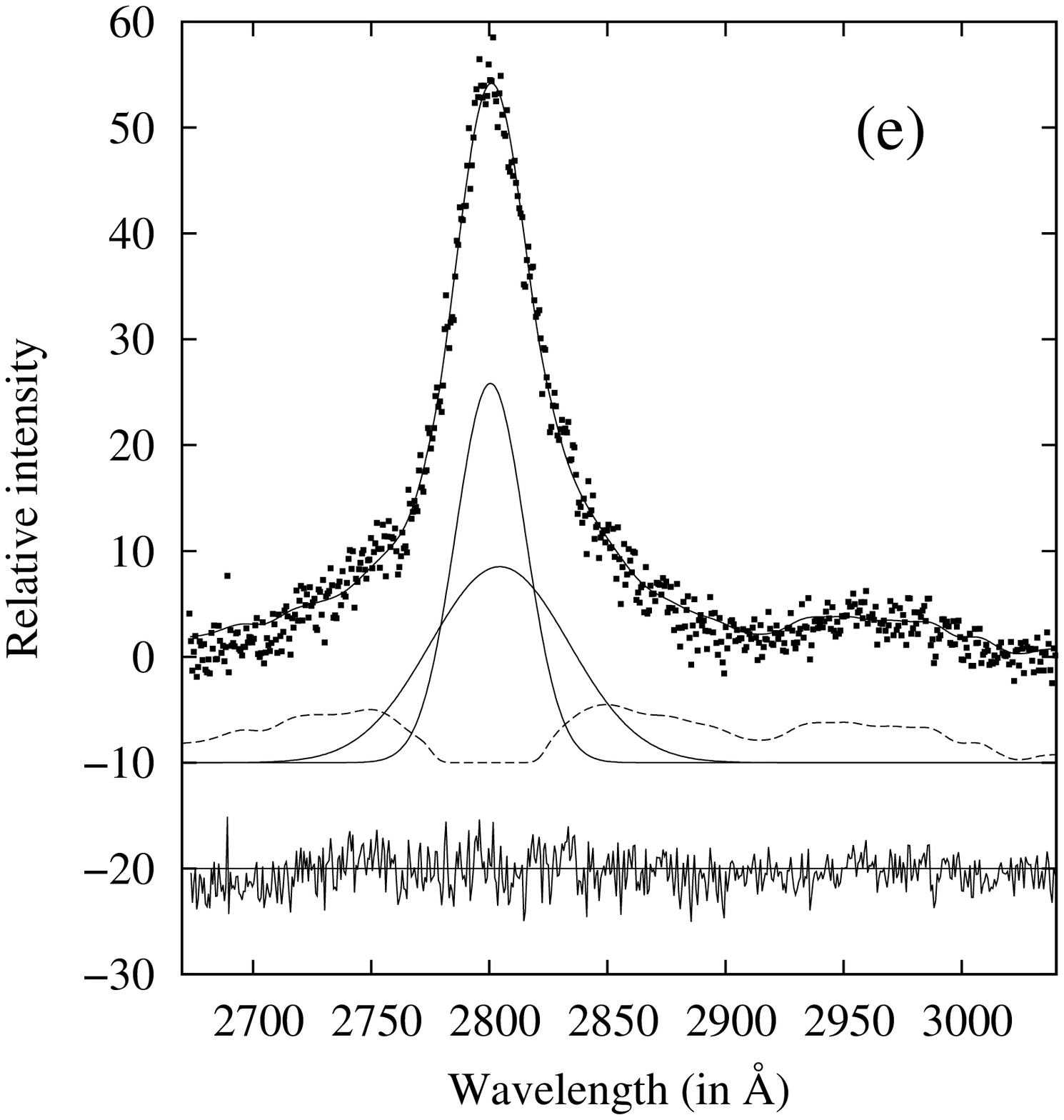}
\includegraphics[width=0.42\textwidth]{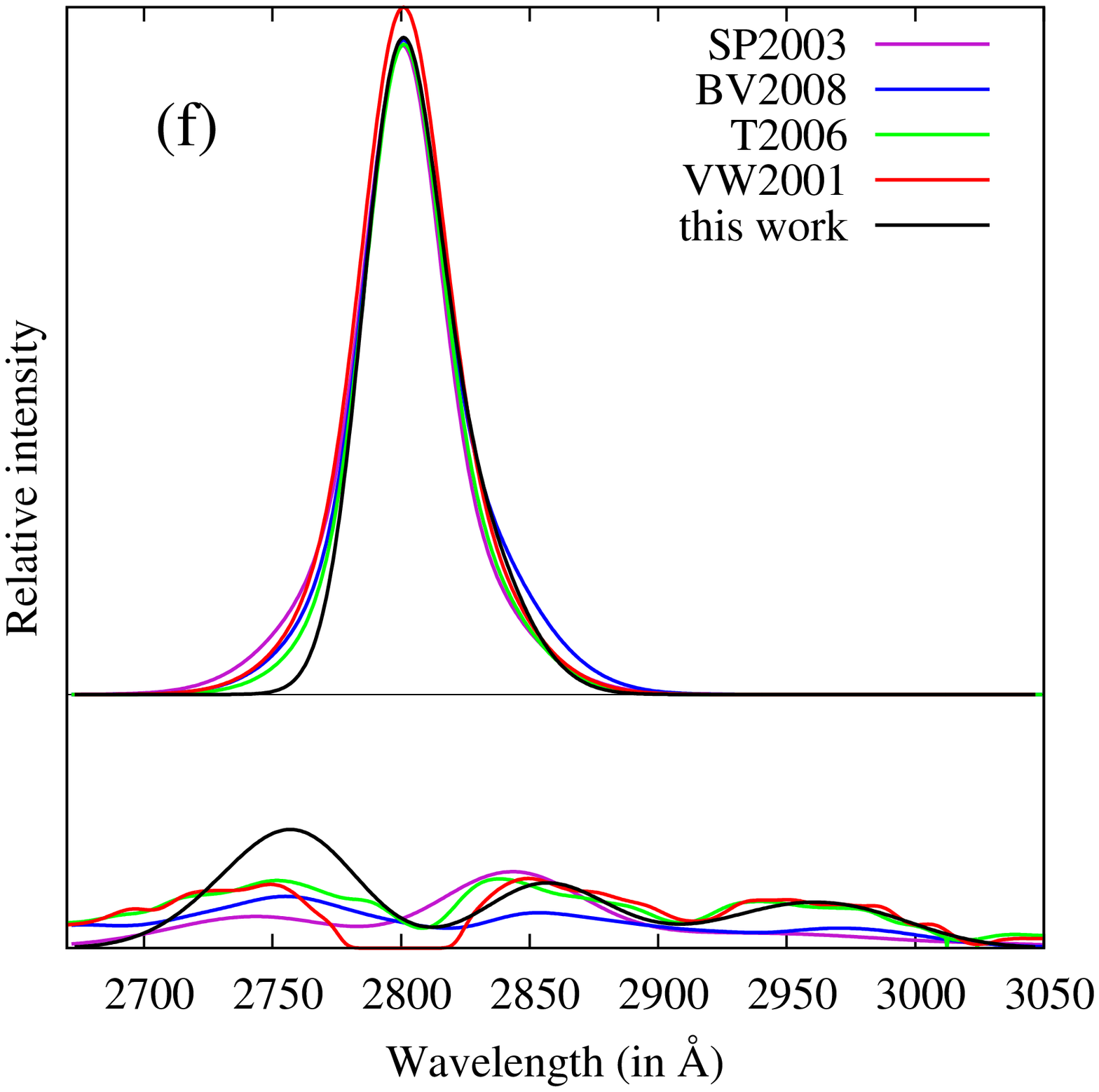}
\caption{Fits of SDSS J135045.66$+$233145.2 with different UV \ion{Fe}{ii} templates: (a) this work, (b) \citet{bv08} (BV2008), (c) \citet{SP03} (SP2003), (d) \citet{ts06} (T2006), (e) \citet{vw01} (VW2001), and (f) comparison of the UV \ion{Fe}{ii} lines (bottom) and \ion{Mg}{ii} profiles (top) obtained from the best fit using different UV \ion{Fe}{ii} models (for colored version of this Figure see electronic version). We obtained the \ion{Mg}{ii} profile with red asymmetry after decomposition with our improved UV \ion{Fe}{ii} model (a).}
\label{figA3}
\end{figure*}

\begin{figure*}
\centering
\includegraphics[width=0.43\textwidth]{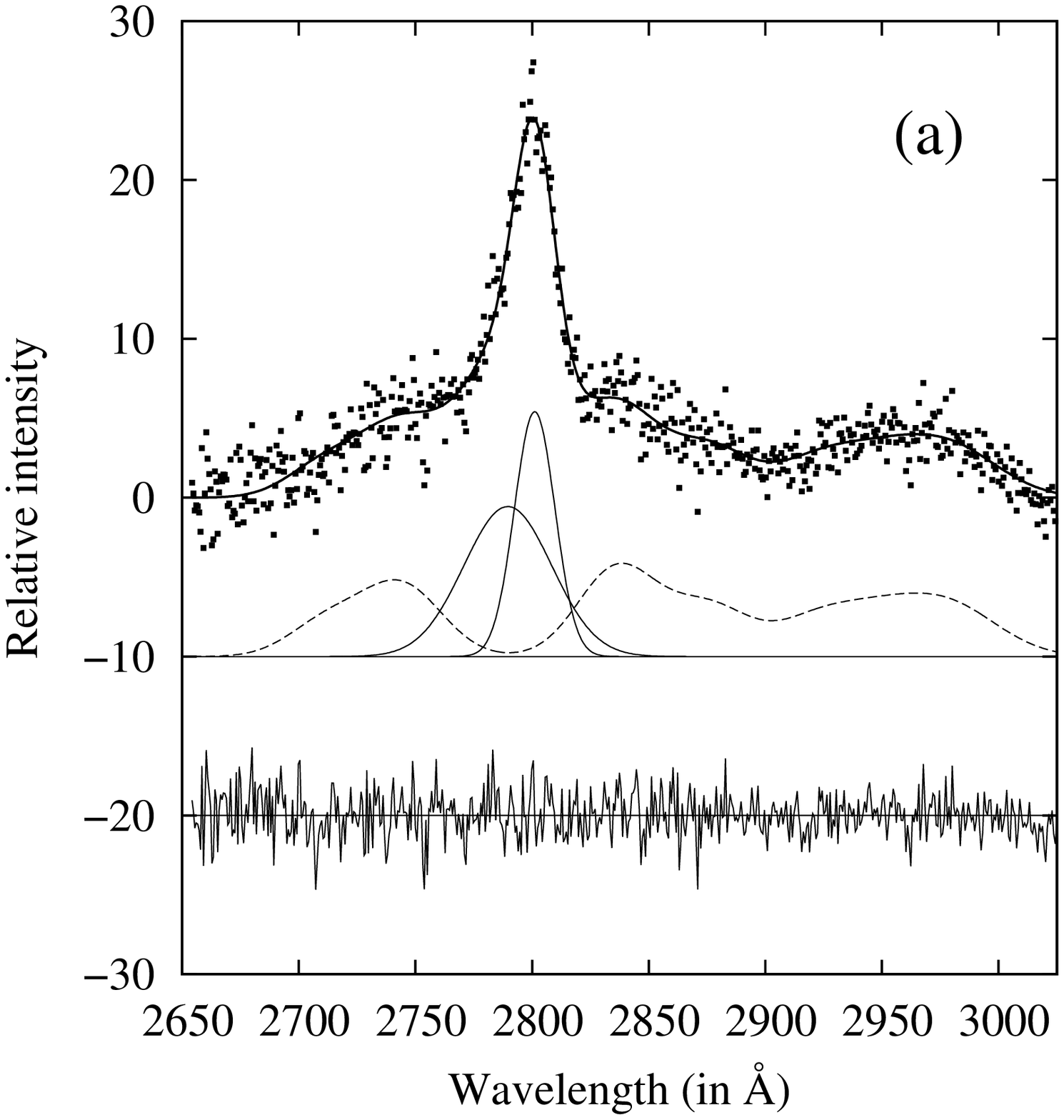}
\includegraphics[width=0.43\textwidth]{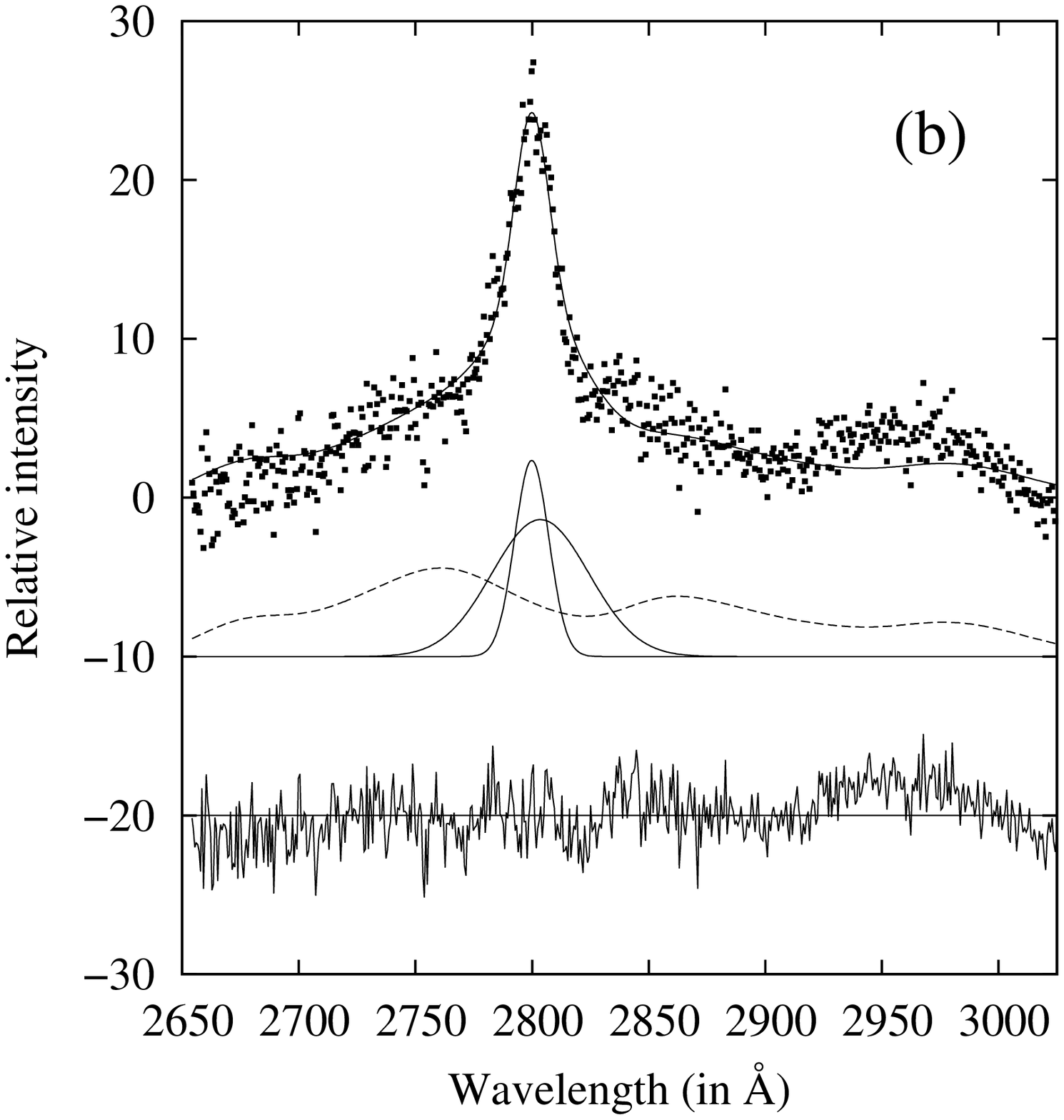}
\includegraphics[width=0.43\textwidth]{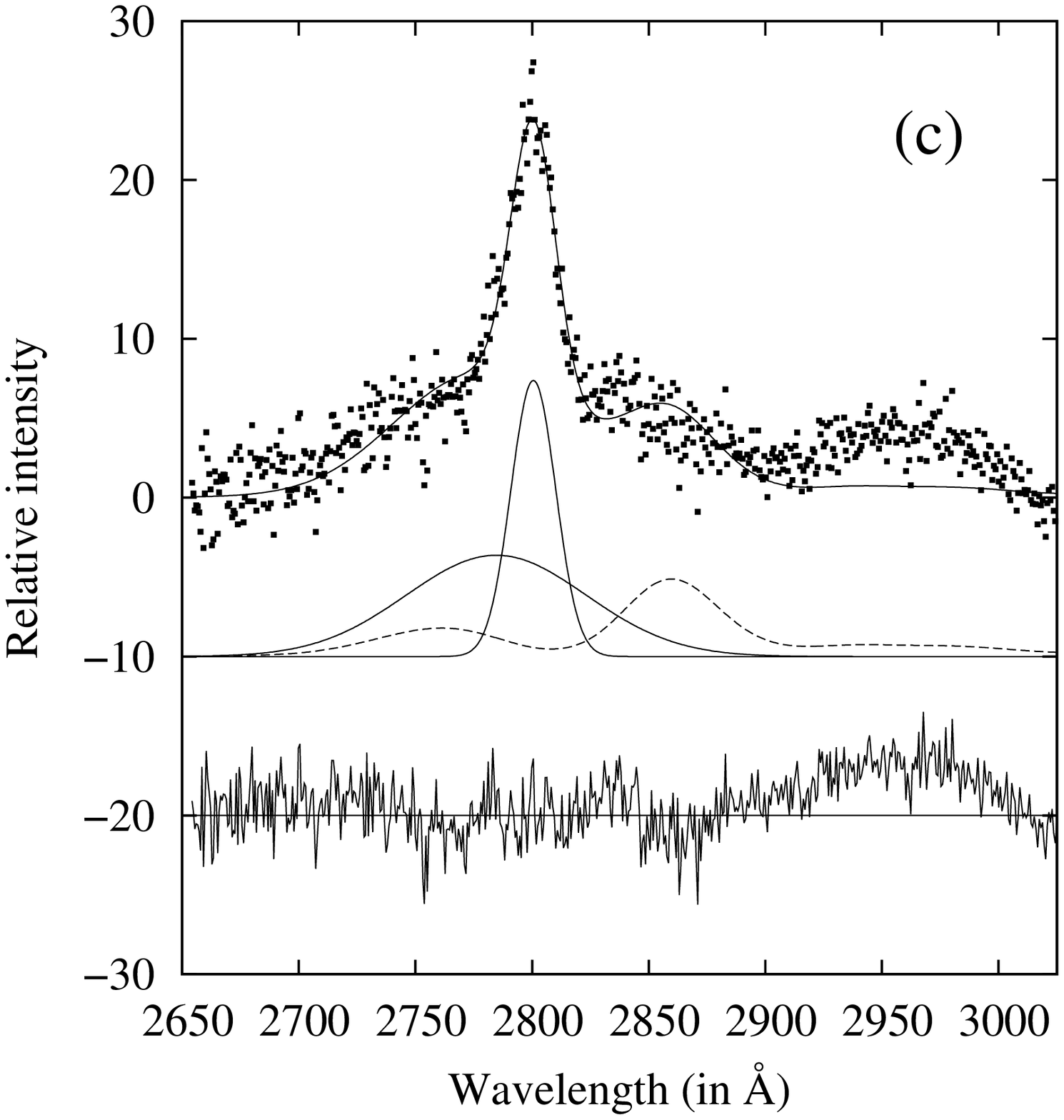}
\includegraphics[width=0.43\textwidth]{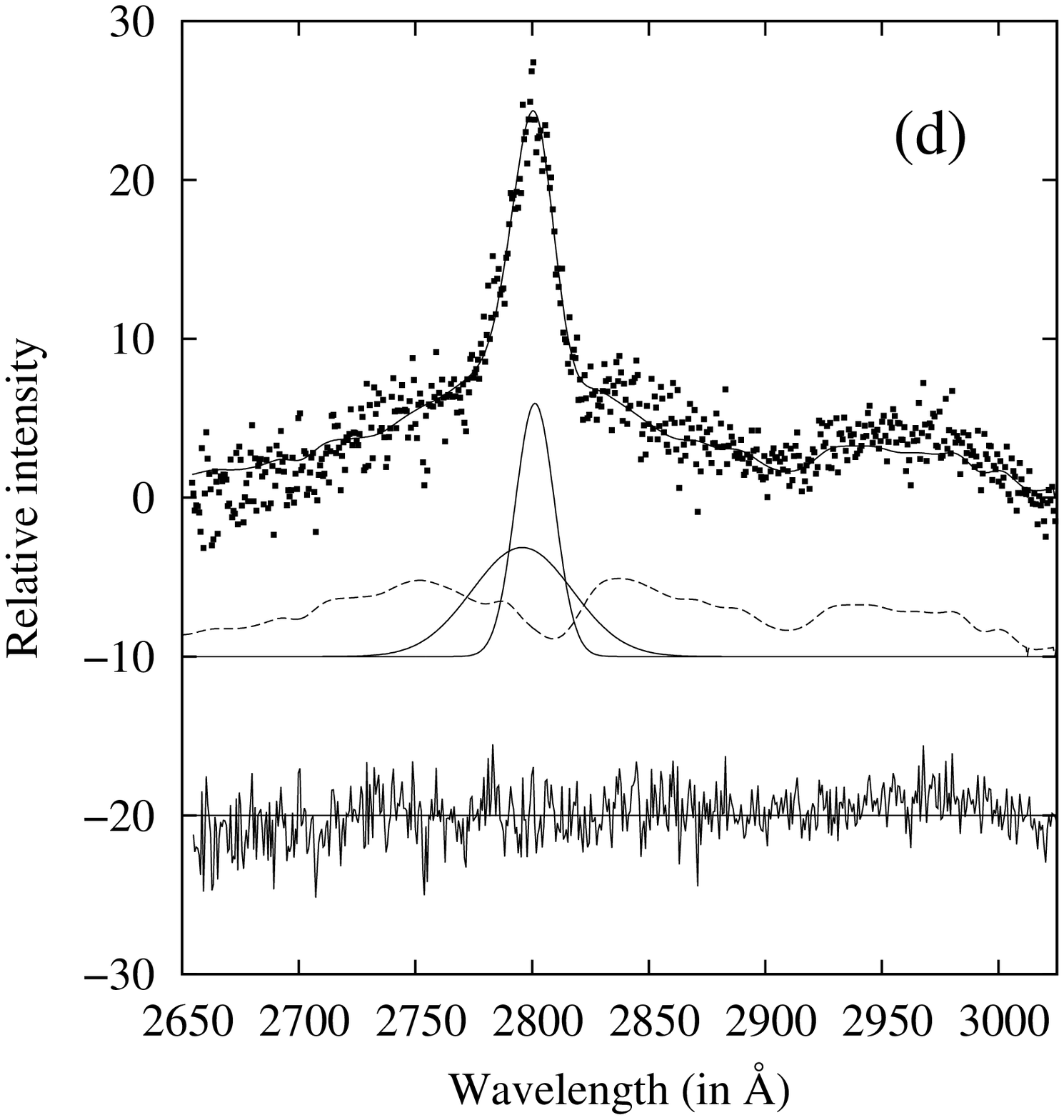}
\includegraphics[width=0.43\textwidth]{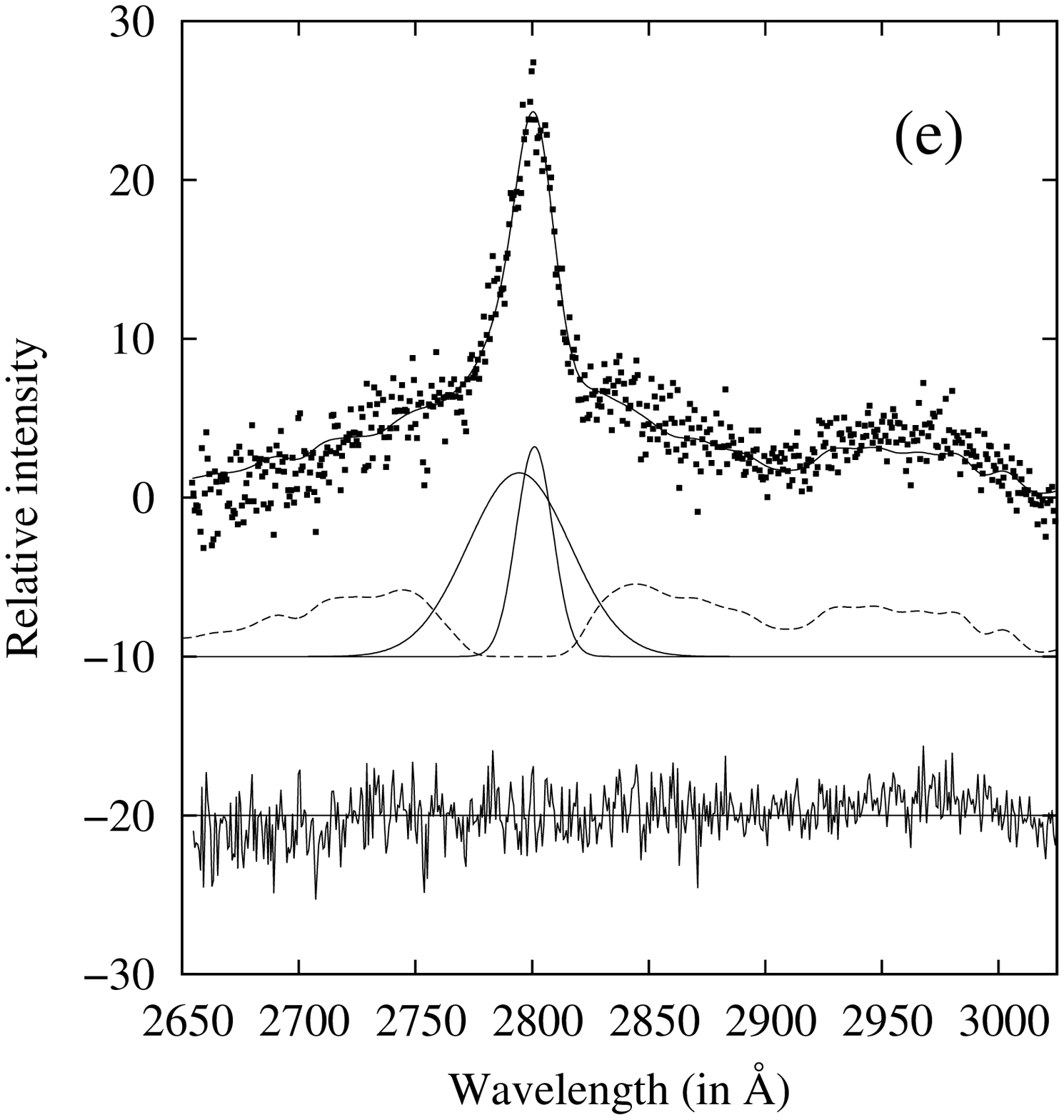}
\includegraphics[width=0.43\textwidth]{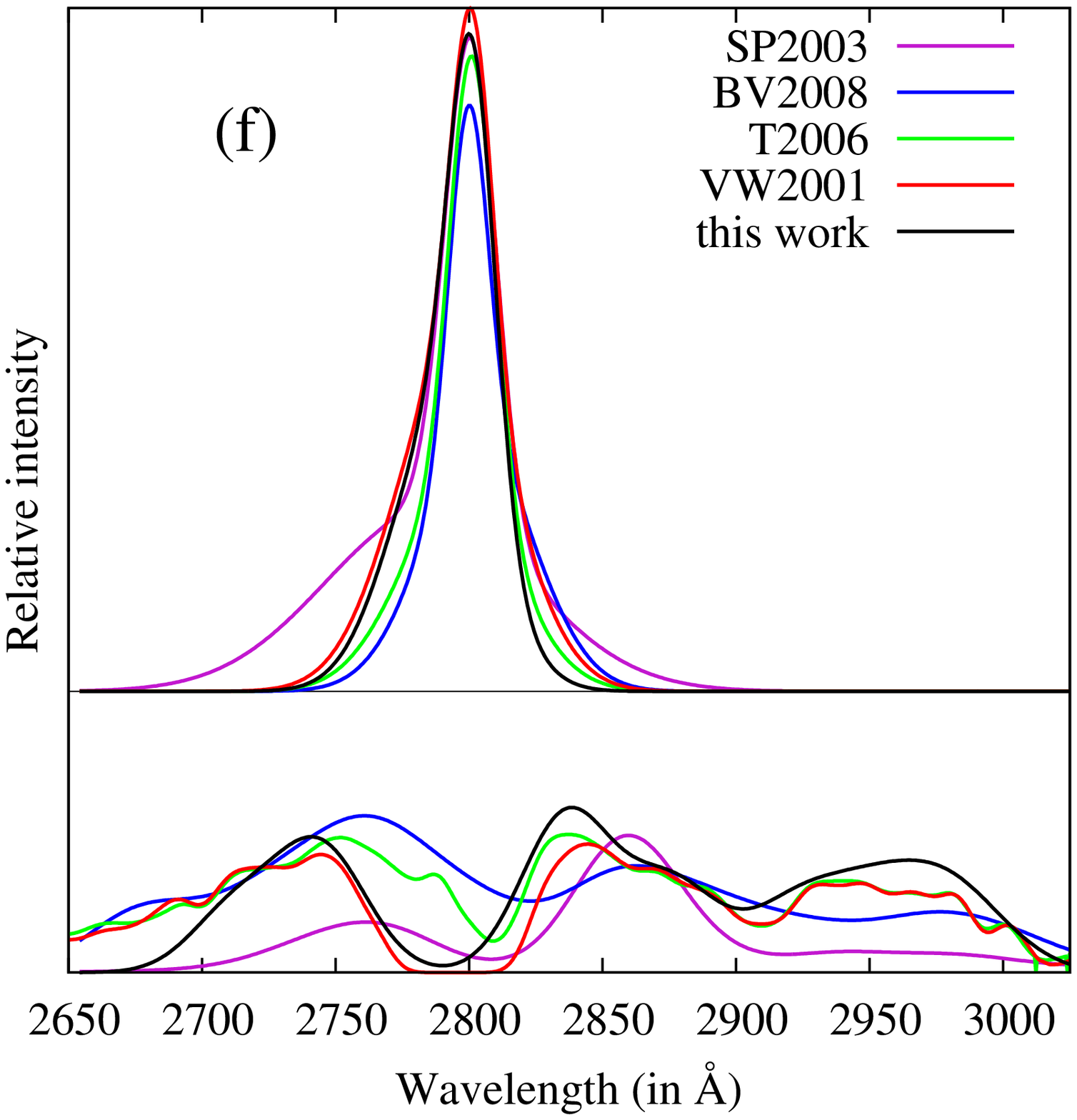}
\caption{ The same as in Fig. \ref{figA3} but for object SDSS J105540.81$+$421241.2, for which we obtained the \ion{Mg}{ii} profile with blue asymmetry with our decomposition model. }
\label{figA4}
\end{figure*}

\begin{figure*}
\centering
\includegraphics[width=0.43\textwidth]{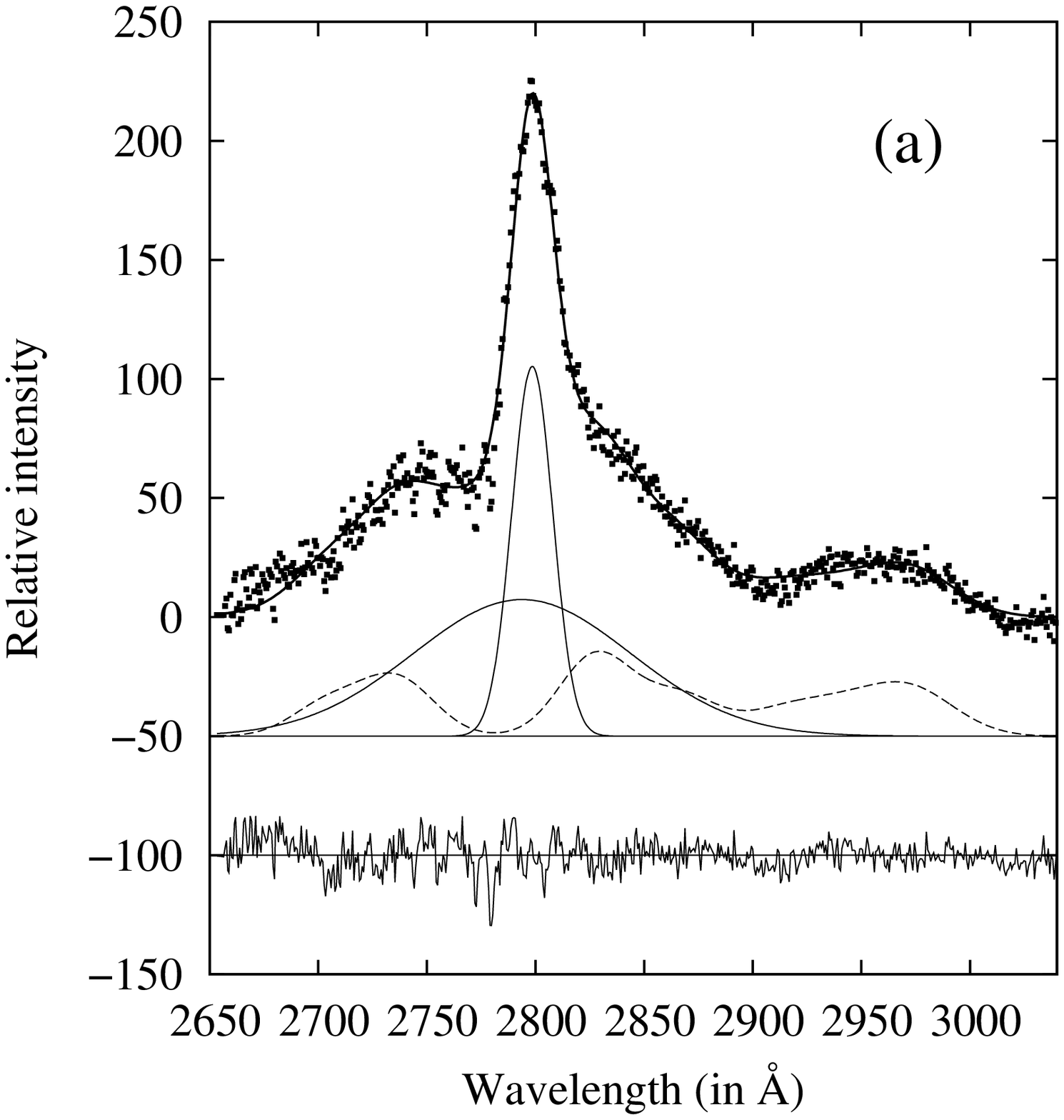}
\includegraphics[width=0.43\textwidth]{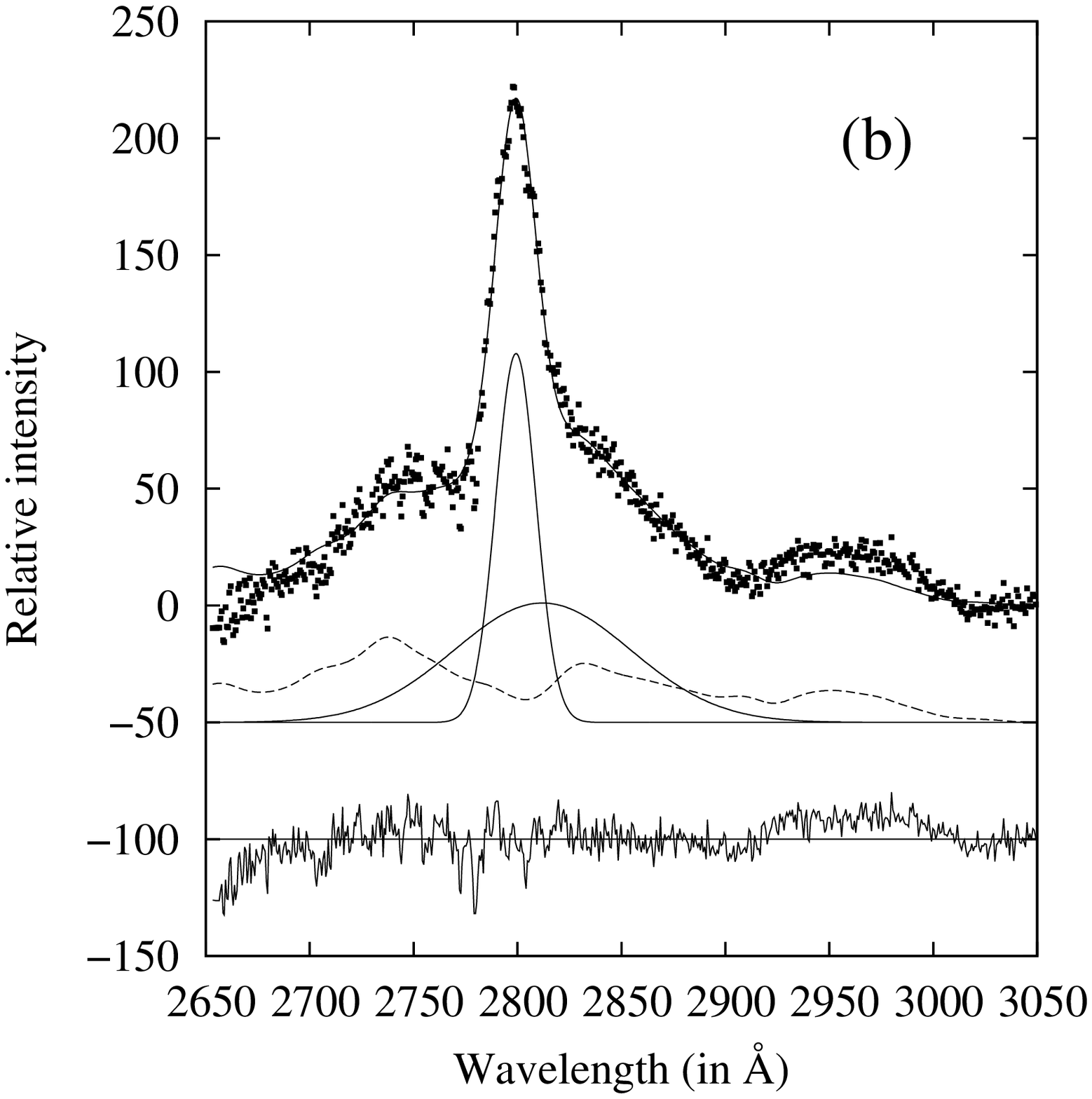}
\includegraphics[width=0.43\textwidth]{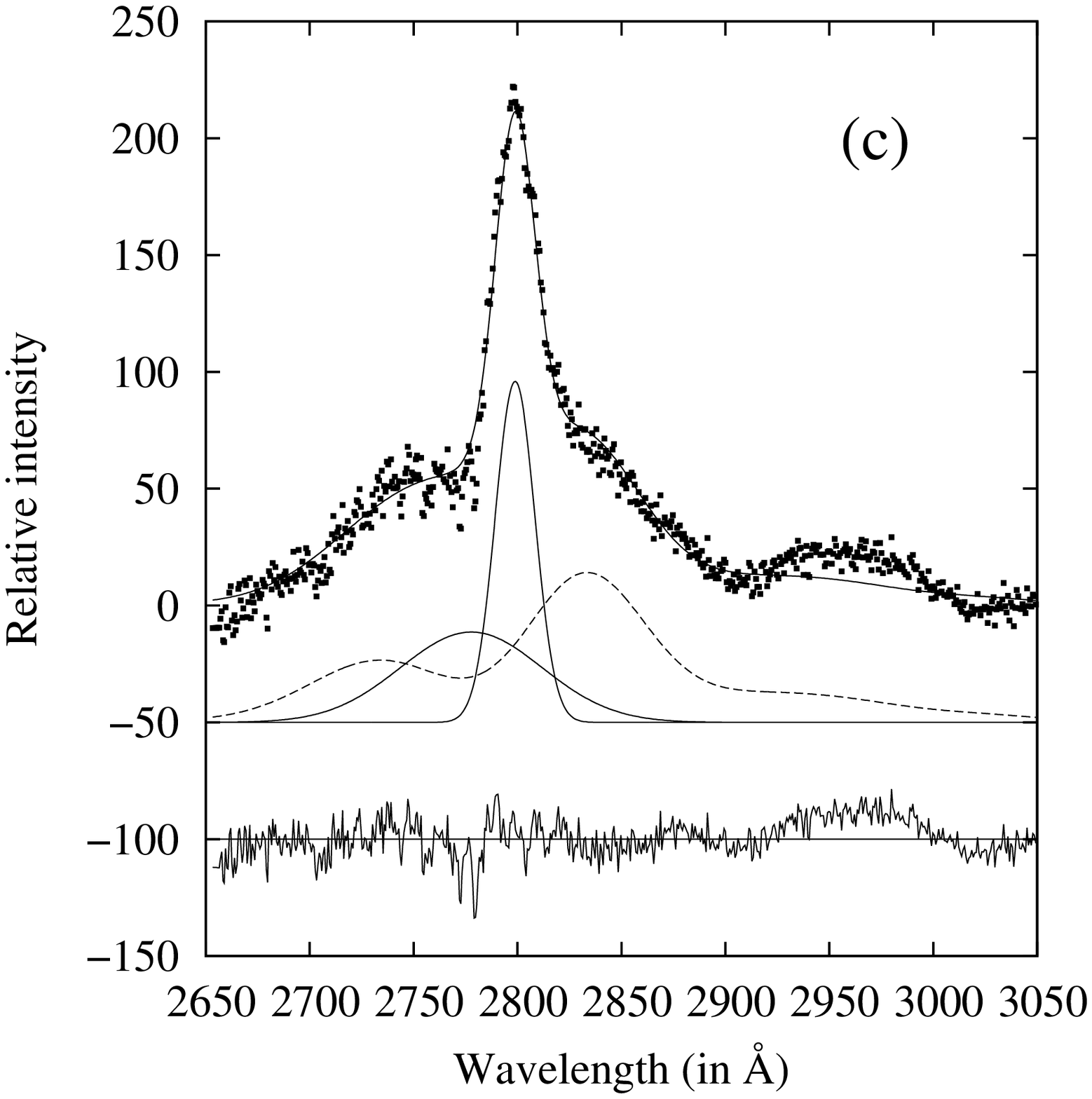}
\includegraphics[width=0.43\textwidth]{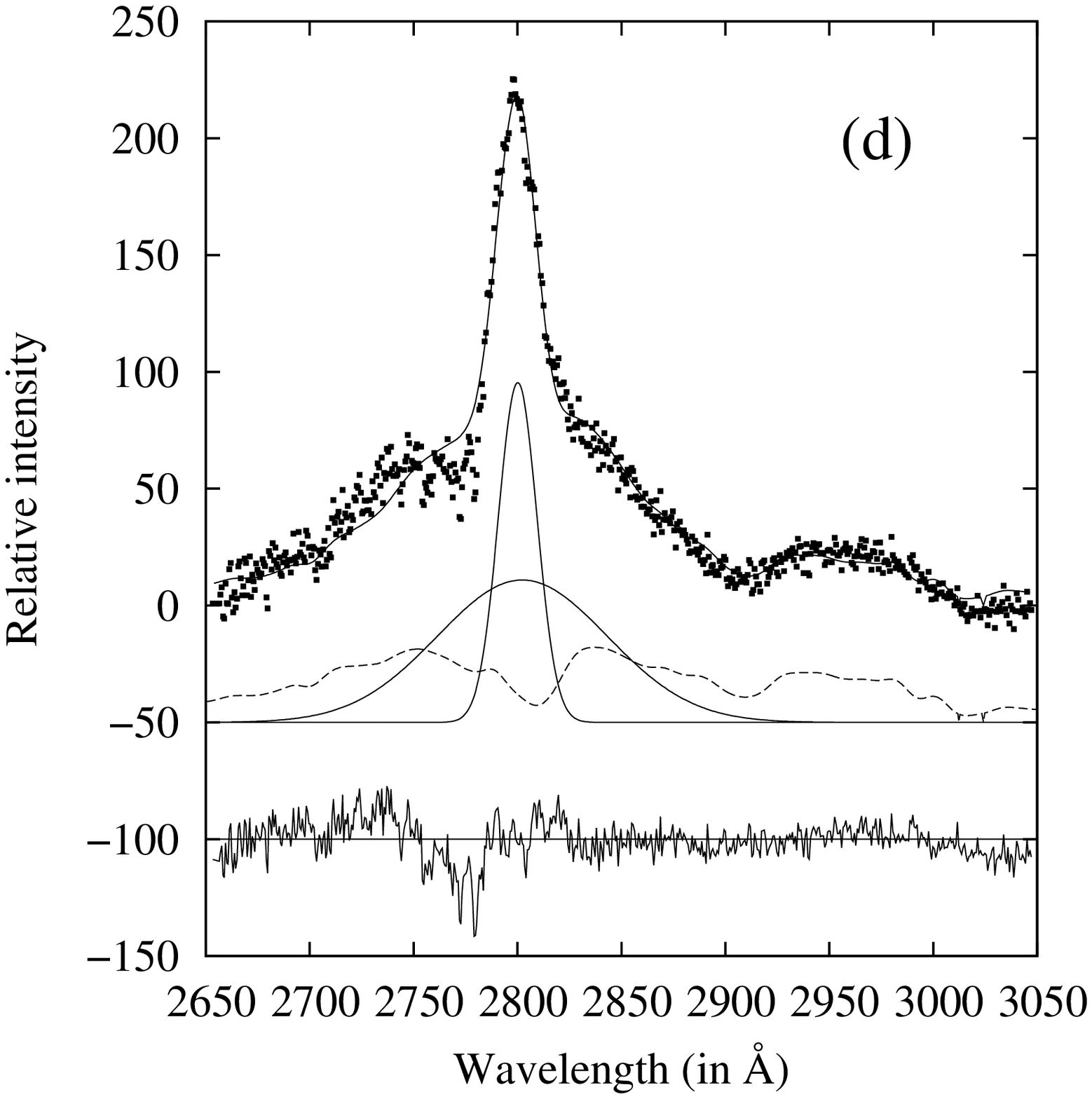}
\includegraphics[width=0.43\textwidth]{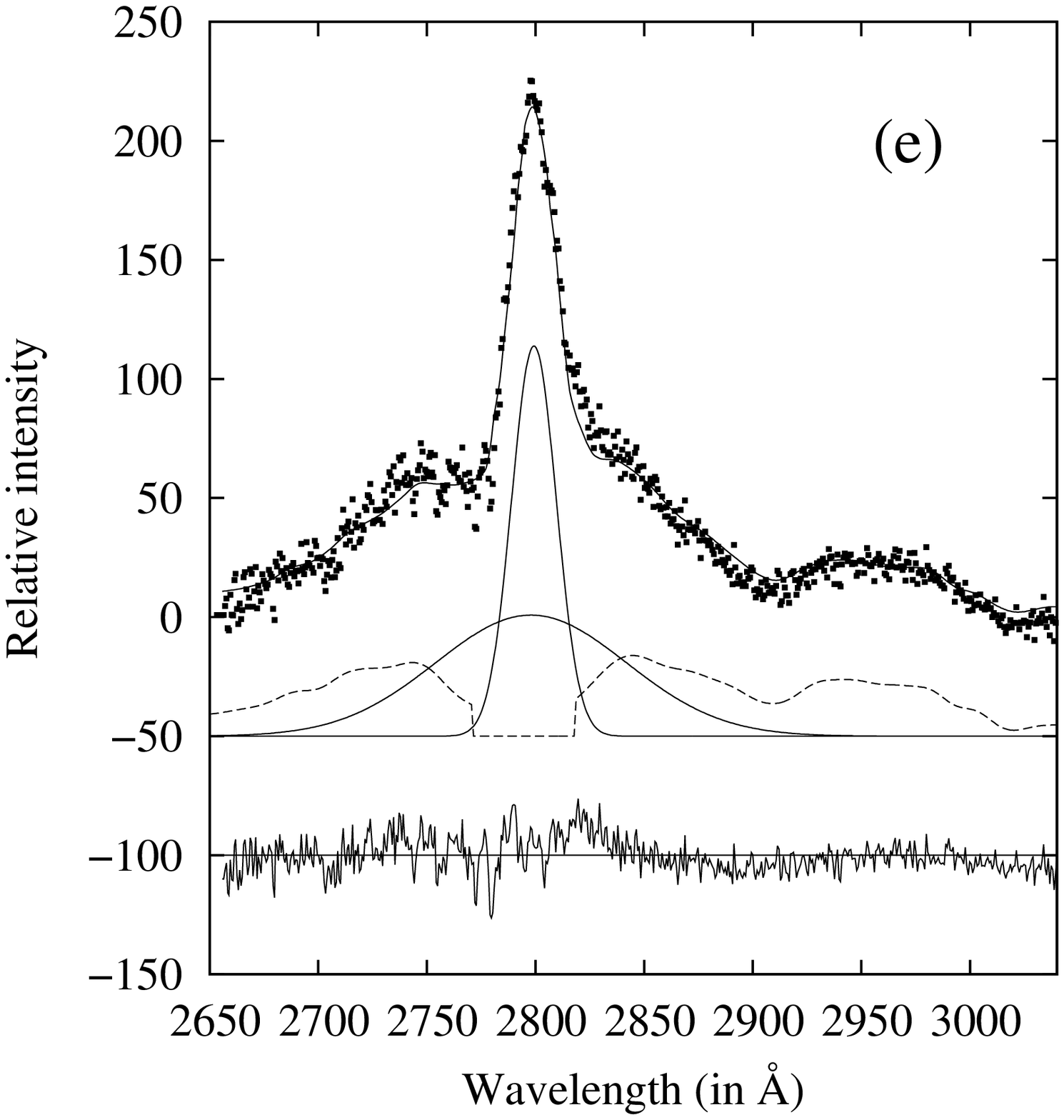}
\includegraphics[width=0.43\textwidth]{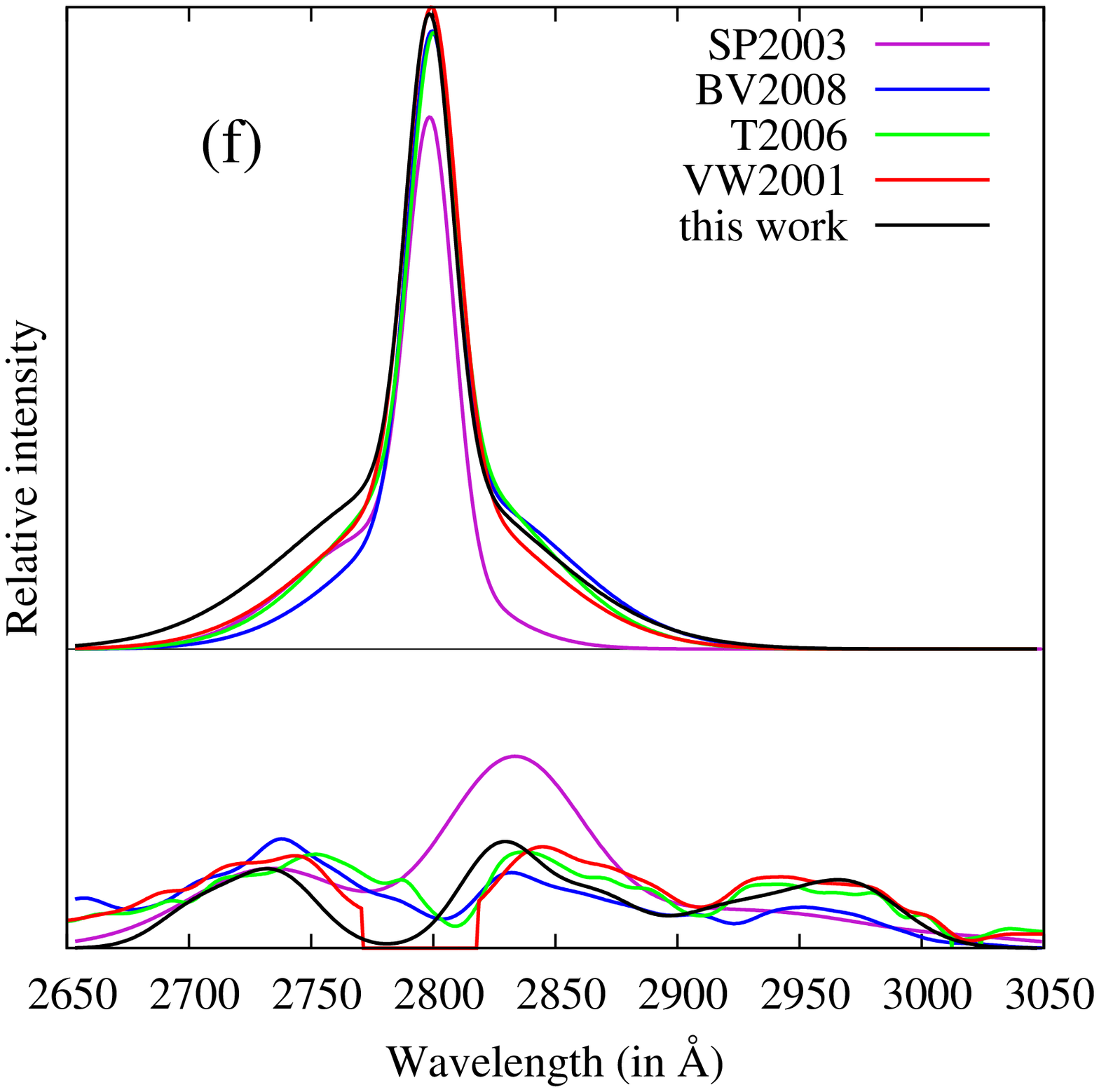}
\caption{ The same as in Fig. \ref{figA3} but for object SDSS J122454.46$+$212246.4, for which we obtained nearly symmetric \ion{Mg}{ii} profile with our decomposition model.}
\label{figA5}
\end{figure*}

To test the different UV \ion{Fe}{ii} templates, we chose three examples of spectra where we obtained different shapes of \ion{Mg}{ii} lines after applying our improved UV \ion{Fe}{ii} model: with red asymmetry in \ion{Mg}{ii} profile (Fig. \ref{figA3}), with blue asymmetry in \ion{Mg}{ii} profile (Fig. \ref{figA4}), and with no significant asymmetry of \ion{Mg}{ii} (Fig. \ref{figA5}). We fitted these examples with different templates, and compared the obtained 
profiles of \ion{Fe}{ii} and \ion{Mg}{ii} lines with different \ion{Fe}{ii} models (see Figs \ref{figA3}, \ref{figA4} and \ref{figA5}).

 It can be seen that all considered UV \ion{Fe}{ii} templates have different shapes, which result with different spectral decomposition of \ion{Mg}{ii}$+$UV \ion{Fe}{ii} lines in all examples, and consequently with different shape of extracted pure \ion{Mg}{ii} profile. The empirical templates of \cite{ts06} and \citet{vw01} are both made using the spectrum of the I Zw 1 galaxy, and therefore these templates are the most similar among the others, 
 and they give the most similar fits of \ion{Mg}{ii}$+$UV \ion{Fe}{ii} lines. The improved UV \ion{Fe}{ii} model presented in this work slightly vary from case to case comparing to the empirical templates, while the theoretical models of \cite{SP03} and \cite{bv08} in some cases have larger incompatibility with other three templates.

Namely, the theoretical models of \cite{SP03} and \cite{bv08} cannot fit well the 
UV \ion{Fe}{ii} bump at $\sim$2900-3000  \AA \ range 
in all given examples (see Fig. \ref{figA3}, \ref{figA4} and \ref{figA5}, case (b) and (c)). The discrepancy is the strongest in the spectrum shown in Fig. \ref{figA4}, with strong UV \ion{Fe}{ii} lines at $\sim$2900-3000  \AA \ range.

The empirical templates of \cite{ts06} and \citet{vw01} fit well all given examples. However, in the case of complex \ion{Mg}{ii}$+$UV \ion{Fe}{ii} shapes 
(see Fig \ref{figA5}), the fits with these templates have small discrepancy (at $\sim$2700-2780  \AA \ range in Fig. \ref{figA5} (d), and at $\sim$2820-2840  \AA \ range in Fig. \ref{figA5} (e)). Additionally, the empirical template of \cite{vw01} has artificial cut from 2770 \AA \ to 2819 \AA \ (under the \ion{Mg}{ii} emission), i.e. the UV \ion{Fe}{ii} emission is set to zero in the range 2770--2819 \AA, that also can artificially change the \ion{Mg}{ii} broad line 
profiles. Therefore, we developed the semi-empirical UV \ion{Fe}{ii} model \citep[an improved template of][]{kp15} which can fit well all these different \ion{Mg}{ii} line profiles (see Fig. \ref{figA3}, \ref{figA4} and \ref{figA5}, case (a)) and is able to account possible 
difference between the multiplet intensities which may be caused by slight different physical conditions in the UV \ion{Fe}{ii} emitting region.
The template is available on-line, through the Serbian Virtual Observatory\footnote{\url{http://servo.aob.rs/}}.

\begin{table*}
\begin{center}

\caption{ The improved UV \ion{Fe}{ii} $\lambda\lambda$ 2650-3050 \AA \ model: the relative intensities of the UV \ion{Fe}{ii} emission lines within multiplets calculated for different temperatures (see Eq. (\ref{eq:1})) and using updated atomic data. The lines with the strength of four order of magnitude lower than the strongest line from multiplet are excluded from model, since their contribution in flux is negligible. \label{tbl-1} }
\begin{tabular}{c c c c c}

\hline
\hline

Wavelength&Transitions&\multicolumn{3}{c} {Relative intensity}\\
& &   T=5000 K & T=10000 K & T=50000 K\\
\hline \\
 & multiplet \ a${\ }^4D$ - z${\ }^6F^o$ (60)&\\
2907.853 & a${\ }^4D_{7/2}$ - z${\ }^6F^o_{5/2}$   &   0.008   &    0.009      &         0.009       \\

2916.148 & a${\ }^4D_{7/2}$ - z${\ }^6F^o_{7/2}$   &   0.006   &    0.006      &        0.006         \\

2926.585  & a${\ }^4D_{7/2}$ - z${\ }^6F^o_{9/2}$  &   1.000   &     1.000     &       1.000          \\

2939.507   & a${\ }^4D_{5/2}$ - z${\ }^6F^o_{3/2}$ &   0.015   &     0.016     &        0.016         \\

2945.264   &  a${\ }^4D_{5/2}$ - z${\ }^6F^o_{5/2}$&   0.005   &    0.005      &         0.005         \\

2953.774 & a${\ }^4D_{5/2}$ - z${\ }^6F^o_{7/2}$   &   0.835   &     0.849     &       0.861           \\

2961.273 & a${\ }^4D_{3/2}$ - z${\ }^6F^o_{1/2}$   &   0.013   &     0.014     &    0.014              \\

2964.659 & a${\ }^4D_{3/2}$ - z${\ }^6F^o_{3/2}$    &   0.012    &      0.013    &       0.013  \\

2970.514 & a${\ }^4D_{3/2}$ - z${\ }^6F^o_{5/2}$   &   0.355   &    0.366      &     0.376              \\

2975.933 & a${\ }^4D_{1/2}$ - z${\ }^6F^o_{1/2}$   &   0.025   &    0.027      &     0.028           \\
2979.353 &  a${\ }^4D_{1/2}$ - z${\ }^6F^o_{3/2}$  &   0.184   &    0.191      &     0.197             \\

\hline  \\

 & multiplet \ a${\ }^4D$ - z${\ }^6P^o$ (61)&\\

2833.369  &a${\ }^4D_{7/2}$ - z${\ }^6P^o_{5/2}$ &  0.010     &        0.010 &  0.011   \\

2837.737  &a${\ }^4D_{5/2}$ - z${\ }^6P^o_{3/2}$ &   0.001    &         0.001&  0.001    \\

2861.168  &a${\ }^4D_{3/2}$ - z${\ }^6P^o_{3/2}$ &   0.028    &         0.032&  0.035     \\

2868.875  &a${\ }^4D_{5/2}$ - z${\ }^6P^o_{5/2}$ &   0.208    &         0.225&  0.240     \\

2874.854  &a${\ }^4D_{1/2}$ - z${\ }^6P^o_{3/2}$   &   0.028    &         0.032  &  0.036    \\

2880.7563  &a${\ }^4D_{7/2}$ - z${\ }^6P^o_{7/2}$ &   1.000    &       1.000  &  1.000    \\

2892.832  &a${\ }^4D_{3/2}$ - z${\ }^6P^o_{5/2}$ &   0.077    &         0.083&  0.088     \\
2917.462  &a${\ }^4D_{5/2}$ - z${\ }^6P^o_{7/2}$ &   0.079    &          0.079&   0.079  \\

\hline \\
 & multiplet \ a${\ }^4D$ - z${\ }^4F^o$ (62)&\\

2692.834 &a${\ }^4D_{7/2}$ - z${\ }^4F^o_{5/2}$ & 0.002    & 0.003& 0.003      \\

2724.884  &a${\ }^4D_{5/2}$ - z${\ }^4F^o_{5/2}$&0.017     & 0.019& 0.021      \\

2730.734 &a${\ }^4D_{3/2}$ - z${\ }^4F^o_{3/2}$ & 0.031    & 0.036&  0.041     \\

2743.197 &a${\ }^4D_{1/2}$ - z${\ }^4F^o_{3/2}$ & 0.444    & 0.515&  0.579     \\

2746.483  &a${\ }^4D_{3/2}$ - z${\ }^4F^o_{5/2}$& 0.544    & 0.613&  0.673     \\

2749.321  &a${\ }^4D_{5/2}$ - z${\ }^4F^o_{7/2}$&0.745     & 0.801&  0.848      \\

2755.736 &a${\ }^4D_{7/2}$ - z${\ }^4F^o_{9/2}$ & 1.000    & 1.000& 1.000      \\

\hline \\
 & multiplet \ a${\ }^4D$ - z${\ }^4D^o$ (63)&\\
2714.413  & a${\ }^4D_{7/2}$ - z${\ }^4D^o_{5/2}$&   0.132     &0.139  &  0.144 \\

2727.539 &a${\ }^4D_{5/2}$ - z${\ }^4D^o_{3/2}$  &  0.120    & 0.131 & 0.140  \\

2736.966  &a${\ }^4D_{1/2}$ - z${\ }^4D^o_{3/2}$ &  0.056    & 0.063 & 0.068  \\

2739.547  &a${\ }^4D_{7/2}$ - z${\ }^4D^o_{7/2}$ &  1.000    & 1.000 & 1.000  \\

2746.982  &a${\ }^4D_{5/2}$ - z${\ }^4D^o_{5/2}$ & 0.519     & 0.544 & 0.565  \\

2749.181  &a${\ }^4D_{3/2}$ - z${\ }^4D^o_{3/2}$ & 0.231     & 0.251 &  0.268 \\

2749.486  &a${\ }^4D_{1/2}$ - z${\ }^4D^o_{1/2}$ &  0.106    & 0.118 &  0.128 \\

2761.813  &a${\ }^4D_{1/2}$ - z${\ }^4D^o_{3/2}$ & 0.053     & 0.058 &  0.062 \\

2768.934   &a${\ }^4D_{3/2}$ - z${\ }^4D^o_{5/2}$&  0.022    &0.023  &  0.024  \\

2772.723   &a${\ }^4D_{5/2}$ - z${\ }^4D^o_{7/2}$&  0.001    &0.001  &  0.001  \\
\hline \\
 & multiplet \ a${\ }^4P$ - z${\ }^4P^o$ (78)&\\
2944.395   &a${\ }^4P_{3/2}$ - z${\ }^4P^o_{1/2}$ & 0.117   & 0.128 & 0.138     \\

2947.655   &a${\ }^4P_{5/2}$ - z${\ }^4P^o_{3/2}$ & 0.189   & 0.200 &  0.210     \\

2964.624   &a${\ }^4P_{1/2}$ - z${\ }^4P^o_{1/2}$ & 0.045   & 0.049 &  0.053     \\

2965.032  &a${\ }^4P_{3/2}$ - z${\ }^4P^o_{3/2}$  &  0.132  &0.140  &  0.146    \\

2984.825   &a${\ }^4P_{5/2}$ - z${\ }^4P^o_{5/2}$ & 1.000    &1.000  &   1.000    \\

2985.545   &a${\ }^4P_{1/2}$ - z${\ }^4P^o_{3/2}$ & 0.665   &0.705  &  0.739    \\

3002.644   &a${\ }^4P_{3/2}$ - z${\ }^4P^o_{5/2}$ & 0.620    &0.620  &  0.620    \\
\hline \\
 & 'I Zw 1 lines' - relative intensity of Gaussiaans &\\
2715 & 0.357  &  \\

2840 & 1  &   \\
\hline \\
\end{tabular}
\end{center}
\end{table*}

\clearpage

\section{Contribution of the 'fountain-like' region and the accuracy of the BH mass measurements}

 We analysed the different parameters of \ion{Mg}{ii} lines, trying to find some \ion{Mg}{ii} line properties which indicate dominant emission from the 'fountain-like' region  and some quantitative constraints of \ion{Mg}{ii} parameters for which \ion{Mg}{ii} line is  reliable (or not reliable) for the single epoch M$_{\rm BH}$ estimation. 

Previously, in Sec. \ref{sec:3.1}  we explored the correlation between the intrinsic shift (z$_{10\%}$) and the width (FW10\%M) that indicates virialization in the \ion{Mg}{ii} emitting region. As it can be seen in Fig. \ref{f02}b, the objects with the \ion{Mg}{ii} lines with approximately z$_{10\%}$ $<$ -800 km s$^{-1}$, or FW10\%M $>$ 12250 km s$^{-1}$, do not follow the linear correlation. In these cases, the 
emission of the 'fountain-like' region is probably dominant in the wings of the \ion{Mg}{ii}.
However, we could not find the strict constraint in the \ion{Mg}{ii} line parameters which would guarantee that the M$_{\rm BH}$ estimated using the \ion{Mg}{ii} FWHM is reliable, or opposite, that it is much different from the M$_{\rm BH}$ estimated using H$\beta$.

Additionally we explored the correlation between the contribution of the 'fountain-like' region to the  FWHM and FWH10\%M of \ion{Mg}{ii}. 

To estimate the contribution of the 'fountain-like' region, we normalized the intensities of H$\beta$ and \ion{Mg}{ii} to 1, then rescaled the \ion{Mg}{ii} to have the same FWHM as H$\beta$, and measured the flux of the difference between these two line profiles [F(\ion{Mg}{ii})-F(H$\beta$)].
 Roughly, it can be assumed that this difference is in correlation with the contribution of the 'fountain like' region to the \ion{Mg}{ii} line profiles. Taking 1$\sigma$ criteria, we divided our data in two subsamples:  one where the disagreement between the BH masses calculated using H$\beta$ and \ion{Mg}{ii} parameters are $<$ 1$\sigma$ ('good BH measurements'), and another where disagreement is $>$1 $\sigma$ ('bad BH measurements'). The 1 $\sigma$ = 0.19 for log[M$_{\rm BH}$(H$\beta$)/M$_{\rm BH}$(\ion{Mg}{ii})] distribution.
 
\begin{figure}

\includegraphics[width=0.45\textwidth]{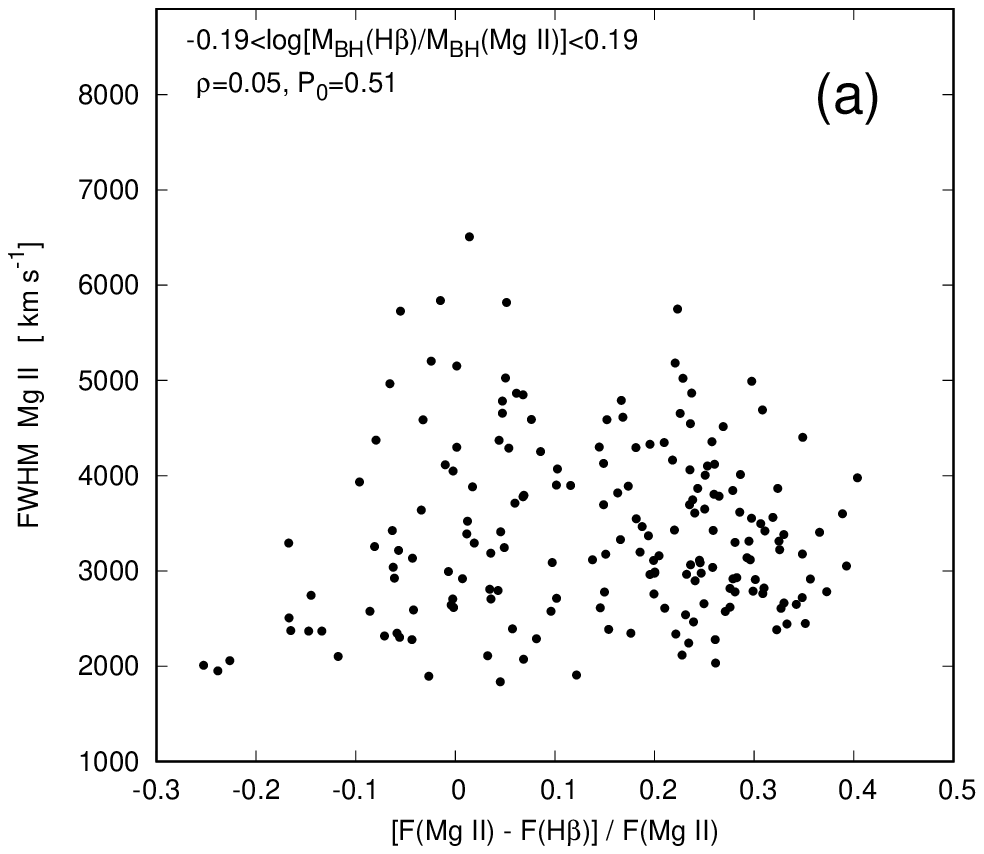}
\includegraphics[width=0.45\textwidth]{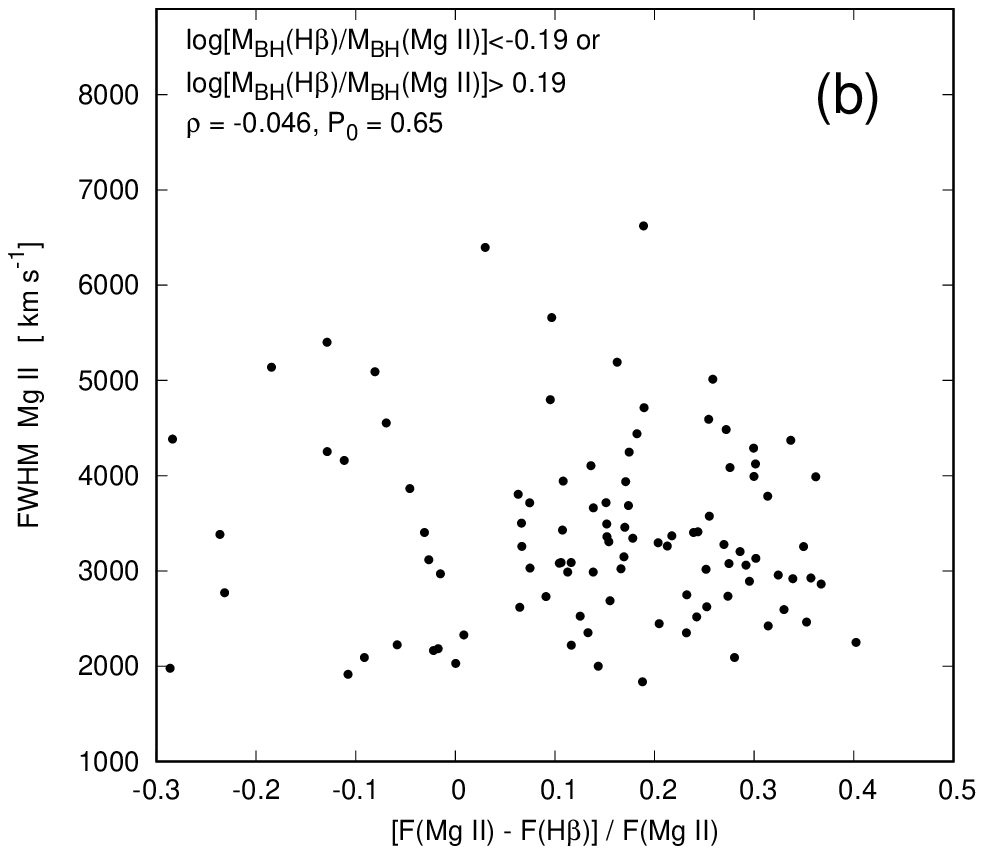}

\caption{ The relationship between the FWHM \ion{Mg}{ii} and [F(\ion{Mg}{ii})-F(H$\beta$)]/F(\ion{Mg}{ii}), where F(\ion{Mg}{ii}) and F(H$\beta$) are fluxes of \ion{Mg}{ii} and H$\beta$ normalized to 1, and fitted to have the same FWHM. Panel (a): for subsample with log[M$_{\rm BH}$(H$\beta$)/M$_{\rm BH}$(\ion{Mg}{ii})] smaller than 1 $\sigma$ ($\sigma$ = 0.19), and Panel (b): larger than 1 $\sigma$.}
\label{figA7}
\end{figure}

\begin{figure}

\includegraphics[width=0.45\textwidth]{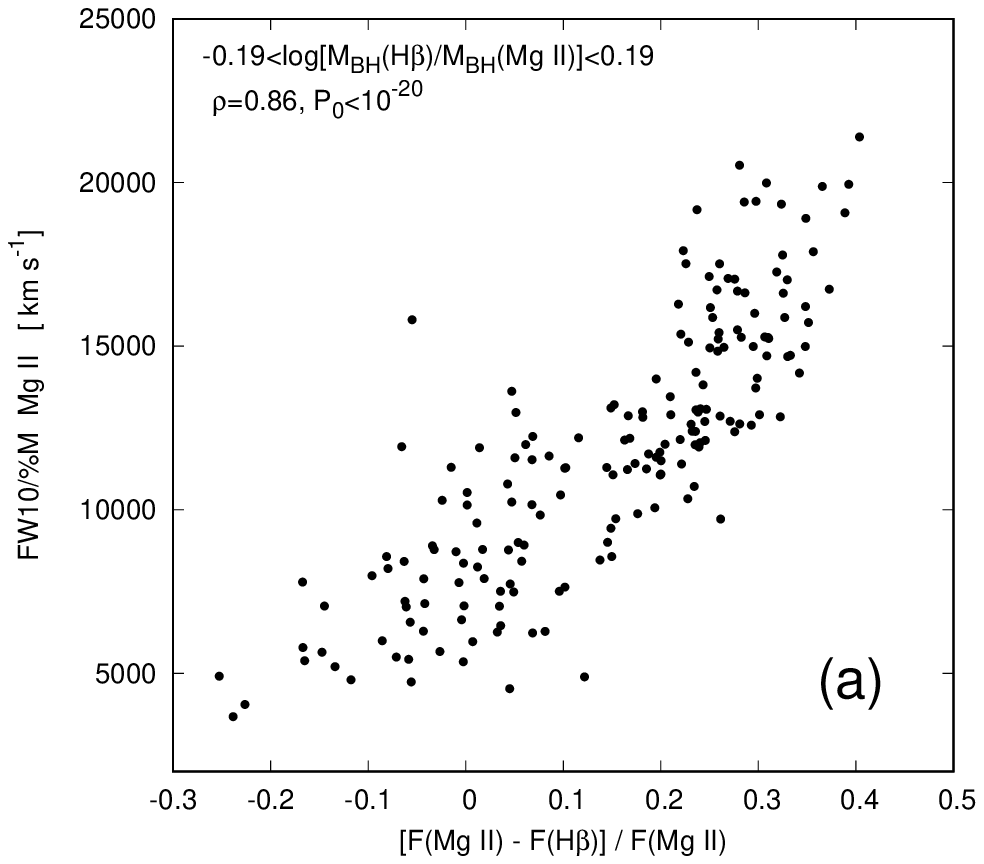}

\includegraphics[width=0.45\textwidth]{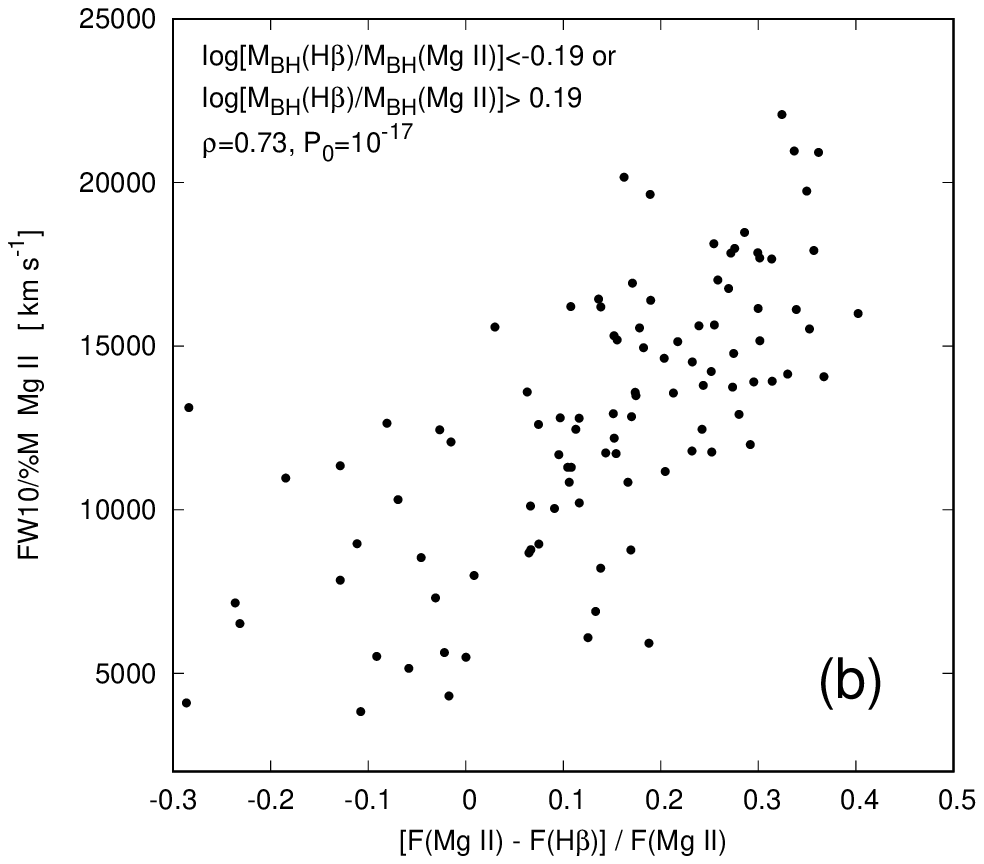}

\caption{ The same as in previous figure but for FW10\%M \ion{Mg}{ii} and the [F(\ion{Mg}{ii})-F(H$\beta$)]/F(\ion{Mg}{ii}).}
\label{figA8}
\end{figure}

 In Figs. \ref{figA7} and \ref{figA8} we present the FWHM and FW10\%M of \ion{Mg}{ii} as a function of the estimated 'fountain-like' emission\footnote{We should note here that negative contribution denotes that the H$\beta$ is broader in wings, and therefore the contribution of the 'fountain-like' is insignificant or contributes to the central part.}. As it can be seen in Fig. \ref{figA7}ab, there is no correlation between the FWHM \ion{Mg}{ii} and the 'fountain-like' contribution, in the both cases: in so-called 'good BH measurements' and 'bad BH measurements'. However, there is a high correlation 
 between FWH10\%M \ion{Mg}{ii} and 'fountain-like' contribution in both cases (see Fig. \ref{figA8}ab). For subsample of 'good BH measurements', the correlation is higher ($\rho$=0.86, $P_0<$10$^{-20}$), compared the one for 'bad BH measurements' ($\rho$=0.73, $P_0$=10$^{-17}$). In the case of 'good BH measurements', the correlation is higher probably because the 'fountain-like' region contributes more to the far line wings (FWH10\%M),  and do not affect the FWHM that is used for BH estimates. In the case of the 'bad BH measurements' the correlation is smaller and scattering of the points is higher, which may indicates that the 'fountain-like' contribution is likely to be more important in the line center, and therefore the mass (and FWHM) measurements may be partly affected by contribution of this region.

\bsp
\label{lastpage}

\end{document}